\documentclass[12pt]{article}

\input{epsf}
\usepackage[dvips]{graphicx}
\usepackage{amsmath}
\usepackage{cite}
\usepackage{color}
\usepackage{amsfonts}
\usepackage{amssymb}
\def\1ad{\mbox{\normalsize $^1$}}
\def\2ad{\mbox{\normalsize $^2$}}
\def\3ad{\mbox{\normalsize $^3$}}
\def\4ad{\mbox{\normalsize $^4$}}

\def\5ad{\mbox{\normalsize $^5$}}
\def\6ad{\mbox{\normalsize $^6$}}
\def\7ad{\mbox{\normalsize $^7$}}
\def\8ad{\mbox{\normalsize $^8$}}

\def\dj{\hbox{d\kern-0.347em \vrule width 0.3em height 1.252ex depth
-1.21ex \kern 0.051em}}

\def\rag{\rangle}
\def\lag{\langle}

\newcommand{\ra}{\rightarrow}

\newcommand{\be}{\begin{equation}}
\newcommand{\ee}{\end{equation}}
\newcommand{\ben}{\begin{equation*}}
\newcommand{\een}{\end{equation*}}
\newcommand{\ba}{\begin{eqnarray}}
\newcommand{\ea}{\end{eqnarray}}
\newcommand{\ban}{\begin{eqnarray*}}
\newcommand{\ean}{\end{eqnarray*}}
\newcommand{\brr}{\begin{array}}
\newcommand{\err}{\end{array}}
\newcommand{\bc}{\begin{center}}
\newcommand{\ec}{\end{center}}

\newcommand{\bea}{\begin{eqnarray}}
\newcommand{\eea}{\end{eqnarray}}
\newcommand{\bean}{\begin{eqnarray*}}
\newcommand{\eean}{\end{eqnarray*}}

\newcommand{\leqsim}{\,\raisebox{-0.6ex}{$\buildrel < \over \sim$}\,}

\newcommand{\bk}{{\mathbf k}}
\newcommand{\bh}{{\mathbf h}}
\newcommand{\bp}{{\mathbf p}}

\newcommand{\by}{{\mathbf y}}
\newcommand{\bx}{{\mathbf x}}
\newcommand{\br}{{\mathbf r}}

\newcommand{\bv}{{\mathbf v}}
\newcommand{\bq}{{\mathbf q}}

\newcommand{\al}{\alpha}

\newcommand{\de}{\delta}
\newcommand{\De}{\Delta}

\newcommand{\ga}{\gamma}

\newcommand{\la}{\lambda}
\newcommand{\Om}{\Omega}

\newcommand{\vph}{\varphi}

\newcommand{\dd}{\partial}
\newcommand{\mr}{\mathrm}

\newcommand{\xfin}{x_{\rm fin}}
\newcommand{\xin}{x_{\rm in}}
\newcommand{\vint}{v_{\rm int}}
\newcommand{\vout}{v_{\rm out}}
\newcommand{\vb}{v_{\rm b}}
\newcommand{\vf}{v_{\rm f}}
\newcommand{\rint}{r_{\rm int}}

\newcommand{\etafin}{\eta_{\rm fin}}
\newcommand{\etain}{\eta_{\rm in}}

\begin{document}

\setcounter{page}{0}
\thispagestyle{empty}

\begin{flushright}
CERN-PH-TH/2007-206\\
SACLAY-T07/142 
\end{flushright}

\vskip 8pt

\begin{center}
{\bf \Large {

Gravitational wave generation from bubble collisions in first-order
phase transitions: an analytic approach
}}
\end{center}

\vskip 10pt

\begin{center}
{\large Chiara Caprini$^{a,b}$, Ruth Durrer$^{a}$ and G\'eraldine
  Servant $^{b,c}$ }
\end{center}

\vskip 20pt

\begin{center}

$^{a}$ {\it D\'{e}partement de Physique Th\'{e}orique, Universit\'{e} de Gen\`{e}ve,}\\
{\it  24 Quai E. Ansermet, CH-1211 Gen\`{e}ve, Switzerland}
\vskip 3pt
\centerline{$^{b}${\it Service de Physique Th\'eorique, CEA Saclay, F91191 Gif--sur--Yvette,
France}}
\vskip 3pt
\centerline{$^{c}${\it CERN Physics Department, Theory Division, CH-1211 Geneva 23, Switzerland}}
\vskip .3cm
\centerline{\tt  chiara.caprini@cea.fr, Ruth.Durrer@physics.unige.ch, geraldine.servant@cern.ch}
\end{center}

\vskip 13pt

\begin{abstract}
\vskip 3pt
\noindent

Gravitational wave production from bubble collisions was calculated in
the early nineties using numerical simulations. In this paper, we
present an alternative analytic estimate, relying on a different
treatment of stochasticity. In our approach, we provide a model
for the bubble velocity power spectrum, suitable for both detonations
and deflagrations. From this, we derive the anisotropic stress and
analytically solve the gravitational wave equation. We provide
analytical formulae for the peak frequency and the shape of the
spectrum which we compare with numerical estimates.
In contrast to the previous analysis, we do not work in the envelope
approximation.
This paper focuses on a particular source of gravitational waves from
phase transitions. In a companion article, we will add
together the different sources of gravitational wave signals from
phase transitions: bubble collisions, turbulence and magnetic fields
and discuss the prospects for probing the electroweak phase transition
at LISA.

\end{abstract}

\vskip 13pt
\newpage

\section{Introduction}

In the next decades, a new science will emerge  from direct detection of gravitational radiation, that will open a qualitatively new way of probing the distant universe.
Ground-based (LIGO \cite{LIGOURL} and VIRGO \cite{VIRGOURL}) and space-based (LISA \cite{LISAURL}) interferometers will reach the  required sensitivity to detect many kinds of distant sources over a range of more than a million in frequency.
Because gravitational waves (GW) penetrate all regions of time and space, with almost no attenuation, GW detectors can explore scales, epochs and new physical effects not accessible in any other way.

Although the first GW detections will come from astrophysical processes, such as merging of black holes, another mission of GW astronomy will be to search for a stochastic background of  GWs of primordial origin.
An important mechanism for generating such a stochastic GW background is a relativistic first-order phase transition \cite{Witten:1984rs,Hogan}.
In a first-order phase transition, bubbles are nucleated, rapidly expand and collide.
The free energy contained in the original vacuum is released
and converted into thermal energy and kinetic energy of the bubble walls and the surrounding fluid. Most of the gravitational radiation comes from the final phase of the transition, from many-bubble collisions and the subsequent MHD turbulent cascades.
The associated GW spectrum encodes information on the temperature of the universe $T_*$ at which the waves were emitted as well as on the {\it strength} of the transition.
The characteristic frequency of the waves corresponds to the physics that produces them.  For cosmological processes, this is close to the Hubble frequency, $H \sim T_*^2/M_{Pl}$.
 Once redshifted to today, this corresponds to
 \be
f \sim 1 \ \mbox{mHz} \ \frac{T_*}{100\mbox{ GeV}} ~.
\ee
Remarkably, for transitions occuring near the electroweak epoch, $f$
is in the frequency range covered by LISA ($10^{-4}-10^{-2}$ Hz).
 It is therefore very exciting that LISA could help probing the nature
 of  the electroweak phase transition, and therefore provide
 information that is complementary to the Large Hadron Collider and
 the future International Linear Collider.

Many types of new physics predict  first-order phase transitions.
Electroweak symmetry breaking in extensions of the Standard Model may be associated with a first-order phase transition (see for example Ref.~\cite{Grojean:2004xa}).
Besides, the last decade has seen the emergence of the ``landscape picture", following developments in String Theory.
Strongly warped regions ({\it throats}) in higher-dimensional space-time are generic features in the string-theory landscape \cite{Hebecker:2006bn} and the phenomenological consequences are only starting to be explored (for instance through the prototype of Randall and Sundrum \cite{Randall:1999ee}). One interesting aspect is the cosmological evolution in these backgrounds. Thanks to holography and the AdS-CFT correspondence, a change in the 5-dimensional metric as the temperature decreases can be understood as a confining phase transition in the dual 4-dimensional gauge theory,
and in models like \cite{Randall:1999ee},
we typically expect  first-order phase transitions at the TeV scale \cite{Creminelli:2001th,Randall:2006py,Hassanain:2007js,Kaplan:2006yi,Nardini:2007me}.
Finally, phenomena such as preheating at the end of inflation could share some common features, as far as gravity wave emission is concerned, with the physics of first-order phase transitions
\cite{Chen:2006xjb,GarciaBellido:2007af,Dufaux:2007pt}.

The GW spectrum resulting from bubble collisions in first order phase transitions was computed
 in the early nineties~\cite{Kosowsky:1992rz,Kosowsky:1991ua,Kosowsky:1992vn,Kamionkowski:1993fg}.  It was realized ten years after the original calculation of~\cite{Kosowsky:1992rz,Kosowsky:1991ua,Kosowsky:1992vn,Kamionkowski:1993fg} that turbulence in the plasma could be a significant source of GW in addition to bubble collisions~\cite{Kosowsky:2001xp, Dolgov:2002ra}. Subsequently, the authors of~\cite{Apreda:2001us} studied the GW signal due to a first order electroweak phase transition in the Minimal Supersymmetric Standard Model (MSSM) and its NMSSM extension. More recently,
 model-independent analysis for the detectability of GW with LISA \cite{Nicolis:2003tg,Grojean:2006bp}, LIGO and BBO
  \cite{Grojean:2006bp} were presented, relying on the formulae derived in \cite{Kosowsky:1992rz,Kosowsky:1991ua,Kosowsky:1992vn,Kamionkowski:1993fg,Kosowsky:2001xp, Dolgov:2002ra}.
The spectrum derived in Ref.~\cite{Kosowsky:1992rz,Kosowsky:1991ua,Kosowsky:1992vn,Kamionkowski:1993fg}  was estimated  using numerical simulations, and no alternative calculation was performed afterwards.
As argued above, we believe this is of high interest and it is time to revisit this question.
 In this paper, we  present an analytical calculation of the stochastic GW background resulting from bubble collisions only\footnote{Turbulent fluid motions triggered by bubble collisions together with magnetic fields are actually additional relevant sources for gravity waves from phase transitions. These effects have been reexamined recently \cite{Caprini:2006jb,Gogoberidze:2007an,Kahniashvili:2008er} and since the subject is not closed,
we will present revisited results from
these contributions  elsewhere \cite{CDS2}.}. Since bubble collisions take place in a thermal bath, and since we want to extend our treatment to deflagrations, we use the energy-momentum tensor of the relativistic fluid in the vicinity of the bubble wall as the GW source, rather than the energy-momentum of the scalar field. The result we find is comparable to that obtained by numerical simulations although the peak frequency is parametrically larger.

A deterministic spherically symmetric expanding bubble does not produce gravitational radiation by itself. The reason is, that the transverse and traceless part of the energy momentum tensor for a radial deterministic distribution of the velocity field is identically zero (as we demonstrate in Appendix \ref{Appen:onesinglebubble}). To produce a non-zero background of GW, one has to account for the fact that, towards the end of the phase transition, the collision of bubbles breaks spherical symmetry and leads to a non-zero tensor anisotropic stress.
In the numerical simulations of Refs. \cite{Kosowsky:1991ua,Kosowsky:1992vn,Kamionkowski:1993fg}, this is accounted for by evaluating the transverse traceless component of an ``incomplete'' energy momentum tensor coming from the portion of bubble wall that remains uncollided at a given time. This energy momentum tensor is not spherically symmetric and has a non-zero tensor anisotropic stress component.
The total tensor anisotropic stress is obtained by summing  all the contributions from single uncollided bubble walls.
Each simulation  provides a given configuration of uncollided bubble walls; bubble nucleation and collision being random processes,  the GW power spectrum is obtained by averaging the results of several simulations. This procedure is valid under the thin-wall ``envelope'' approximation, {\it i.e.} when the transition is strong and the bubble front evolves as a detonation.

In the analytical evaluation which we present here, the situation is quite
different. The GW production comes not only from the bubble wall, but
from the entire fluid velocity profile in the vicinity of the phase
discontinuity. If the non-zero fluid velocity shell contracts to a
surface with vanishing thickness, no gravitational waves are produced.
Therefore, we are not working in the thin wall approximation, and
this is why we are able to apply our results also to the case of
deflagrations.

The  paper is organized as follows. In Section \ref{GW1} we review the general procedure for calculating the relic energy density stored in a stochastic background of gravitational waves.
Section \ref{section:Modelsource} describes our model of the GW source, the calculation of the bubble velocity power spectrum and the anisotropic stress power spectrum.
In Section \ref{timedep} we define the time dependence of the phase transition parameters. In Section~\ref{evaluation} the calculation of the GW spectrum, applicable both for detonations and deflagrations is presented. In Section \ref{section:comments}, we make some comments 
on our analytical approach.
In the last section we collect our final results and compare them with
the existing formulae used in in the literature. Some technical
aspects related to the calculation of the velocity power spectrum are
collected in Appendix \ref{Appen:velocity}. Appendix
\ref{Appen:largeandsmall} is a discussion on the behaviour of the
small and large scale tails of the GW power spectrum.  While the
existing literature provides approximate expressions for the peak
amplitude and peak frequency  of the signal, there is no justification
for the shape of the spectrum. Our analytical approach provides a
rationale for  it based on simple dimensional arguments.

\section{Gravitational wave power spectrum: general remarks}
\label{GW1}

Our goal is to estimate the gravitational wave energy density generated by bubbles during a first-order phase transition. This kind of cosmological source leads to a stochastic background of GW, which is isotropic, stationary, unpolarized and therefore characterized entirely by its frequency spectrum \cite{michele}. We consider a Friedmann universe with flat spatial sections.
The tensor metric perturbations are defined by
\be
 ds^2 = a^2[-d\eta^2 +(\de_{ij} +2h_{ij})dx^idx^j]~.
 \ee
The gravitational wave energy density is then given by
\be
{\rho_{GW}}(\eta)=\frac{\langle \dot{h}_{ij}(\bx) \dot{h}_{ij}(\bx) \rangle}{8\pi G a(\eta)^2}~.
\label{definition}
\ee
The over-dot denotes derivative with respect to conformal time and
$\langle ...\rangle$ denotes both 
time averaging over several periods of oscillation
and ensemble average for a stochastic background.
The variables ${\bf x }$ and later also $\bf r$ denote comoving
distances, $\eta$ and later $\tau$, $\zeta$ denote comoving time.
The density parameter is always scaled to today, $\Om_X (\eta)\equiv
\rho_X(\eta)/\rho_c(\eta_0)$, where the index $_0$ indicates the
present time. For relativistic species we have therefore
$\Om_X(\eta)=\Om_X(\eta_0)/a^4(\eta)$; we normalize $a(\eta_0)=1$ and
sometimes denote the present value of a density parameter simply by
$\Om_X(\eta_0) \equiv \Om_X$; likewise, $\rho_c=\rho_c(\eta_0)$.
$\mathcal{H}=\dot{a}/a$ denotes the conformal Hubble parameter.
The radiation energy density today is taken to be $\Om_{\rm
rad}(\eta_0)h^2\equiv \Om_{\rm rad}h^2 =4.2 \times 10^{-5}$~\cite{WMAP}.

We define the statistically homogeneous and isotropic gravitational wave energy density spectrum by
\be
\langle \dot{h}_{ij}(\bk,\eta)  \dot{h}_{ij}^*(\bq,\eta) \rangle=\delta(\bk-\bq)|\dot{h}|^2(k,\eta)~,
\label{GWpowspec}
\ee
 where ${\bf k }$ is the comoving wave vector.
The gravitational wave energy density, normalized to the critical energy density is:
\be
\Omega_{GW}(\eta) =\frac{\rho_{GW}(\eta)}{\rho_c}=\int_0^\infty {dk} 
\frac{k^2  |\dot{h}|^2(k,\eta) }{ 2 (2 \pi)^6 G \rho_c a^2} ~,
\ee
where the factor $(2\pi)^{-6}$ comes from the Fourier transform convention.
We want to estimate the present day gravitational wave energy spectrum, in other words the gravitational wave energy density per logarithmic frequency interval,
\be
\left. \frac{d \Omega_{GW} (k)}{d \ln k}\right|_{\eta_0}\equiv
\frac{k^3  |\dot{h}|^2(k,\eta_0) }{ 2 (2 \pi)^6 G \rho_c}~.
\label{GWenergy}
\ee
In an expanding radiation-dominated universe, ${h}_{ij}(\bk,\eta)$ is the solution of the wave equation
\be
\ddot{h}_{ij}(\bk,\eta)+\frac{2}{\eta}
\dot{h}_{ij}(\bk,\eta)+k^2h_{ij}(\bk,\eta)=8\pi G a^2(\eta)
\Pi_{ij}(\bk,\eta) ~.
\label{waveequation}
\ee
$\Pi_{ij}(\bk,\eta)$ is the tensor part of the anisotropic stress, the transverse-traceless component of the energy momentum tensor that generates tensor perturbations $h_{ij}$ of the metric:
\be
\Pi_{ij}(\bk,\eta)=(P_{il} P_{jm}  -\frac{1}{2} P_{ij}  P_{lm} ) T_{lm}(\bk,\eta)~,
\label{projected}
\ee
where $P_{ij}={\delta}_{ij}- \hat{k}_i \hat{k}_j$ is the transverse
projector and $T_{lm}(\bk,\eta)$ are the spatial components of the
energy momentum tensor. As will
be discussed in the next section, the anisotropic stress is a
stochastic variable for the generation process under consideration. It
accounts for the intrinsic randomness of bubble nucleation and collision.

Our source of gravitational radiation is active for an interval of time corresponding to the duration of the phase transition, which is much shorter than one Hubble time \cite{guth1,turner1}. We can therefore neglect the expansion of the universe while the source is still active, and rewrite Eq.~(\ref{waveequation}) as
\be
{h}_{ij}^{''}(x)+h_{ij}(x)=\frac{8\pi G a^2_*}{k^2} \Pi_{ij}(x) ~,
\label{reducedwe}
\ee
where $x=k\eta$, $'$ denotes derivative with respect to $x$ and $a_*$ is the scale factor at the time of the phase transition. The dependence of $h_{ij}(\bk,\eta)$ on directions of the wave-vector enters only in the polarization of the wave and is irrelevant for our discussion. As will become clear at the end of this section, in Eq.~(\ref{Pipowerspectrum}), this is due to statistical  homogeneity and isotropy of the source. We assume that the source turns on at time $\eta_{\rm in}$ and turns off at time $\eta_{\rm fin}$. The solution of (\ref{reducedwe}) is
\be
{h}_{ij}(x\le x_{\rm fin})=\frac{8\pi G a_*^2}{k^2} \int_{x_{\rm in}}^{x} dy \,  {\cal G}(x,y)\Pi_{ij}(y)~,
\ee
where $y=k\tau$ ($\tau$ denotes conformal time) and ${\cal G}=\sin(x-y)$ is the Green function satisfying ${\cal G}(x,x)=0$ and ${\cal G}'(x,x)=1$. Once the source is no longer active, we have to match the above solution with the solution of the free wave equation during radiation domination
\bea
& & {h}_{ij}^{''}(x)+\frac{2}{x}h_{ij}'(x)+h_{ij}(x)=0 \label{freeweeq}\\
& & h_{ij}(x>x_{\rm fin})=A_{ij}\frac{\sin(x-\xfin)}{x}+B_{ij}\frac{\cos(x-\xfin)}{x}~.
\label{freewe}
\eea
The matching procedure gives the coefficients
\bea
B_{ij}&=&\frac{8\pi Ga_*^2}{k^2}\xfin\int_{\xin}^{\xfin} dy \sin(\xfin-y)\Pi_{ij}(y)~, \nonumber\\
A_{ij}&=& \frac{B_{ij}}{\xfin}+ \frac{8\pi Ga_*^2}{k^2}\xfin\int_{\xin}^{\xfin} dy \cos(\xfin-y)\Pi_{ij}(y)~.
\eea
In order to simplify the equations, we neglect the first term in $A_{ij}$ which gives a subdominant contribution to the GW spectrum in the range of frequencies we are interested in.
In fact, this term contributes in a sizable way only for modes larger than the horizon,
\be
\xfin=k\eta_{\rm fin}\leq 1~, ~~k\leq 1/\eta_{\rm fin} \simeq \mathcal{H}_* ~,
\ee
where $\mathcal{H}_*$ denotes the conformal Hubble factor at the time of the phase transition, and is assumed to be constant from $\eta_{\rm in }$ to $\eta_{\rm fin}$ since the phase transition lasts for a time much shorter than one Hubble time. We will see that the GW spectrum grows very steeply at large scales (as $k^3$) and peaks at a scale corresponding to the maximal size of the bubbles, which is typically much smaller than the horizon. Therefore, we are mainly interested in the sub-horizon part of the spectrum and in order to evaluate it we can safely neglect the term ${B_{ij}}/{\xfin}$. Using definition (\ref{GWpowspec}) and solution (\ref{freewe}) we finally find ($z=k\zeta$)
\bea
|h'(k,x)|^2&=&\frac{1}{2x^2}\left(\lag A_{ij}A^*_{ij}\rag+\lag B_{ij}B^*_{ij}\rag\right)\nonumber\\
&=& \left(\frac{8\pi Ga_*^2}{k^2}\right)^2\frac{\xfin^2}{2x^2}
\int_{\xin}^{\xfin} dy \int_{\xin}^{\xfin} dz \cos(z-y)\Pi(k,y,z)
\label{h'2}
\eea
In the double integral above, we have combined the products of two Green's functions into the simpler term $\cos(z-y)$. Moreover, we have introduced the unequal time correlator of the tensor anisotropic stress in Fourier space,
\be
\lag \Pi_{ij}(\bk,\tau)\Pi^*_{ij}(\bq,\zeta)\rag =
\de(\bk-\bq)\Pi(k,k\tau,k\zeta)~.
\label{Pipowerspectrum}
\ee	
The delta function is due to the statistical homogeneity of the source, and because of statistical isotropy the power spectrum of the anisotropic stress only depends on the wave number. Note that for the matching we have used the free wave propagation equation (\ref{freeweeq}), which is valid in an expanding, radiation dominated universe with $a(\eta)\propto\eta$. Hence, solution (\ref{h'2}) for $\eta>\eta_*$ implicitly assumes that the number of relativistic degrees of freedom is constant. We come back to this issue in section \ref{evaluation}.

To summarize, in order to determine the spectrum of the gravitational
radiation Eq.~(\ref{GWenergy}), we  have to calculate the power
spectrum of the anisotropic stress evaluated at different times.  This
requires computing the correlator of the energy momentum tensor.
The next section is devoted to a calculation of $\Pi(k,y,z)$
(Eq.~\ref{Pipowerspectrum}). For this we need a model of the
energy momentum tensor that sources the gravitational waves.

\section{Model of the GW source}
\label{section:Modelsource}

We now develop a model for the stochastic source of gravitational radiation. We are dealing with a cosmological first order phase transition taking place in a thermal bath \cite{Steinhardt,Kamionkowski:1993fg}. The cosmic fluid  of the initial metastable  phase supercools
until the nucleation of bubbles of the final phase can start.
The initial high-temperature phase or false vacuum is typically but not necessarily the symmetric phase. However, in the remaining of the paper, we will use the term ``symmetric'' for the initial phase and ``broken'' for the final phase.
The phase transition ends when the entire universe has been converted to the broken phase by bubble percolation. We are only interested in the last stages of bubble growth. Towards the end of the phase transition, the bubbles can be considered simply as spherical combustion fronts moving at constant velocity \cite{Ignatius}. Any memory of the initial shape of the bubbles, driven by the scalar field dynamics, is lost and the problem can be reduced to a purely hydrodynamical description. The bubbles are modeled as spherically symmetric configurations of fluid velocity. The velocity field is a stochastic variable, following the intrinsic stochasticity of the nucleation process.

\subsection{Anisotropic stress power spectrum: general remarks}

 Since we are interested only in the anisotropic stress, we start with the spatial, off-diagonal part of the energy momentum tensor of the cosmic fluid, quantifying the spatial components of the kinetic
 stress-energy tensor of a bubble configuration \cite{Kamionkowski:1993fg}:
\be
T_{ab}(\bx,\tau)=(\rho+p)\frac{v_a(\bx,\tau)v_b(\bx, \tau)}{1-v^2(\bx,\tau)} ~.
\label{Tab}
\ee
$\bf v$ is the velocity of the fluid in the frame of the bubble center, and $v=||{\bf v}||$. 
We want to calculate the anisotropic stress power spectrum given in Eq.~(\ref{Pipowerspectrum}). In order to simplify the calculation, we neglect the spatial dependence of the fluid enthalpy density $w=\rho+p$ and of the gamma factor $\gamma^2=1/(1-v^2)$. This assumption is necessary in order to be able to proceed analytically. It supposes that the only stochastic variables in the problem are the fluid velocity components $v_a(\bx,\tau)$, and that the spatially dependent $\gamma$ factor can be approximated by $\ga(\bx) \simeq \langle\ga\rangle \equiv \ga$. 
The consequences of this assumption cannot be quantified exactly. 
However, we know that $\langle v^2\rangle$
 varies smoothly from $v_f^2 (r_{\rm int}/R)^2$ to $v_f^2$ (see Eq.~\ref{lab:correlation}) where $r_{\rm int}$ and $R$ are defined in Eq.~\ref{velocity}. A
 conservative choice is to always set $\langle v^2\rangle$ to its smallest value,
 and this is what we will do in Eqs.~(\ref{rhokin}) and (\ref{Omkin}). Under these assumptions, we can write the Fourier transform,
\be
T_{ab}(\bk,\tau)=\frac{w(\tau)}{1-v^2(\tau)}\int d^3p \, v_a(\bk-\bp,\tau)v_b(\bp,\tau) ~.
\ee
With this expression, the power spectrum of the energy momentum tensor involves the four-point function of the velocity distribution:
\bea
\lefteqn{\lag T_{ab}(\bk,\tau) T^*_{cd}(\bq,\zeta) \rag =}\nonumber \\
 & &\frac{w(\tau)w(\zeta)}{(1-v^2(\tau))(1-v^2(\zeta))}
\int d^3p \int d^3h \lag v_a(\bk-\bp, \tau)v_b(\bp,\tau)v_c(\bq-\bh,\zeta)
v_d(\bh,\zeta)\rag~.
\label{Tpowspec}
\eea
There is in principle no reason why our stochastic velocity field should have a Gaussian distribution. However, we have to make some assumptions in order
to calculate analytically the four-point function in the above expression. As one
often does, we assume that Wick's theorem, which is strictly valid
only for Gaussian random variables, gives a good enough
approximation to the four-point function. It certainly gives a better
estimate than, for example, the simple product of expectation
values. Applying it we find
\bea
\lefteqn{
\langle T_{ab}(\bk,\tau)  T^*_{cd}(\bq,\zeta) \rangle=\frac{w(\tau)w(\zeta)}{(1-v^2(\tau))(1-v^2(\zeta))}
\delta(\bk-\bq) } \nonumber \\
& &\times \int d^3p \left[ \hat{C}_{ac}(p,\tau,\zeta)  \hat{C}_{bd}(|\bk-\bp|,\tau,\zeta) +
\hat{C}_{ad}(p,\tau,\zeta)  \hat{C}_{bc}(|\bk-\bp|,\tau,\zeta) \right]~,
\eea
where
\be
\hat{C}_{ac}(p,\tau,\zeta) =\int d^3r C_{ac} ({\bf
  r},\tau,\zeta)e^{i\bp\cdot\br} \ \ \ \ , \ \ \ {C}_{ac}({\bf
  r},\tau,\zeta)=\langle v_a(\bx,\tau) v_c(\bx+\br,\zeta) \rangle~.
\label{Cac}
\ee
 The correlation between point $\bx$ and point $\by=\bx+\br$ is a
 function of $\br$ only because of statistical homogeneity.

In our approach, the ensemble average in Eq.~(\ref{definition}) is now
traced back into a correlator for the bubble velocities. This is where
the stochasticity of the process is encoded. Since the velocity field
is statistically homogeneous and  isotropic, its power spectrum has
the general form (see the next subsection 
\ref{velocity_power})
\be
\hat{C}_{ac}(p,\tau,\zeta) = F(p,\tau,\zeta) \delta_{ac} + G(p,\tau,\zeta) \hat{p}_a \hat{p}_c~.
\label{Cachom}
\ee
The power spectrum of the tensor part of the anisotropic stress is calculated using the
definition~(\ref{Pipowerspectrum}) by applying the transverse traceless
projector as in Eq.~(\ref{projected}):
\bea
& &\lag \Pi_{ij}(\bk, \tau)\Pi^*_{ij}(\bq,\zeta)\rag=\mathcal{P}_{abcd} \lag T_{ab}(\bk,\tau) T^*_{cd}(\bq,\zeta) \rag \nonumber \\
& &\mathcal{P}_{abcd}=\left(P_{ia} P_{jb}  -\frac{1}{2} P_{ij}  P_{ab} \right)(\bk)\left(P_{ic} P_{jd}  -\frac{1}{2} P_{ij}  P_{cd}\right)(\bq)
\eea
A somewhat lengthy calculation yields
\bea
\Pi(k,\tau,\zeta)&=&\frac{w(\tau)w(\zeta)}{(1-v^2(\tau))(1-v^2(\zeta))}
\int d^3p \ [  4 F(p)  F(|\bk-\bp|) + 2 (1-\beta^2) F(p)  G(|\bk-\bp|)
  \nonumber \\
&+&2(1-\lambda^2) G(p)  F(|\bk-\bp|)  +(1-\lambda^2)(1-\beta^2) G(p)
  G(|\bk-\bp|) ]
 \label{Pik}
 \eea
where $\lambda=\hat{k}\cdot\hat{p}$ and
$\beta=\hat{k}\cdot\widehat{k-p}$ and we have suppressed the time
variables $\tau$ and $\zeta$ in $F$ and $G$.

The problem is now reduced to the determination of the functions $F$
and $G$ which define the power spectrum of the fluid velocity via
Eqs.~(\ref{Cac},\ref{Cachom}). For this, we need a model of the
fluid velocity which we discuss next.

\subsection{Velocity profile of bubbles}
\label{velocity_profile}
Since we are only interested in the last stage of the phase transition, we consider the hydrodynamics of bubble growth at late times, when a steady state solution is reached. The bubble wall, in other words the combustion front where the phase transition is happening, is moving at constant velocity. In the hydrodynamical description of the combustion, the front is treated as a surface of discontinuity. Energy and momentum must be conserved across the front and all the entropy production is confined to it \cite{Steinhardt,Ignatius}. Elsewhere, the fluid is in a state of thermal equilibrium. The energy momentum tensor of the burnt (broken) and unburnt (symmetric) phases is simply that of two perfect fluids (see Eq.~\ref{Tab}). There are two kinds of solutions to this hydrodynamical problem, detonations and deflagrations. These are classified following the characteristics of the fluid flow in the rest frame of the combustion front \cite{Landau,Courant}.

\begin{figure}[htb!]
\begin{center}
\includegraphics[height=9cm,width=9cm]{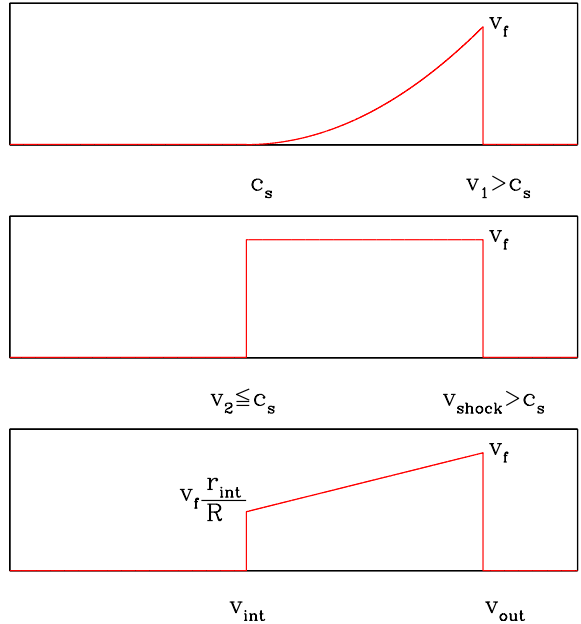}
\caption{\label{fig1}This figure shows the qualitative profile of the velocity of the broken phase fluid in the frame of the bubble center, for detonations (top panel), planar deflagrations (middle panel) and  the approximation given in Eq.~(\ref{velocity}) (bottom panel). The horizontal axis shows $r/t$ where $t$ denotes the time after bubble nucleation ($t=0$) and $r$ is the distance from the bubble center.}
\end{center}
\end{figure}

In detonations, the incoming velocity of the symmetric phase fluid into the front is supersonic $v_1>c_s$  in the rest frame of the front. The outgoing velocity of the broken phase fluid out of the front can be supersonic $v_2>c_s$ for weak detonations, or equal to the speed of sound for Jouguet detonations $v_2=c_s$. The case of strong detonations $v_2<c_s$ is forbidden \cite{Landau}. Although weak detonations are possible \cite{Laine}, in the following we concentrate for simplicity on the case of Jouguet detonations. This is the case analyzed in \cite{Kamionkowski:1993fg}, for which the dynamics of the bubble growth is completely determined in terms of the phase transition strength.
Since both fluid phases are relativistic, they have the sound speed $c_s=1/\sqrt{3}$.
In the rest frame of the bubble center, the velocity of the bubble front $\vb$ is supersonic, corresponding to $\vb=v_1>c_s$: the symmetric phase fluid is therefore at rest, and the front is followed by a rarefaction wave in the broken phase fluid. The rarefaction wave brings the fluid motion to rest towards the center of the bubble. Near the detonation front, the broken phase fluid velocity $\vf$ in the rest frame of the center of the bubble is simply given by the Lorentz transformation
\be
\vf=\frac{v_1-v_2}{1-v_1v_2}
\label{vf}
\ee
The velocity profile of the broken phase fluid for a Jouguet
detonation has been studied in detail in
Refs.~\cite{Landau,Steinhardt,Kamionkowski:1993fg} and is shown
schematically in the first panel of Fig.~\ref{fig1}. As customary, we
show the velocity profile as a function of the parameter $r/t$. Here
$t=0$ is the time of bubble nucleation and $r$ denotes the distance
form the bubble center\footnote{Throughout this section, for
  simplicity we use a generic time variable $t$; we switch back to
  comoving time in paragraph \ref{subsec:unequal}.}. We remind that
this situation corresponds to the steady state solution at late times,
long after the nucleation time. The velocity of the broken phase fluid
goes to zero in the interior of the bubble at a distance from the
center corresponding to $c_s t$ \cite{Landau}.

 For deflagrations, on the other hand, the incoming velocity of the symmetric phase fluid into the front is subsonic  $v_1<c_s$. The outgoing velocity of the broken phase fluid out of the front can be subsonic $v_2<c_s$ for weak deflagrations, or equal to the speed of sound for Jouguet deflagrations $v_2=c_s$. Strong deflagrations are again impossible \cite{Landau}. In the rest frame of the bubble center, the velocity of the bubble front corresponds in this case to $\vb=v_2$. The front moves at subsonic velocity $\vb=v_2\leq c_s$ and is therefore preceded in the symmetric phase by a shock wave. Inside the combustion bubble (broken phase) and outside the shock wave (symmetric phase) the fluid is at rest; in between, the fluid moves outwards. In the case of planar deflagrations, it does so at constant velocity given by $\vf$ (Eq.~\ref{vf}) \cite{Landau,Ignatius}.
The qualitative features of the velocity profile for a planar deflagration are shown in the second panel of Fig.~\ref{fig1} \footnote{In Appendix A of Ref.~\cite{Kamionkowski:1993fg} it is shown that, in the case of spherical deflagrations, the velocity profile actually decreases between $v_2$ and $v_{\rm shock}$. We do not account for this behaviour here, since in any case we are forced to introduce an approximate form for the velocity profile.}.

For our analytic calculation, we want to simplify the real velocity profile both for detonations and deflagrations. We assume a velocity profile which grows linearly within a shell near the bubble wall, as shown in the last panel of Fig.~\ref{fig1}. We have normalized the velocity profile at the outer boundary (bubble wall or shock front) to the correct value $\vf$ for detonations and deflagrations. This is because the biggest contribution to the GW energy density comes from the highest velocity region. Therefore, our approximated profile does reproduce the most relevant feature as far as GW generation is concerned. The boundaries of the shell, defined as $\vint$ and $\vout$ in Fig.~\ref{fig1} are left as free parameters, in order to allow for an approximated description of both detonations and deflagrations.

\begin{figure}[htb!]
\begin{center}
\includegraphics[height=6cm,width=6cm]{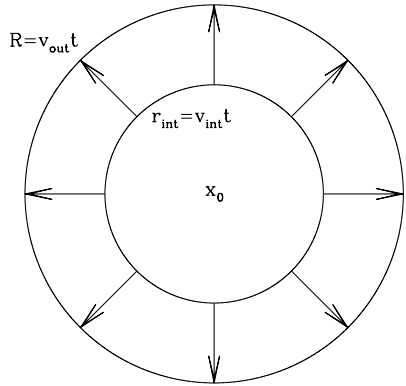}
\caption{\label{bul} A schematic drawing of the non-zero velocity region, corresponding to the bubble (for detonations) or to the shock front (for deflagrations).}
\end{center}
\end{figure}

The simplified profile is
\bea
v_a(\bx,t)=\left\{\begin{array}{ll}
(\vf/R)\,(\bx-\bx_0)_a& {\rm for~~}~r_{\rm int}=\vint t<|\bx-\bx_0|<R=\vout t ~,\\
0 & {\rm otherwise.} \end{array}\right.
\label{velocity}
\eea
Here $\bx_0$ is the position of the bubble center. The radii of the shell's  inner and outer boundaries are respectively  $r_{\rm int}=\vint t$ and $R=\vout t$, where $t$ is much later than the nucleation time $t=0$. In the case of Jouguet detonations, the inner boundary is in the broken phase and corresponds to $\vint=c_s$. For deflagrations, it corresponds to the bubble wall,
$\vint=v_2=\vb$. The radius of the outer boundary is $R=\vout t$ and for detonations it is associated to the bubble wall velocity  $\vout=v_1=\vb$, while for deflagrations  to the shock front $\vout=v_{\rm shock}$.
To summarize:
\begin{center}
\begin{tabular}{p{4cm}p{1cm}p{4cm}}
{Detonations} && {Deflagrations} \\
$\vint=c_s$ & & $\vint=v_2=\vb$ \\
$\vout=v_1=\vb ~.$ &&$\vout=v_{\rm shock} ~.$
\end{tabular}
\end{center}
The numerical values of $\vint$, $\vout$ and $\vf$ will be crucial in
determining the amplitude of the GW signal and will be discussed in
Sections \ref{subsection:DETO} and \ref{GWdef}. A schematic drawing of
the bubble (or shock front) is given in Fig.~\ref{bul}.

\subsubsection{Velocity power spectrum}
\label{velocity_power}

\begin{figure}[htb!]
\begin{center}
\includegraphics[height=5cm]{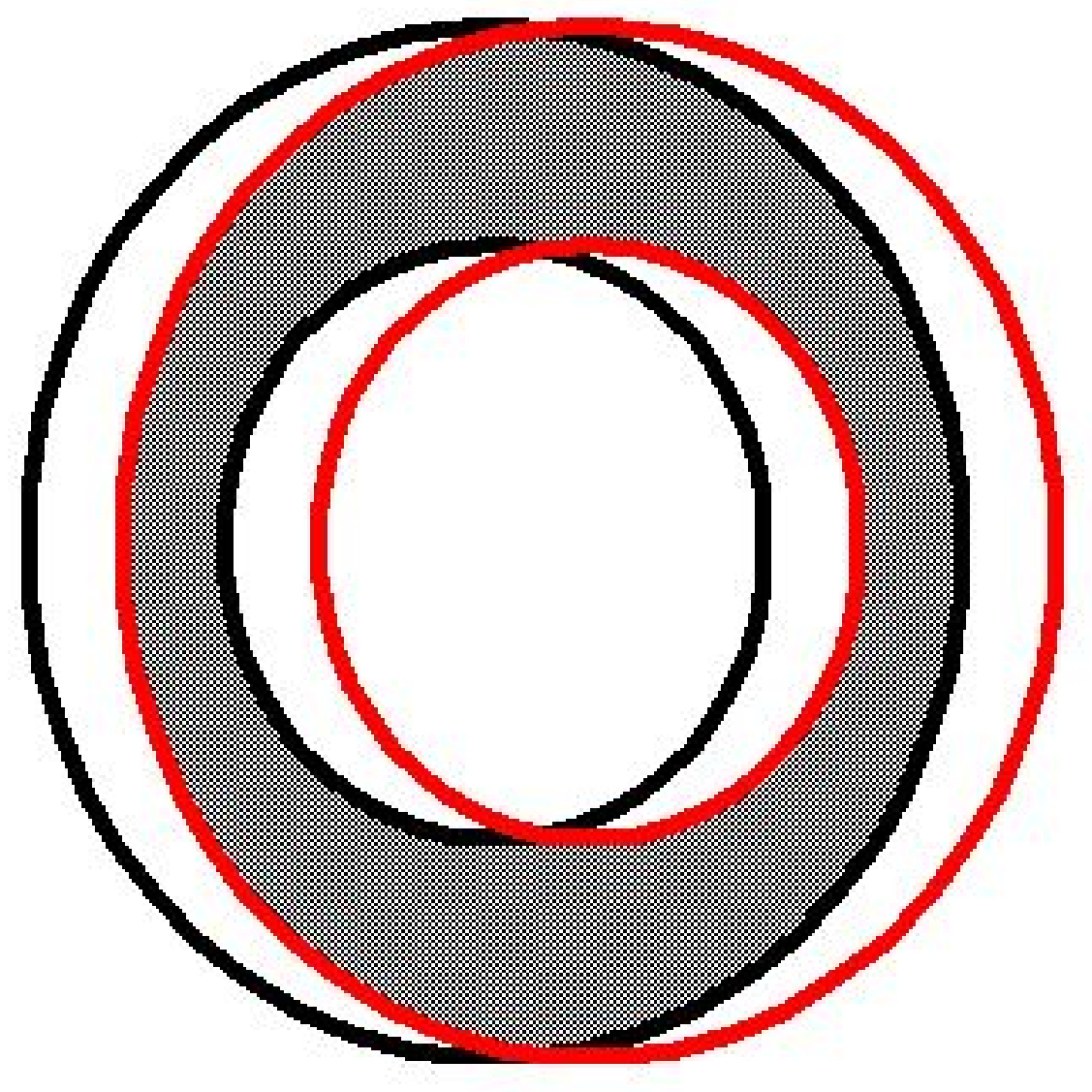}
\includegraphics[height=5cm]{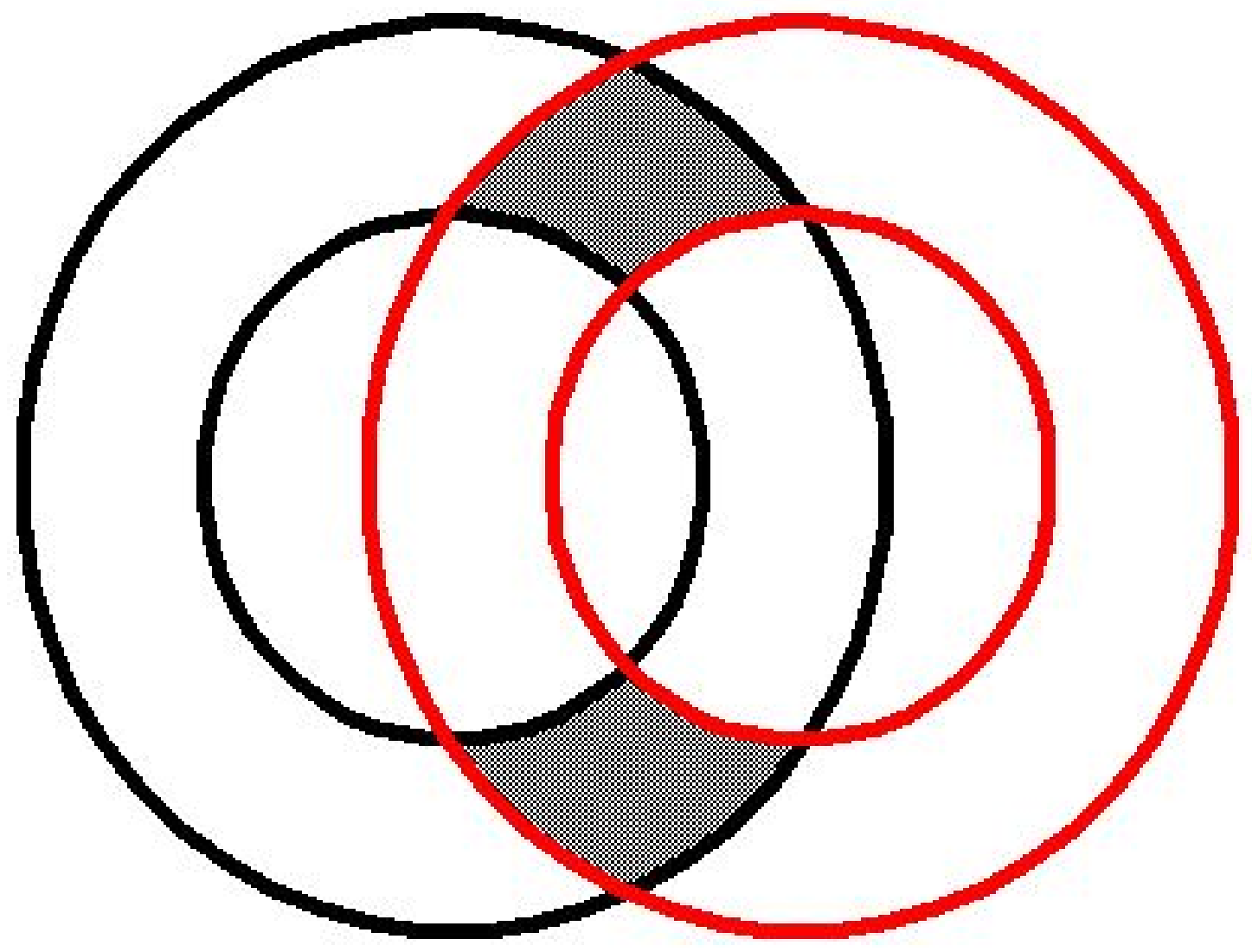}
\includegraphics[height=5cm]{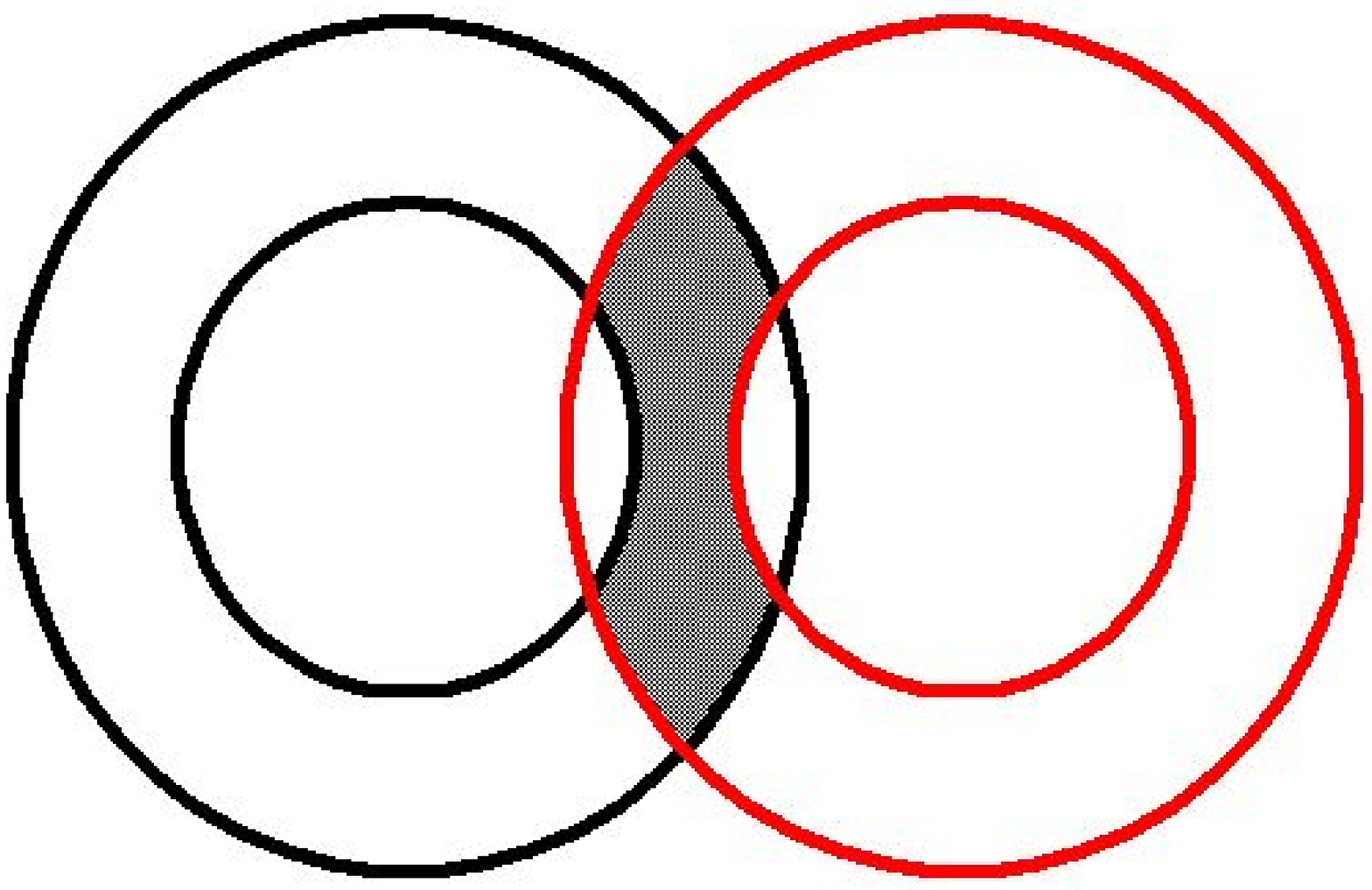}
\includegraphics[height=5cm]{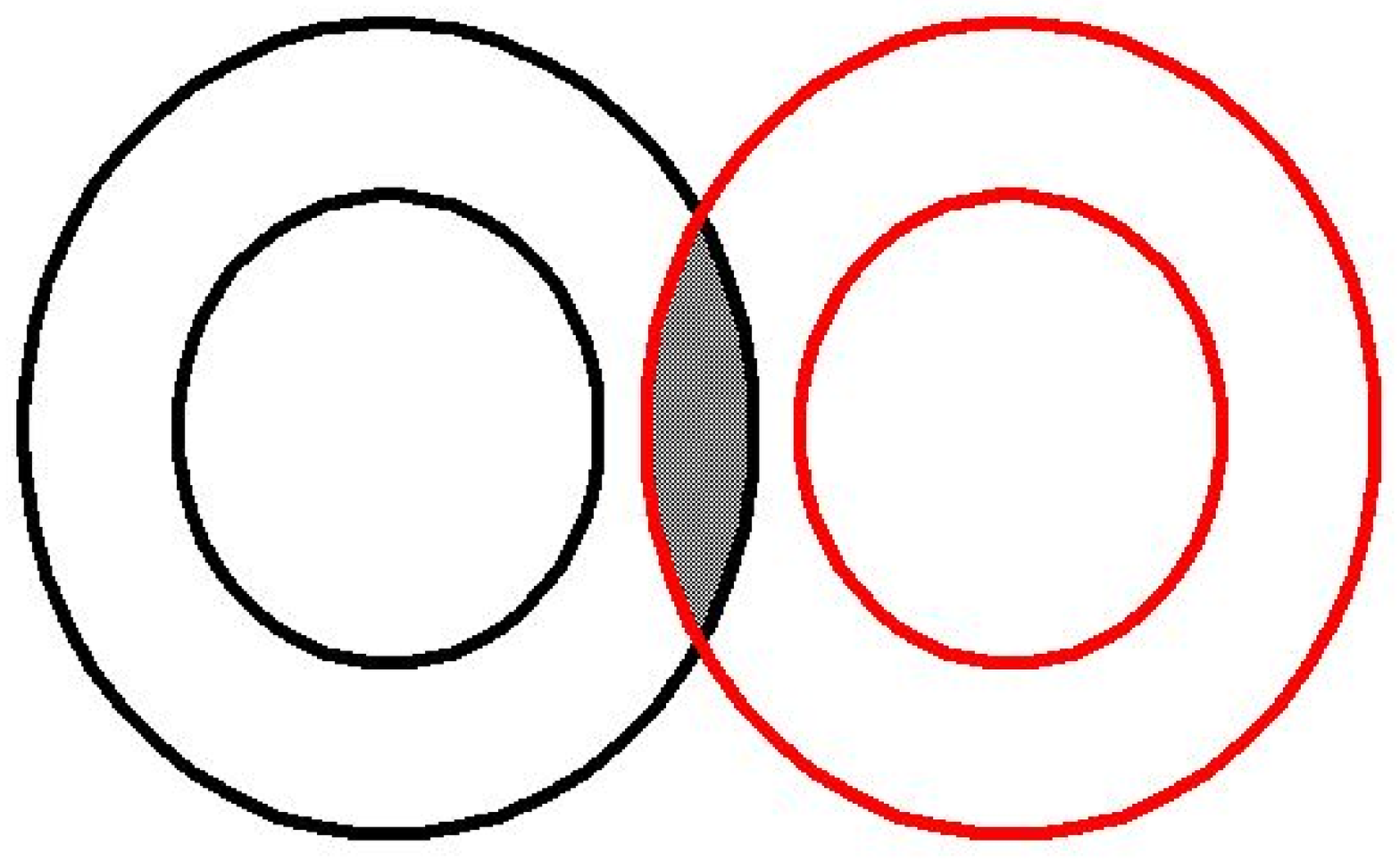}
\caption{\label{fig2}
This figure shows how the intersection volume $V_i$ changes as a
function of the separation between $\bx$ and $\by$, $r=|\bx-\by|$,
where $\bx$ and $\by$ are located at the centers of the shells. The
upper left, upper right, lower left and  lower right plots
respectively correspond to $0\leq r \leq R-\rint$,
$R-\rint \leq r \leq 2\rint$, $2\rint<r<R+\rint$ and  $R+\rint\leq
r\leq 2R$. Therefore, this figure does not depict bubble collision (in
our approach we do not actually collide bubbles). The shaded volume
does not represent the volume of intersection between two different
shells, but it accounts for all possible positions of the center of the
bubble to which two given points $\bx$ and $\by$ belong.}
\end{center}
\end{figure}

Given the velocity profile, we can proceed to calculate the velocity power spectrum. We start by evaluating the two-point correlation function at equal time for fixed positions $\bx$ and $\by$ defined in Eq.~(\ref{Cac}). The position of the bubble center $\bx_0$ defined in Eq.~(\ref{velocity}) is the stochastic variable. Therefore, in the region of non-zero velocity we have:
\be
\lag v_i(\bx,t)v_j(\by,t) \rag=\frac{\vf^2}{R^2}\lag (\bx-\bx_0)_i (\by-\bx_0)_j\rag
\label{lab:correlation}
\ee
where we remind that $\vf$ is the maximal value of the fluid velocity in the rest frame of the bubble center.
For the velocity correlation function not to be zero, $\bx$ and $\by$ must be separated by a distance $|\bx-\by|<2R$ and they have to be in the same bubble (or shock wave, in the case of deflagrations). Moreover,
they have to be in the shell where the fluid velocity (\ref{velocity}) is not zero. These conditions are satisfied provided that the center $\bx_0$ of the bubble they belong to is in a volume $V_i$ given by the intersection of two shells centered in $\bx$ and $\by$, which have inner radius $\rint$ and outer radius $R$ (see Fig.~\ref{fig2}). Therefore, the correlation function is given by the mean over all the possible center positions $\bx_0$ within this intersection volume, multiplied by the probability $\sigma(t)$ that there actually is a bubble center in this region:
\be
\langle v_{i}(\bx,t)  v_{j}(\by,t) \rangle=\sigma(t)\,\frac{\vf^2}{R^2}\,\frac{1}{V_i}\int_{V_i}d^3x_0 (\bx-\bx_0)_i (\by-\bx_0)_j~.
\label{correlationintegral}
\ee
As customary in cosmology, here we use the ergodic assumption: ensemble averages are equivalent to space averages. The probability of having a bubble center in the intersection region is simply given by
\be
\sigma(t)=\phi(t)\frac{V_i}{V_c}
\label{prob}
\ee
where $\phi(t)$ is the fraction of volume occupied by bubbles at time $t$, and $V_c$ is the volume of the region where $\bx_0$ can be, in order for $\bx$ or $\by$ to be in the same bubble: the total volume of the two overlapping spheres in Fig.~\ref{fig2}. Setting $\br=\bx-\by$ we find
\be
V_c=\frac{2\pi}{3}\left(2R^3+\frac{3}{2}R^2r-\frac{r^3}{8}\right)~.
\label{Vc}
\ee

The tensorial structure of the two-point correlation function of a
statistically homogeneous and isotropic field is known: the
correlation function can only depend on the distance between $\bx$ and
$\by$. We choose an orthonormal basis with $\widehat{\bx-\by}
\parallel \hat{\bf e}_2$. The off-diagonal components of the integral in
Eq.~(\ref{correlationintegral}) are zero by symmetry, and therefore we
find
\bea
I_{ij}(r,R,\rint)&=&\int_{V_i}d^3x_0 (\bx-\bx_0)_i (\by-\bx_0)_j\\
\langle v_{i}(\bx,t)  v_{j}(\by,t) \rangle&=&\phi(t)\,\frac{\vf^2}{R^2}\,\frac{1}{V_c}\,I_{ij}(r,R,\rint)
\label{vcorrfunct}\\
I_{ij}(r,R,\rint)&=&I_{11}\,\de_{ij}+(I_{22}-I_{11})\,\hat{r}_i\hat{r}_j
\eea
(since $I_{11}=I_{33}$). The functions $I_{11}$ and $I_{22}$ have to
be calculated by performing the necessary integration in the four
volume regions $V_i$ shown in Fig.~\ref{fig2}. The details of the
calculations are given in Appendix B.

The velocity power spectrum is then obtained by Fourier transforming the two point correlation function
(\ref{vcorrfunct}) with respect to the variable $\br$. Remembering the
definitions~(\ref{Cac},\ref{Cachom}) one finds the general expressions
($\mathcal{F}$ denotes the Fourier transform)
\bea
\langle v_{i}(\bk,t)  v_{j}^*(\bq,t) \rangle&=&
\de(\bk-\bq) \hat{C}_{ij}(k,t)=\de(\bk-\bq)[F(k,t)\de_{ij}+G(k,t)\hat{k}_i\hat{k}_j] \label{powv}\\
F(k,t)&=&\phi(t)\frac{\vf^2}{R^2}\left[ \mathcal{F}\left(\frac{I_{11}}{V_c}\right)-
\frac{1}{k}\frac{d}{dk}\mathcal{F}\left(\frac{I_{22}-I_{11}}{r^2V_c}\right)\right]\\
G(k,t)&=&\phi(t)\frac{\vf^2}{R^2}\left[\frac{1}{k}\frac{d}{dk}\mathcal{F}\left(\frac{I_{22}-I_{11}}{r^2V_c}\right)
-\frac{d^2}{dk^2}\mathcal{F}\left(\frac{I_{22}-I_{11}}{r^2V_c}\right)\right]~,
\eea
where we remind that $\phi(t)$ is defined below Eq.~(\ref{prob}).

\begin{figure}[htb!]
\begin{center}
\includegraphics[height=6.75cm]{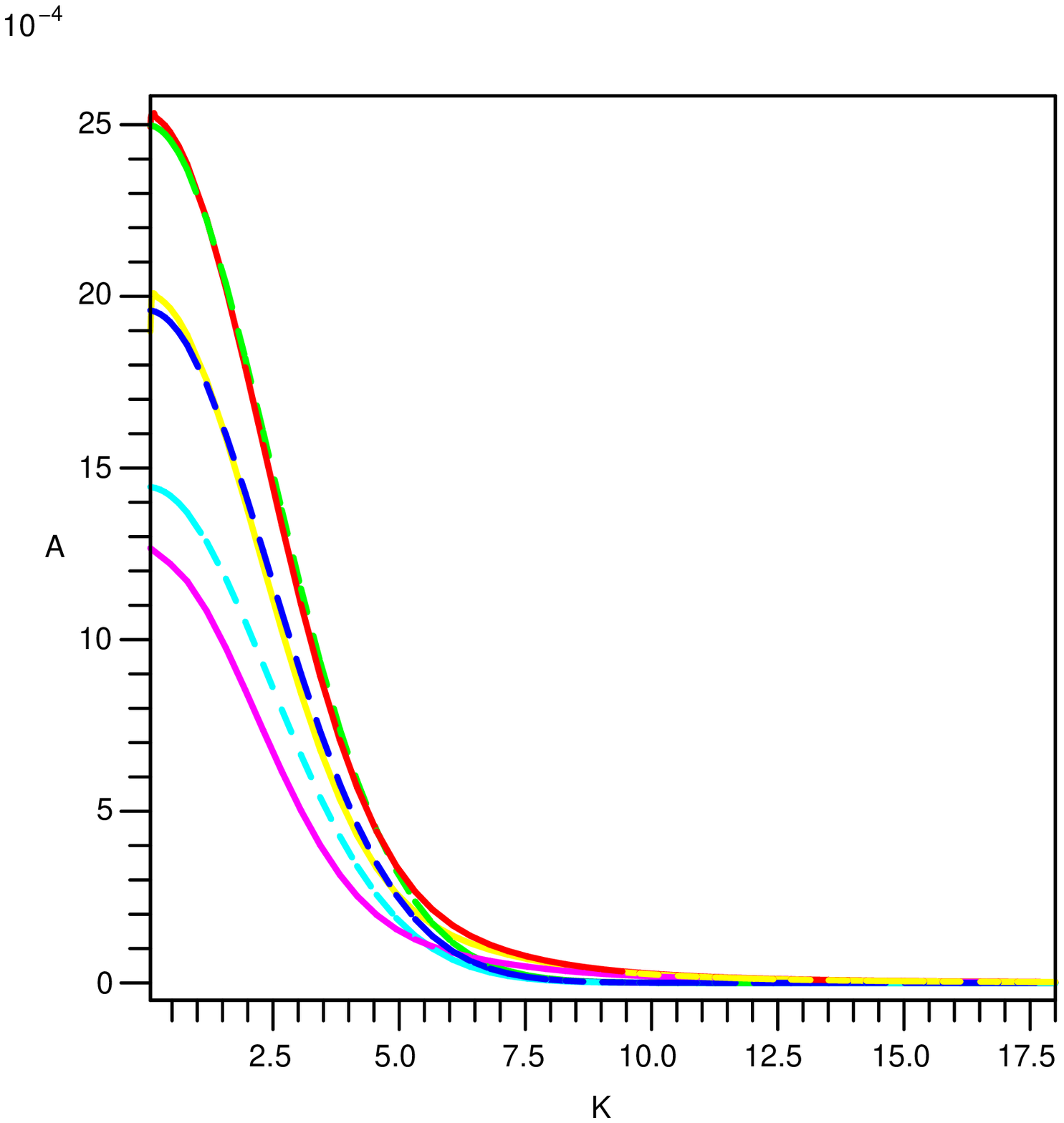}
\includegraphics[height=6.75cm]{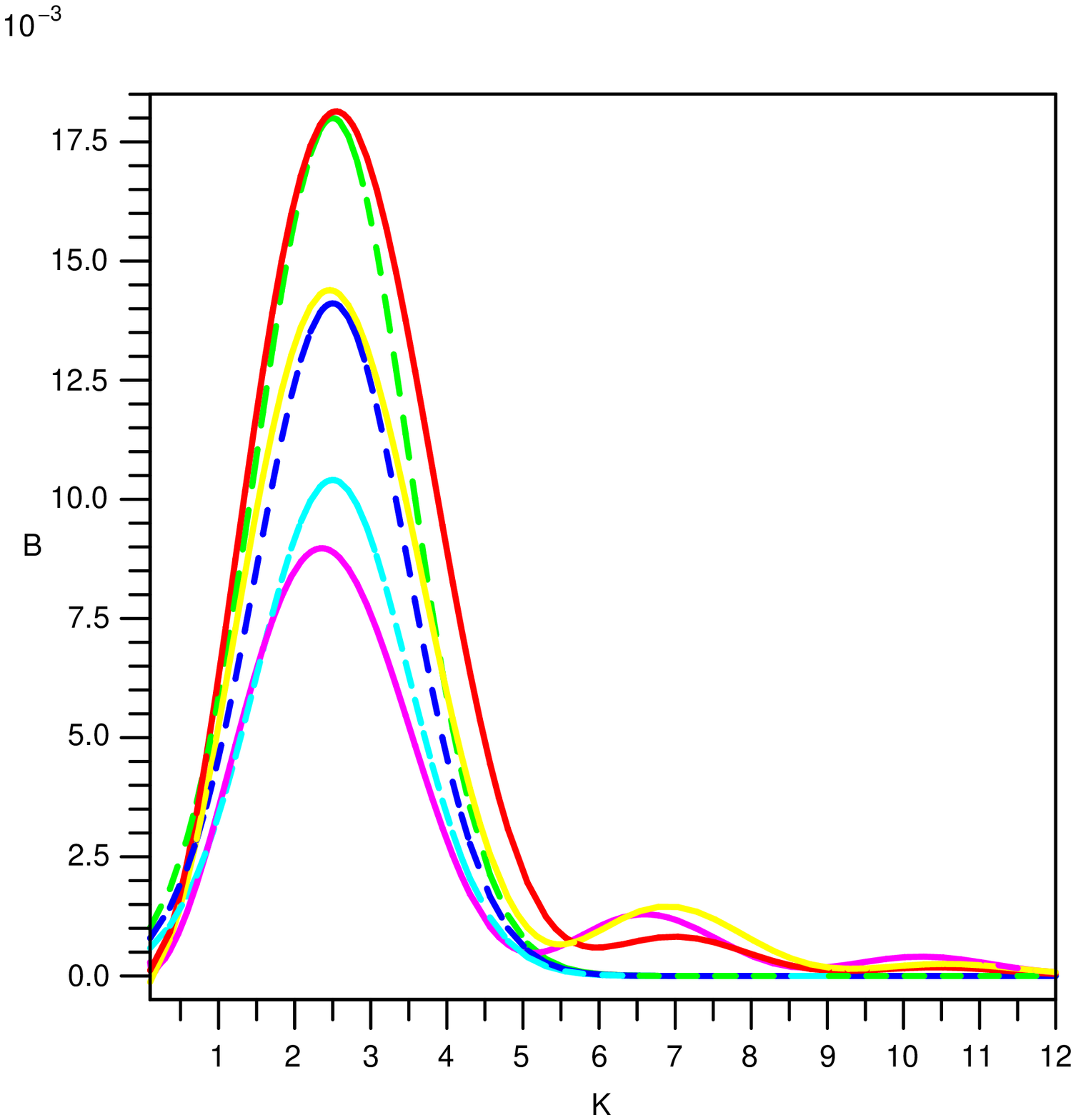}
\includegraphics[height=6.75cm]{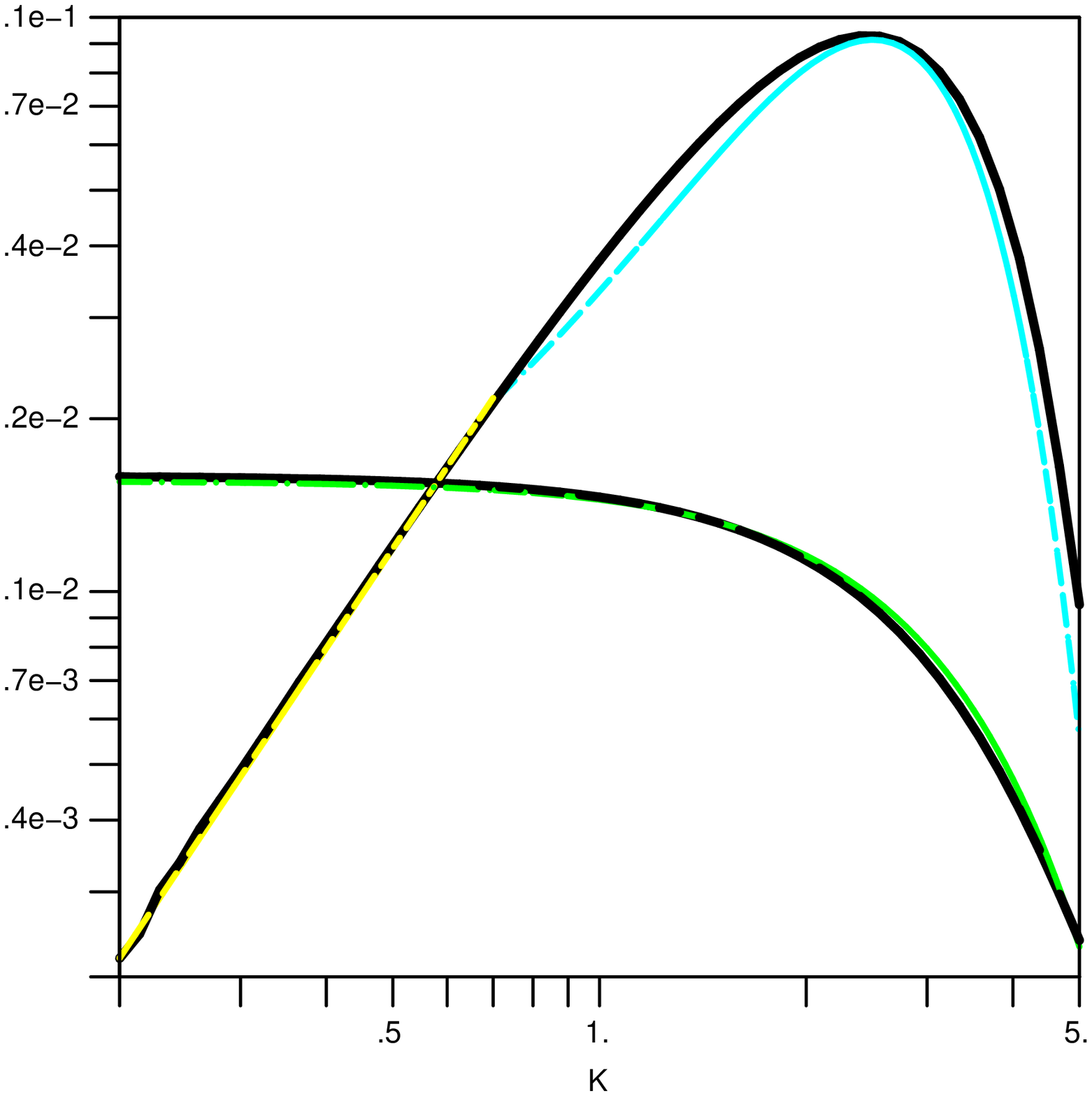}
\caption{\label{fig3}Velocity power spectrum.
 The top left panel shows the function $A(K)$ determining the diagonal
 part of the velocity power spectrum and the fit given in Eq.~(\ref{Ak})
 for different values of $s=\vint/\vout=\rint/R$. The solid lines from
 top to bottom (red, yellow and pink) are the correct functions and
 the  dashed lines (green, blue and cyan) are the fits for $s=0$, $0.6$ and
 $0.75$ respectively. The top right panel shows the function $B(K)$
 and the fit given in Eq.~(\ref{Bk}), again for the same values of
 $s$. The approximations overestimate for $s>0.74$ by about 16\%. The
 lower panel shows again $A(K)$ and $B(K)$ and the fits of
 Eqs.~(\ref{Ak},~\ref{Bk}) for  $s=0.6$. The
 flatter curve is $A(K)$ and the dashed line is its fit given in
 Eq.~(\ref{Ak}). We note the white noise behaviour of $A(K)$ for small
 values of $K$.
The more peaked solid line is $B(K)$ and the dashed
 line the fit given in  Eq.~(\ref{Bk}). At small $K$, $B(K)$
 grows like $K^2$ while $A(K)$ is constant. At large $K$ both
 functions decay like $K^{-4}$.}
\end{center}
\end{figure}

We define the new dimensionless variable $K=kR$ and the fraction
\be
s=\vint/\vout=\rint/R~.
\label{definitionofs1}
\ee
 Note that $1-s = (R-\rint)/R$ is the relative thickness of the shell,
 and will contribute to the amplitude of the GW signal. In our approach,
 the GW signal will vanish in the limit of vanishing thickness.
 We perform the Fourier transform and obtain the following expression for the velocity power spectrum:
\be
\langle v_{i}(\bk,t)  v_{j}^*(\bq,t) \rangle = \de(\bk-\bq) \,4\pi\phi(t)\vf^2R(t)^3
[A(K)\de_{ij}+B(K)\hat{k}_i\hat{k}_j] \label{finalspectrum}\ee
\bea
A(K)&\simeq&  0.0025(1-s^3)\left\{ \begin{array}{ll}
 \exp{(-K^2/12)} & \mbox{if } \quad K\le 4.5 \\
 \exp{(-4.5^2/12)}\left(\frac{4.5}{K}\right)^4 & \mbox{if
 } \quad K\ge 4.5~, \end{array} \right.   \label{Ak} \\
 && \nonumber\\
B(K)&\simeq&  0.0180 (1-s^3)\left\{ \begin{array}{ll}
 \exp{(-(0.7-2.5)^2/2)}\left(\frac{K}{0.7}\right)^2
&  \mbox{if } \quad K\le 0.7 \\ 
 \exp{(-(K-2.5)^2/2)} & \mbox{if } \quad 0.7\le K\le 4.5  \\
\exp{(-(4.5-2.5)^2/2)}\left(\frac{4.5}{K}\right)^4 & \mbox{if
 } \quad K\ge 4.5~. \end{array} \right.  \label{Bk}
\eea
The last two equations are good fits to the real  functions
$A(K),~B(K)$ which are shown in Fig.~\ref{fig3}. $A(K)$
behaves as white noise for $K\leqsim 1$ while $B(K) \propto K^2$. On
small scales (large values of $K$) both $A(K)$ and $B(K)$ decay like
$K^{-4}$. In the region where $A(K)$ and $B(K)$ are maximal, 
our approximations  overestimate the real functions for values of
$s>0.65$ (we will comment on this in Section \ref{subsection:DETO}).
In the following analysis we will consider the maximal value of
$s=0.74$: for this value the approximations overestimate by about 16\%. 

The behaviour of $A(K)$ and $B(K)$ can be predicted from general
considerations. First of all, the correlation function
(\ref{correlationintegral}) vanishes for $r > 2R$ : the
characteristic scale of the correlation function is the diameter of
the bubbles. We therefore expect this scale to show up in the power
spectrum at wave-number $k\simeq 2\pi/2R$. This is indeed what
happens, since the functions $A(K)$ and $B(K)$ change their behaviour
at approximately $K\simeq 2.5 \simeq \pi$. Moreover, since the
correlation function is a function with compact support, its Fourier
transform, the power spectrum, must be analytic in $\bk$. We remind that an
analytic function can be developed in power series around any point of
its domain. The term $A(K)\de_{ij}$ of Eq.~(\ref{Ak})  is analytic for
$K\rightarrow 0$, if and only if $A(K)\propto K^n$ and $n$ is an even
integer $n\geq 0$. This justifies the white noise behaviour observed
at large scales for $A(K)$. On the other hand, analyticity of the term
$B(K)\hat{k}_i\hat{k}_j$ for $K\rightarrow 0$ is satisfied if and only
if $B(K)\propto K^n$ with $n$ an even integer and $n\geq 2$. This
justifies why $B(K)$ increases as $K^2$ at large scales.

\subsubsection{Unequal time correlation function}
\label{subsec:unequal}

Up to now we have evaluated the velocity correlation function at equal
times. We note however, from Eq.~(\ref{Cac}), that we actually need the
correlation function evaluated at different (comoving) times $\tau$,
$\zeta$. The velocity in point $\bx$ at time $\tau$ can be correlated
with the velocity in point $\by$ at time $\zeta$. Consider, for
example, $\tau<\zeta$. In this case, the unequal time correlation
function is not zero if the  velocity shell in the bubble includes
$\bx$ at time $\tau$, and grows to include $\by$ at time
$\zeta$. According to the approach outlined above, evaluating the
correlation function at different times means performing the volume
integral of Eq.~(\ref{correlationintegral}) within regions $V_i$ given
by the intersection of spheres of different radii
(cf. Fig.~\ref{fig2}). This integral is too complicated to be done
analytically: therefore, within our analytical approach, we first try
to simply approximate the unequal time correlation function with the
one at equal time calculated in the previous subsection. This is a
reasonable approximation, provided that the region of non-zero
velocity at $\tau$ overlaps with the region of non-zero velocity at
$\zeta$. If this is not the case, we simply set the unequal time
correlation function to zero.

Reintroducing $\etain$ as the time of nucleation we find that, in the
limiting case, the inner boundary of the non-zero velocity shell in the
bubble at time $\zeta$ equals the outer boundary at time $\tau$ if
$\zeta=(\tau-\etain)/s+\etain$, where 
we remind that $s=\rint/R$. Symmetrizing among the two times, we have then
\bea
\langle v_{i}(\bx,\tau)  v_{j}(\by,\zeta) \rangle &\simeq&
\langle v_{i}(\bx,\tau)  v_{j}(\by,\tau) \rangle \Theta(\zeta-\tau)
\Theta((\tau-\etain)/s+\etain-\zeta)+\nonumber\\
&+&\langle v_{i}(\bx,\zeta)  v_{j}(\by,\zeta) \rangle
\Theta(\tau-\zeta)\Theta((\zeta-\etain)/s+\etain-\tau)
\label{uneqtime}
\eea
where $\Theta(\tau)$ is the Heaviside function. We choose arbitrarily
to set the time appearing in the equal time correlator corresponding
to the smaller of the two times. 

In Section \ref{evaluation}, we derive the gravitational wave spectra
obtained using this approximate form of the unequal time correlation
function and discuss its shortcomings. We will eventually propose
another method, which consists in giving an approximate form directly
for the unequal time anisotropic stress power spectrum, rather than
for the velocity correlation function. As we will see, proceeding in
this way we can  have better control  over the positivity of the power
spectrum, and obtain more reliable results.

\subsection{Anisotropic stress power spectrum: calculation}
\label{sec3.3}
We now have everything we need to evaluate the anisotropic stress
power spectrum of our source, using Eq.~(\ref{Pik}). 

\begin{figure}[htb!]
\begin{center}
\includegraphics[height=7cm]{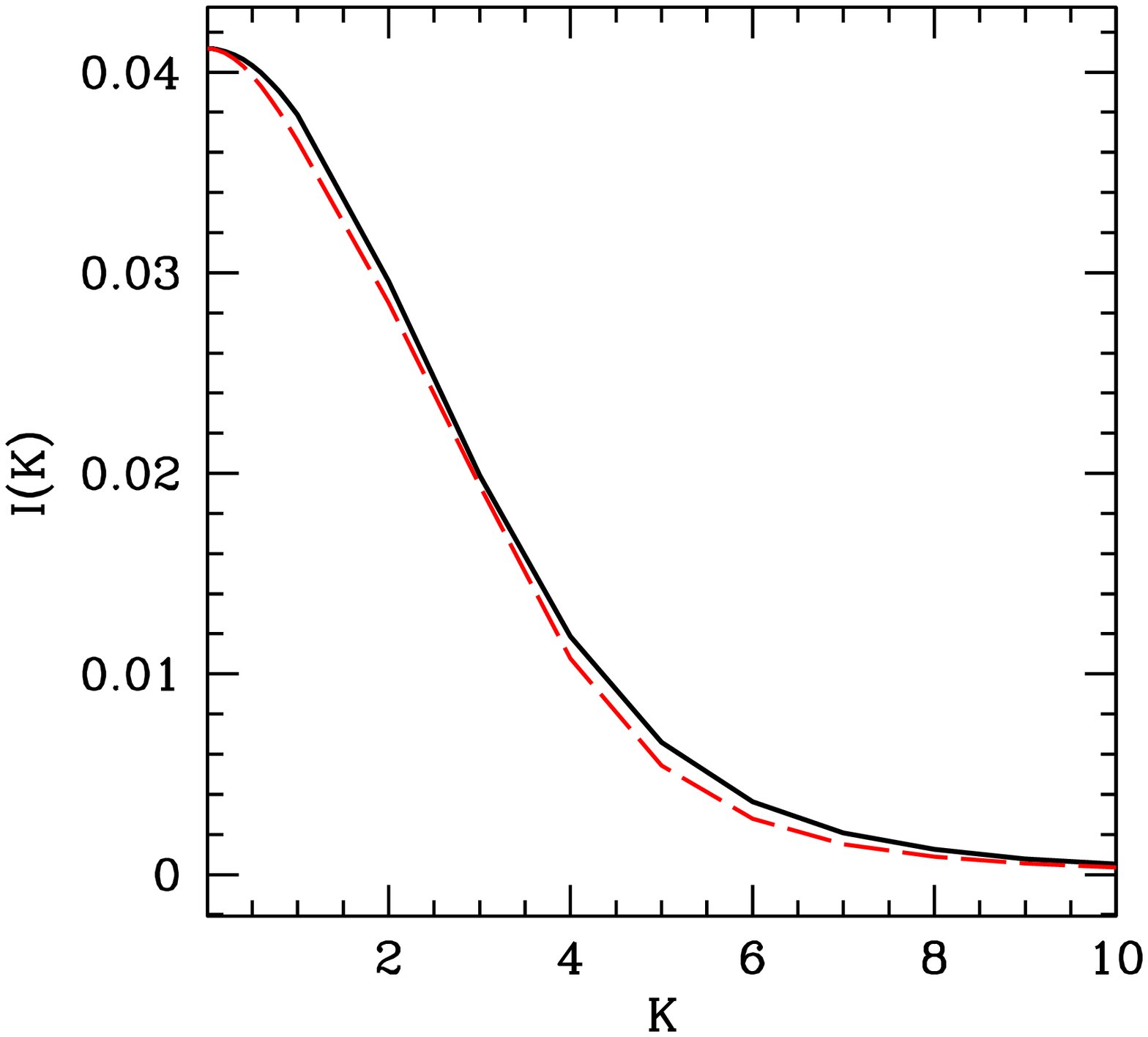}
\includegraphics[height=7cm]{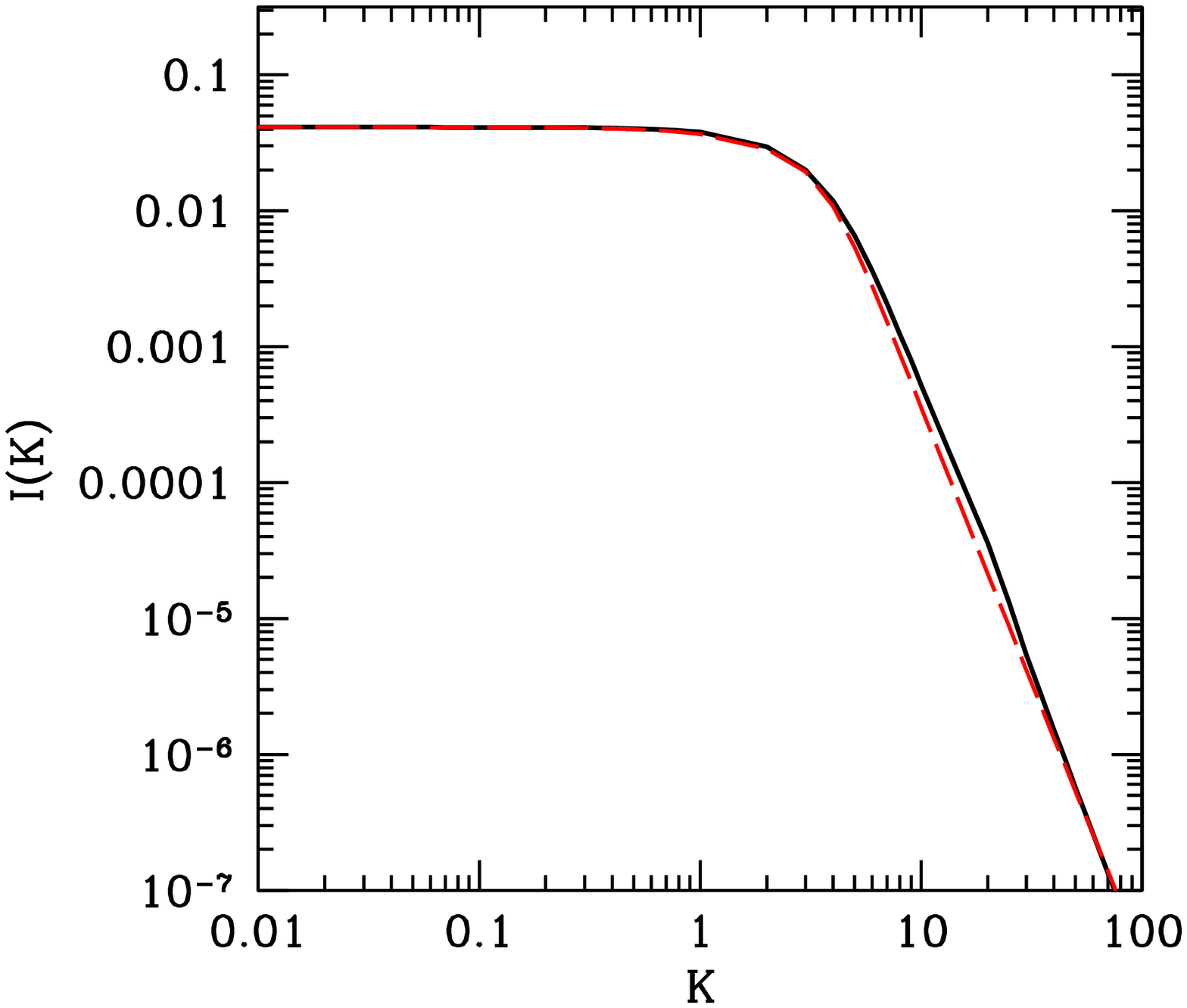}
\caption{\label{fig5} These two figures show the exact integral in
  Eq.~(\ref{Pint}) in solid (black), and the fit ${\cal I}(K)$ given in
  Eq.~(\ref{alk}) in dashed (red). In the right panel we clearly see
  the white noise behavior for small values of $K$ and the $K^{-4}$ behavior for large values of $K$, as expected (see
  discussion above and in Appendix~\ref{Appen:largeandsmall}).}
\end{center}
\end{figure}

The unequal time correlation function Eq.~(\ref{uneqtime}), together
with the equal time velocity power spectrum given in
Eq.~(\ref{finalspectrum}), lead to 
\bea
&&\hspace{-0.8cm}\Pi(k,\tau,\zeta)=\frac{w(\tau)w(\zeta)}{(1-v^2(\tau))(1-v^2(\zeta))} 
\Big\{ [4\,\pi\,\phi(\tau)\,\vf^2]^2 \,
R(\tau)^3 \, \Theta(\zeta-\tau)
\Theta\left(\frac{1}{s}(\tau-\etain)
 + \etain-\zeta\right) \nonumber \\
&&\hspace{-0.8cm}\times \int d^3P\,[  4 A(P,\tau) A(|{\bf K}-{\bf P}|,\tau) +
   2A(P,\tau)B(|{\bf K}-{\bf P}|,\tau)(1-\la^2)  \nonumber \\ 
   && \hspace{-0.8cm} 
+\,2B(P,\tau)A(|{\bf K}-{\bf P}|,\tau)(1-\beta^2) +B(P,\tau)\,B(|{\bf K}-{\bf
  P}|,\tau)(1-\beta^2)(1-\la^2) ]  
  \nonumber \\
&&\hspace{-0.8cm} +\,{\rm symmetric~} \tau\leftrightarrow\zeta\Big\} ~,
\label{Pint}
\eea
with $P=pR(\tau)$, $K=kR(\tau)$, $\lambda=\hat{k}\cdot\hat{p}$, and $\beta=\hat{k}\cdot\widehat{k-p}$. 
Using the functions $A(K)$ and $B(K)$
given in Eqs.~(\ref{Ak}, \ref{Bk}), we can perform the above integral. 
A good approximation to the integral, after factorizing out the
$s$-dependence as $(1-s^3)^2$, is given by  
\be
{\cal I}(K)=0.0412\,\frac{1+\left(\frac{K}{3}\right)^2}{1+
  \left(\frac{K}{2}\right)^2 + \left(\frac{K}{3}\right)^6}\,.
\label{alk}
\ee
This fit, together with the exact integral, is shown in
Fig.~\ref{fig5}. The function is flat on large scales and changes
slope at $K\simeq 3$. A white noise behavior at large scales 
 is expected, since the anisotropic stress power spectrum is
the convolution of the velocity power spectrum: this simply means that the
anisotropic stress is not 
correlated at distances larger than the source correlation scale
$2R$. In Appendix~\ref{Appen:largeandsmall}, we derive this in details. 
As one sees in Fig.~\ref{fig5}, Eq.~(\ref{alk}) is a very good
approximation to the numerical integral. On small scales the
convolution  decays like $A(K)$ and $B(K)$, hence like $K^{-4}$. This can be 
understood as follows: when $K\gg K_{\max}$, where $K_{\max}$ denotes
the wave number at which $A(K)$ and $B(K)$ peak, the main contribution
to the convolution integral comes from the region $|{\bf P}-{\bf K}|\simeq  K_{\max} \ll  
K$. The value of a typical term in this region is about
$A(K)A(K_{\max})$, and the phase space volume is $ K_{\max}^3$: hence
we expect $\mathcal{I}(K) \simeq  K_{\max}^3A(K_{\max})A(K)$ for
$K\gg K_{\max}$ and analogous for the contributions containing $B(K)$. 
 
Before we are ready to insert the expression (\ref{Pint}) for
$\Pi(k,\tau,\zeta)$  in Eq.~(\ref{h'2}) to evaluate the gravitational
radiation power spectrum, we need to determine the time dependence of
$R(\eta)$ and $\phi(\eta)$. 

\section{Time dependence of the phase transition parameters}
\label{timedep}

We now investigate the actual time dependence of some parameters introduced previously, such as the fraction of volume occupied by bubbles at time $\eta$, $\phi(\eta)$, which we need in Eq.~(\ref{prob}), or the bubble radius\footnote{In this section we
specify what we actually take for the bubble radius, and we will
denote it by $\bar{R}(\eta)$. We remind that, in the case of deflagrations, this does not coincide with $R$, which is the position of the shock front, but with $\rint$.}. In this section we closely follow Ref.~\cite{turner1} in the modeling of the first order phase transition.

The rate of bubble nucleation of the broken phase bubbles is defined as
$\Gamma(\eta)=\mathcal{M}^4 a_*^4\ e^{-S(\eta)}$, where $\mathcal{M}$ is the energy scale of the phase transition and $S(\eta)$ the tunneling action. We Taylor expand the action at first order around a fixed time $\etafin$: the time at which the transition ends. Defining $\tilde\beta\equiv -dS/d\eta|_{\etafin}$, one can rewrite the nucleation rate as $\Gamma(\eta)=\Gamma(\etafin)\exp{(\tilde{\beta}(\eta-\etafin))}$. Note that we define $\tilde\beta=a_*\beta$ in terms of comoving time, differently from the usual convention. The probability that a given point remains in the false vacuum at time $\eta$ is given by
\be
p(\eta)=e^{-I(\eta)}
\ee
where $I(\eta)$ is the fraction of volume occupied by broken phase bubbles at time $\eta$ without considering bubble overlap \cite{turner1,guth2,guth3}. Assuming that the universe remains static for the entire duration of the phase transition, and assuming a constant velocity for the bubble expansion $\vb$, $I(\eta)$ is simply given by
\be
I(\eta)=\frac{4\pi}{3}\int_{\etain}^{\eta} d\tau \, \Gamma(\tau) \,\vb^3\,(\eta-\tau)^3\simeq 8\pi \,\frac{\vb^3}{\tilde\beta^4}\,\Gamma(\eta)
\ee
The quantity $\phi(\eta)$ in Eq.~(\ref{prob}) is given by
$\phi(\eta)=1-p(\eta)$.  The times $\etain$ and $\etafin$ are defined
such that $p(\etain)\simeq 1$ and  $p(\etafin)\simeq 0$. Following
\cite{turner1,Kosowsky:1992vn}, we choose a number $M\gg 1$ and define
$\etafin$ as $\Gamma(\etafin)=\tilde\beta^4M/8\pi\vb^3$, so that
$p(\etafin)=\exp{(-M)}\simeq 0$. In the same way, we choose a number
$m\ll 1$ such that $\tilde\beta(\etafin-\etain)=\ln{(M/m)}$ and
$p(\etain)=\exp{(-m)}\simeq 1$. This gives the duration of the phase
transition
\be
\etafin-\etain=\tilde\beta^{-1}\ln{\frac{M}{m}} ~.
\label{withlog}
\ee

In order to evaluate the mean bubble radius, we consider the number of bubbles which have a given radius $\delta$ at time $\eta$. Calling $\eta_\de$ the nucleation time of a bubble with radius $\de$ at time $\eta$, one has:
\be
\left.\frac{dN}{d\de}\right|_\eta=\frac{\Gamma(\eta_\de)p(\eta_\de)}{\vb} ~.
\ee
The shape of this distribution is shown in Fig.~5 of Ref.~\cite{turner1}. For each $\eta$, it has a maximum at the value $\bar{R}(\eta)=\frac{\vb}{\tilde\beta}\ln{I(\eta)}$: this value defines the mean radius of the bubbles at time $\eta$. Calling $\bar{\eta}$ the time at which $p(\bar{\eta})=1/e$, so that $I(\bar{\eta})=1$, we set
\bea
\bar{R}(\eta)=\left\{\begin{array}{ll}
0 & {\rm for}~\etain<\eta<\bar{\eta}~, \\
\frac{\vb}{\tilde\beta}\ln{(I(\eta))} &{\rm for}~\bar{\eta}<\eta<\etafin ~. \nonumber\end{array}\right.
\eea
The condition $I(\bar{\eta})=1$ defines $\bar{\eta}=\etafin-\tilde\beta^{-1}\ln{M}$.

As already explained in Sec.~\ref{velocity_profile}, bubbles can be
treated as combustion fronts moving at constant velocity only at times
much later than nucleation time. Therefore, we identify $\etain\equiv
\bar{\eta}$ in the evaluation of the emitted gravitational
radiation. This leads to $\etafin=\etain+\tilde\beta^{-1}\ln{M}$. We
further decide to neglect the logarithms and simply set
\bea
\etafin-\etain&\simeq&\tilde\beta^{-1}\\
\bar{R}(\eta)&\simeq&\vb(\eta-\etain)~. 
\eea
If we do not neglect the logarithms, we have to replace the final
bubble size $v_b\tilde\beta^{-1}$ by $v_b\tilde\beta^{-1}\ln(M/m)$. Neglecting the
logs, we identify the duration of the phase transition with $\tilde\beta^{-1}$,
and the radius of the bubbles at time $\eta$ with its mean value
$\bar{R}(\eta)$. We do not account for the possibility of having
bubbles of different sizes at a given time.

\section{Evaluation of the gravitational wave spectrum}
\label{evaluation}

We now want to evaluate the gravitational wave energy density per
logarithmic  frequency interval (\ref{GWenergy}), which is given in
terms of the GW power spectrum (\ref{h'2}). According to
Eq.~(\ref{h'2}), the latter evolves 
as $x^{-2}\propto \eta^{-2}$; however, this behavior is strictly valid
only in a radiation dominated universe with a constant number of relativistic 
degrees of freedom ({\it c.f.} Sec.~\ref{GW1}) and thus lacks generality. 
To estimate the GW power spectrum at $\eta>\eta_*$, we therefore
evaluate it at the end of the phase transition: $|h'|^2(k,\eta_*)$,
for which no assumptions have been made concerning the number of
relativistic degrees of freedom. Then, we simply use its
radiation-like evolution: 
\be
\frac{d \Omega (k,\eta)}{d \ln k}=\frac{d \Omega (k,\eta_*)}{d \ln k}
\left( \frac{a_*}{a}\right)^4 ~.
\label{evolOm}
\ee
From Eqs.~(\ref{GWenergy}) and (\ref{h'2}), reminding that $x=k\eta$ we have 
\bea
\frac{d \Omega (k,\eta_*)}{d \ln k}&=&
\frac{k^5 |h'|^2(k,\eta_*) }{ 2 (2 \pi)^6 G \rho_c a_*^2}\\
|h'|^2(k,x_*)&=&\frac{1}{2}\left(\frac{8\pi Ga_*^2}{k^2}\right)^2
\int_{\xin}^{\xfin} dy \int_{\xin}^{\xfin} dz \cos(z-y)\Pi(k,y,z)
\label{h'2*}
\eea
where in the last equality we have set $\xfin\simeq x_*$. In the above
equation, we further 
have to substitute expression (\ref{Pint}) for the anisotropic stress
source. As already discussed in Sec.~\ref{GW1}, since the source is
active for an amount of time much shorter than one Hubble time, we
neglect the expansion of the universe while gravitational waves are
produced. Therefore, in Eq.~(\ref{Pint}) we set the enthalpy density
to a constant, denoted by $w_*=w(\tau)\simeq w(\zeta)$. Moreover, we
eliminate the time dependence of the $\gamma=\sqrt{1-v^2}$ factors by
substituting $v(\tau), v(\zeta)$ with the constant fluid velocity
$s\vf$, corresponding to the fluid velocity of the inner boundary of
the shell ({\it cf.} Eq.~(\ref{velocity}) and Fig.~\ref{bul}),
remembering that $s=\rint/R$. We explain the reasons for this choice
below. We remind that the double integration in Eq.~(\ref{h'2*}) is in
time, with the notation $y=k\tau$, $z=k\zeta$. With ${\cal I}(K)$ given in
Eq.~(\ref{alk}), we finally obtain using Eq.~(\ref{evolOm})
\bea
&&\frac{d \Omega (k,\eta)}{d \ln k}= \frac{4}{\pi^2}\,
\frac{G}{\rho_c}\,\frac{a_*^6}{a^4}\,w_*^2\,
\frac{\vf^4(1-s^3)^2}{(1-(s\vf)^2)^2}\, k
\int_{\xin}^{\xfin} dy \int_{\xin}^{\xfin} dz \cos{(z-y)}\nonumber \\
&& \times\left[\phi^2(\tau)R^3(\tau)\Theta(\zeta-\tau)
\Theta\left(\frac{1}{s}(\tau-\eta_{\rm in})+\eta_{\rm in}-\zeta\right)
 {\cal I}(kR(\tau)) +{\rm symmetric~}y\leftrightarrow z \right]
\label{domega}
\eea

Let us first investigate the pre-factor in the above expression. 
The enthalpy density is
\be
w_*=\frac{4}{3}\rho_{\rm rad}^*~~~~~~~~\rho_{\rm rad}^*=
\left(\frac{g_0}{g_*}\right)^{\frac{1}{3}}\,\frac{\rho_{\rm rad}^0}{a_*^4}~,
\ee
where
$\rho_{\rm rad}^*$ denotes the radiation energy density in the universe,
$g_0=3.36$ and $g_*$ denote the effective number of relativistic degrees of
freedom  today and at the time of the phase transition respectively. We
also define a dimensionless parameter estimating the
amount of kinetic energy present in the source, with respect to the
radiation energy density. From the definition (\ref{Tab}) of the
energy-momentum tensor:
\bea
\rho_{\rm kin}^*&=&\frac{4}{3}\rho_{\rm
  rad}^*\frac{(s\vf)^2}{1-(s\vf)^2} ~, \label{rhokin}\\
\frac{\Om_{\rm kin}^*}{\Om_{\rm rad}^*}&=& 
\frac{\rho_{\rm kin}^*}{\rho_{\rm rad}^*}
=\frac{4}{3}\frac{(s\vf)^2}{1-(s\vf)^2}~.
\label{Omkin}
\eea
We have chosen to define the above parameter in terms of the fluid
velocity at the inner boundary of the velocity shell $s\vf$, which
always satisfies $s\vf<c_s$ both for detonations and
deflagrations. This ensures that
${\Om_{\rm kin}^*}/{\Om_{\rm rad}^*}<1$ to remain consistent with an
isotropic FRW universe. 

Summarizing, we can rearrange the pre-factor in Eq.~(\ref{domega}) in
terms of the above defined parameters, of the conformal Hubble factor
$\mathcal{H}_*=a_* H_*$ and of $\Om_{\rm rad}$, as: 
\be
\frac{4}{\pi^2}\,\frac{G}{\rho_c}\,\frac{a_*^6}{a^4}\,w_*^2\,
\frac{\vf^4}{(1-(s\vf)^2)^2}\,(1-s^3)^2=
\frac{3}{2\pi^3}\left(\frac{g_0}{g_*}\right)^{\frac{1}{3}}\frac{\Om_{\rm rad}
\,\mathcal{H}_*^2}{a^4} \left(\frac{\Om_{\rm kin}^*}{\Om_{\rm rad}^*}\right)^2 
\frac{(1-s^3)^2}{s^4}~.
\ee
The gravitational wave energy density is therefore proportional to the
square of the kinetic energy density of the source, as one would
expect. 

In the double integral of Eq.~(\ref{domega}), we define the new
integration variable $u=kR(y)=\vout(y-\xin)$, and the new
dimensionless parameter 
\be
Z=\frac{k\vout}{\tilde{\beta}}\,. 
\ee
Following the discussion in Sec.~\ref{timedep}, the function
$\phi(\tau)$ becomes 
\be
\phi(\tau)=1-\exp{(-\exp{(1+\tilde{\beta}(\tau-\eta_{\rm fin}))})} ~.
\ee
We finally obtain for the gravitational wave energy density spectrum today
\bea
\frac{d \Omega (k,\eta_0)h^2}{d \ln k}&\simeq& \frac{3}{2\pi^3}
\left(\frac{g_0}{g_*}\right)^{\frac{1}{3}}\Om_{\rm rad}h^2 
\left(\frac{\Om_{\rm kin}^*}{\Om_{\rm rad}^*}\right)^2\,
\left(\frac{\mathcal{H}_*}{\tilde{\beta}}\right)^2 
\frac{(1-s^3)^2}{s^4} \nonumber \\
&&\times\frac{2\vout}{Z^2} \left\{ \int_0^{sZ} du \, (1-e^{-e^{(1+u/Z)}})^2\,
u^3 {{\cal I}(u)}\sin{\left(\frac{u}{\vout}\frac{1-s}{s}\right)}
\right.\nonumber \\ 
&&  \quad\qquad +\left. \int_{sZ}^Z du\,  (1-e^{-e^{(1+u/Z)}})^2\,
u^3 {{\cal I}(u)} \sin{\left(\frac{Z-u}{\vout}\right)} \right\}~.
\label{psGW}
\eea
We recover the known result, that the amplitude of the gravitational
wave spectrum is proportional to the square of the ratio between the
duration of the phase transition and the Hubble time,
$\mathcal{H}_*/\tilde{\beta}$ \cite{Witten:1984rs,turner1990}. If the
parameter $s=1$, this means that the fluid motions vanish everywhere,
and therefore no gravitational radiation is produced.
The divergence in $s\rightarrow 0$ is only apparent, due to our
definition of the kinetic energy density parameter in Eq.~(\ref{Omkin}).

Expression (\ref{psGW}) gives in all generality the power spectrum of
gravitational radiation produced by spherically symmetric, stochastic
configurations of fluid motions (such as broken phase bubbles in a
phase transition), characterized by a velocity distribution which
increases linearly in a shell of inner radius $r_{\rm int}=\vint t$ and
outer radius $R=\vout t$. The integral determines the shape of the
spectrum, and depends on the values of $s$ and of the velocity
$\vout$. These parameters should be chosen according to the physical
situation under consideration. 

We expect the large scale part of the GW spectrum to increase as
$k^3$. As explained in details in Appendix~C, this is a simple
consequence of the fact that the source has a finite correlation scale
(corresponding in our case to the length-scale $2R$). It is therefore
a generic behaviour for causal sources. Conversely, the small scale
part of the spectrum depends on the details of the source correlation
function. This part of the spectrum is in principle affected by our choice of a
linear growth for the velocity profile in the non-zero velocity shell
(cf. Sec.~\ref{velocity_profile}). If we could account for the correct
velocity profile, the power law dependence of the GW power spectrum at
large $k$ might be different from what we find. On the other hand, the
frequency at which the GW power spectrum peaks can again be predicted
by general considerations. In fact, the velocity power spectrum given
in Eq.~(\ref{finalspectrum}) has a characteristic wave-number
corresponding to the bubble diameter, $k\simeq \pi/R\simeq
2.5/R$. Also the anisotropic stress power spectrum changes slope at about
$k\simeq \pi/R\simeq 3/R$. For the gravitational radiation power
spectrum we therefore expect a peak at 
approximatively the same wave-number $k\simeq \pi/R(\eta_\mr{fin})$
given by the mean bubble size at the end of the phase transition.

\subsection{Unequal time approximations}

The expression (\ref{psGW}) has been derived using the approximation
Eq.~(\ref{uneqtime})  for the unequal time correlation function of the
velocity field. Under this approximation, the anisotropic stress power
spectrum at different times takes the form given in Eq.~(\ref{Pint}),
which has been substituted in Eq.~(\ref{h'2*}) in order to evaluate
the GW power spectrum. According to the definition of the GW spectrum
$|h'(x)|^2$, the time integral appearing in Eq.~(\ref{h'2*}) should
give a positive result. Therefore, the unequal time anisotropic stress
power spectrum should be, by definition, a positive kernel, such that 
\be
\int_{\xin}^{\xfin} dy \int_{\xin}^{\xfin} dz \,\, \mathcal{G}(x,y)
\mathcal{G}^*(x,z)\Pi(k,y,z)\geq 0~, 
\ee
where $\mathcal{G}(x,y)$ generically denotes the Green function of the
wave equation (\ref{reducedwe}). Even though the approximated form for 
the unequal time correlation function for the velocity field
Eq.~(\ref{uneqtime}) seems reasonable from a physical point of view,
it does not lead to a positive kernel for $\Pi(k,y,z)$ as it
should. Expression (\ref{psGW}) is  not always  positive and therefore
it is  unacceptable. To avoid this problem, we now define approximations
directly for the unequal time correlator $\Pi(y,z)$ which are positive
by construction. We first discuss two cases which are physically less
motivated, but are very useful for comparison: the completely
incoherent and completely coherent approximations. Finally, we discuss
a third and better motivated approximation for $\Pi(y,z)$ that still
gives a positive result. We believe that this last approximation is
the one that best recovers the true result. Apart from
being strictly positive, it has the advantage to be very close to the physically motivated approximation
for the unequal time correlators of the velocity field.
As we shall see, the peak position of the GW spectra is very similar
in all cases, while the peak amplitude varies by up to one order of magnitude.

We consider first a \emph{totally incoherent} source, that is
uncorrelated for unequal times $\tau\neq \zeta$. We take the Ansatz
(cf. Eq.~\ref{Pint}) 
\bea
\lag \Pi_{ij}(\bk, \tau)\Pi^*_{ij}(\bq,\zeta)\rag=\de(\bk-\bq)
\Pi(k,\tau,\tau)\frac{\de(\tau-\zeta)}{\tilde{\beta}}\label{Pitotin}\\
\Pi(k,\tau,\tau)=\frac{w_*^2}{(1-(s\vf)^2)^2}[4\pi \phi(\tau) \vf^2]^2 
R(\tau)^3(1-s^3)^2{\cal I}(K(\tau))~.
\eea
We multiply the $\de$- function by the characteristic time
$1/\tilde{\beta}$ in order to maintain the correct dimensions. Under
this assumption, the oscillating term in Eq.~(\ref{h'2*}),
$\cos(x-y)$, integrated with the delta function, becomes simply 1 and
we recover a positive result. Using the same notations as in
Eq.~(\ref{psGW}), the GW spectrum from totally incoherent bubbles is
(see Fig.~\ref{fig6}) 
\bea
\left. \frac{d \Omega (k,\eta_0)h^2}{d \ln k}\right|_{\rm incoh.}
&\simeq& \frac{3}{2\pi^3}\left(\frac{g_0}{g_*}\right)^{\frac{1}{3}}
\Om_{\rm rad}h^2
\left(\frac{\Om_{\rm kin}^*}{\Om_{\rm rad}^*}\right)^2\,
\left(\frac{\mathcal{H}_*}{\tilde{\beta}}\right)^2
\frac{(1-s^3)^2}{s^4} \nonumber \\
&\times&\frac{1}{Z} \int_0^{Z} du \, (1-e^{-e^{(1+u/Z)}})^2\,
u^3 {{\cal I}(u)}~.
\label{psGWinco}
\eea

The opposite situation is given by a \emph{totally coherent} source,
that is correlated for every time $\tau$ and $\zeta$. In this case,
the Ansatz for the anisotropic stress power spectrum is 
\bea
\lag \Pi_{ij}(\bk,
\tau)\Pi^*_{ij}(\bq,\zeta)\rag=\de(\bk-\bq)\sqrt{\Pi(k,\tau)}
\sqrt{\Pi(k,\zeta)}  \nonumber \\
\sqrt{\Pi(k,\tau)}=\frac{w_*}{1-(s\vf)^2} 4\pi \phi(\tau) 
\vf^2 R(\tau)^{3/2} (1-s^3) \sqrt{{\cal I}(K(\tau))}
\eea
We rewrite the oscillating term $\cos(x-y)$ using the duplication
formula. This allows us to split the double integral into a sum of two
positive terms. The GW spectrum is in this case (see Fig.~\ref{fig6}) 
\bea
\left. \frac{d \Omega (k,\eta_0)h^2}{d \ln k}\right|_{\rm coh.}
&\simeq& \frac{3}{2\pi^3}\left(\frac{g_0}{g_*}\right)^{\frac{1}{3}}
\Om_{\rm rad}h^2
\left(\frac{\Om_{\rm kin}^*}{\Om_{\rm rad}^*}\right)^2\,
\left(\frac{\mathcal{H}_*}{\tilde{\beta}}\right)^2 
\frac{(1-s^3)^2}{s^4} \nonumber \\
&\times&\left\{ \left[\frac{1}{Z} \int_0^{Z} du \, (1-e^{-e^{(1+u/Z)}})\,
u^{3/2} \sqrt{{{\cal I}(u)}} \cos\left( \frac{u}{\vout}\right)
\right]^2 \right. \nonumber\\
&+&\left.  \left[\frac{1}{Z} \int_0^{Z} du \, (1-e^{-e^{(1+u/Z)}})\,
u^{3/2} \sqrt{{{\cal I}(u)}} \sin\left( \frac{u}{\vout}\right) 
\right]^2\right\}   ~.
\label{psGWco}
\eea

The positive part of the result based on the approximated unequal time
correlation function for the velocity field, is in between these two
extreme cases: the Heaviside functions introduce correlation only
among times $\tau$ and $\zeta$ which are sufficiently close to each
other. We therefore expect also the correct result for the GW spectrum
to be in between the limiting cases described above. Moreover, for
sufficiently large scales the details of the time correlation do not
matter, and we expect that the GW spectra derived using these
different approximations become comparable at these scales. As we will
see below, this is indeed the case ({\it c.f.} Fig.~\ref{fig6}).

Finally we introduce top-hat unequal time correlations directly in the
anisotropic stress power spectrum, rather than in the velocity
correlation function. This approximation is also an
intermediate case between the completely coherent and
incoherent ones, and it is constructed to always give a positive
result for the GW power spectrum as we will discuss. We make the
following symmetric Ansatz:
\bea
\lag \Pi_{ij}(\bk, \tau)\Pi^*_{ij}(\bq,\zeta)\rag&=&\de(\bk-\bq)
[\Pi(k,\tau)\Theta(k\zeta-k\tau)\Theta(x_c-(k\zeta-k\tau)) \nonumber\\
&+&\Pi(k,\zeta)\Theta(k\tau-k\zeta)\Theta(x_c-(k\tau-k\zeta))] ~,\\
\Pi(k,\tau)&=&\frac{w_*^2}{(1-(s\vf)^2)^2}[4\pi \phi(\tau) \vf^2]^2 
R(\tau)^3(1-s^3)^2{\cal I}(K(\tau)) ~.
\eea
Under this assumption, we set the correlation to zero for modes with a
time separation larger than $x_c/k$, where $x_c$ is a positive, dimensionless
parameter of order unity (to be specified later). Therefore, the
anisotropic stresses at different times are correlated if the time
separation is less than about one wavelength. Physically, this just
means that longer wavelengths are correlated over a longer time.  
This corrects the lack of correlation that resulted from our very
first approximation. The GW power spectrum becomes:
\bea
& &\left. \frac{d \Omega (k,\eta_0)h^2}{d \ln k}\right|_{\rm mix}
\simeq \frac{3}{2\pi^3}\left(\frac{g_0}{g_*}\right)^{\frac{1}{3}}\Om_{\rm rad}h^2
\left(\frac{\Om_{\rm kin}^*}{\Om_{\rm rad}^*}\right)^2\,
\left(\frac{\mathcal{H}_*}{\tilde{\beta}}\right)^2 \frac{(1-s^3)^2}{s^4} \\
& &\times\frac{2}{Z^2} \int_0^{Z} du \, (1-e^{-e^{(1+u/Z)}})^2\,
u^3 {{\cal I}(u)} \int_0^Z dv \cos\left(\frac{u-v}{\vout}\right)
\Theta(v-u)\Theta\left(x_c-\frac{v-u}{\vout}\right)\nonumber
\label{psGWmix1}
\eea
which is apparently positive for $x_c\leq \pi/2$. Resolving the
Heaviside functions, we have to consider two separate cases. For
$Z<x_c\vout$ we find the spectrum (see  Fig.~\ref{fig6})
\bea
&& \hspace*{-2cm} \left. \frac{d \Omega (k,\eta_0)h^2}{d \ln k}
\right|_{\rm mix}
\simeq \frac{3}{2\pi^3}\left(\frac{g_0}{g_*}\right)^{\frac{1}{3}}
\Om_{\rm rad}h^2
\left(\frac{\Om_{\rm kin}^*}{\Om_{\rm rad}^*}\right)^2\,
\left(\frac{\mathcal{H}_*}{\tilde{\beta}}\right)^2 \frac{(1-s^3)^2}{s^4} \nonumber \\
&& \qquad \times\frac{2\vout}{Z^2} \int_0^{Z} du \, (1-e^{-e^{(1+u/Z)}})^2\,
u^3 {{\cal I}(u)} \sin\left(\frac{Z-u}{\vout}\right)~, \quad
Z<x_c\vout \,,
\label{psGWmix2}
\eea
and for $Z>x_c\vout$ we find
\bea
&& \hspace*{-2cm} \left. \frac{d \Omega (k,\eta_0)h^2}{d \ln k}\right|_{\rm mix}
\simeq \frac{3}{2\pi^3}\left(\frac{g_0}{g_*}\right)^{\frac{1}{3}}
\Om_{\rm rad}h^2
\left(\frac{\Om_{\rm kin}^*}{\Om_{\rm rad}^*}\right)^2\,
\left(\frac{\mathcal{H}_*}{\tilde{\beta}}\right)^2
\frac{(1-s^3)^2}{s^4}  \nonumber \\
&& \times\frac{2\vout}{Z^2} \left\{\sin(x_c)\int_0^{Z-x_c\vout} du \,
 (1-e^{-e^{(1+u/Z)}})^2\,
u^3 {{\cal I}(u)} \right. \nonumber\\
&&+ \left. \int_{Z-x_c\vout}^Z du \, (1-e^{-e^{(1+u/Z)}})^2\,
u^3 {{\cal I}(u)} \sin\left(\frac{Z-u}{\vout}\right) \right\}~, \quad
Z> x_c\vout \, .
\label{psGWmix3}
\eea
The GW spectrum remains positive for $0<x_c<\pi$ since 
in the range $Z<x_c\vout$, $\sin\big(\frac{Z-u}{\vout}\big)>0$,
and in the range $Z>x_c\vout$,  both $\sin(x_c)$ and
$\sin\big(\frac{Z-u}{\vout}\big)$ are positive. In the limit
$x_c\ra 0$ the result tends to zero. A reasonable value is
$\pi/2<x_c<\pi$. In Fig.~\ref{fig7} we show how the
power spectrum depends on the value of $x_c$.
We now present the results obtained for the different approximations
discussed above, in the cases of detonations and deflagrations.

\subsection{GW from detonations}
\label{subsection:DETO}

As explained in Sec.~\ref{velocity_profile}, in the case of
detonations $R(t)=\vout t$ is the outer radius of the
bubbles. Therefore, $\vout=\vb=v_1$. We restrict ourselves to the case of
Jouguet detonations, so that $\vint=c_s=1/\sqrt{3}$. The value of the
bubble wall velocity in Jouguet detonations is given in
Refs.~\cite{Steinhardt,Kamionkowski:1993fg} in terms of the ratio
\be
\al=\rho_{\rm vac}/\rho^*_{\rm rad}
\ee
where $\rho^*_{\rm rad}$ denotes the radiation energy density in the universe:
\be
\vb=\frac{c_s+\sqrt{\al^2+2\al/3}}{1+\al}~.
\label{vbdet}
\ee
The outer maximal fluid velocity is given by the Lorentz transformation (see Eq.~\ref{vf})
\be
\vf=\frac{\vb-c_s}{1-\vb c_s}=\frac{\sqrt{3}(\sqrt{3\al^2+2\al}-\al)}{2+3\al-\sqrt{3\al^2+2\al}} ~,
\label{vfdet}
\ee
where for the last equality we have substituted (\ref{vbdet}) and $c_s=1/\sqrt{3}$. Moreover, the parameter $s=\vint/\vout$ takes the form
\be
s=\frac{c_s}{\vb}=\frac{1+\al}{1+\sqrt{3\al^2+2\al}} ~.
\label{sdet}
\ee
For detonations\footnote{To be in the detonation regime means that
  there is essentially no friction. In other words, interactions
  between the bubble wall (the Higgs field) and the particles in the
  thermal bath can be neglected. This is a strong assumption which can
  work if the released latent heat is indeed very large. Model-independent studies (such as Ref.~\cite{Grojean:2006bp}) have used Eq.~(\ref{vbdet})
for the bubble wall velocity. However, when considering a particular
model, one should compute friction effects and therefore derive a
more realistic value for the bubble wall velocity,  following for
instance the procedure of Ref.~\cite{Moore:2000wx}. This requires to
compute the bubble wall profile. }, it is
therefore sufficient to specify $\al$, and all the quantities
necessary to evaluate the integrals in
Eqs.~(\ref{psGW},\ref{psGWinco},\ref{psGWco},\ref{psGWmix2},\ref{psGWmix3})
are determined. Knowing in addition the duration of the phase
transition $\mathcal{H}_*/\tilde{\beta}$ fully determines the
gravitational wave signal. We choose a broad range of values for
$\al$: $\al=0.1,1/2,1,10$.  $\al>0.1$ corresponds to  $s<0.74$. As
discussed in section \ref{velocity_power}, for  $s=0.74$ our
approximations for the velocity power spectrum overestimate the true result by
about 16\%.
\begin{figure}[htb!]
\begin{center}
\includegraphics[height=8cm]{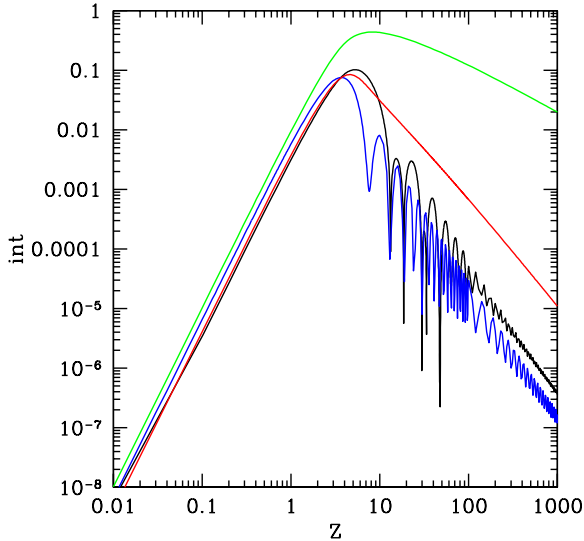}
\caption{\label{fig6} Integrals determining the GW spectra as a
  function of $Z=k \vout/\tilde{\beta}$ for different approximations,
  in the detonation case with $\al=1/2$.
The top (green), bottom (blue, oscillating) and middle (red) curves are
  respectively the incoherent  (Eq.~\ref{psGWinco}),  coherent
  (Eq.~\ref{psGWco}) and top-hat in wave number
  (Eqs.~(\ref{psGWmix2},\ref{psGWmix3})  with $x_c=0.9\pi$)
  approximations. The middle line with the spikes (black)  is the
  absolute value of the approximation with top-hat in the velocity correlation
  (Eq.~\ref{psGW}). The spikes represent the passages through zero of
  this unphysical spectrum. All the spectra are comparable at large scales
  and have a similar peak frequency. The incoherent approximation
  overestimates the peak amplitude by  nearly an order of magnitude.}
\end{center}
\end{figure}

Let us first discuss the shape of the spectrum. We have four different
approximations for the unequal time correlator: the top-hat unequal time
correlator for the velocity field in real space, Eq.~(\ref{psGW}), the
incoherent case Eq.~(\ref{psGWinco}), the coherent case
Eq.~(\ref{psGWco}), and the top-hat unequal time correlator for the
anisotropic stress in Fourier space,
Eqs.~(\ref{psGWmix2},\ref{psGWmix3}). They give rise to comparable
spectra. In Fig.~\ref{fig6} we show the result of the integrals as a
function of $Z=k \vout/\tilde{\beta}$. We have fixed the values
$\al=1/2$ and $x_c=0.9\pi$. The spectra increase like $k^3$ at large
scales, and have comparable amplitudes. This is expected, since the
details of the time correlations should not matter at sufficiently
large scales and the source is uncorrelated in real space
(\textit{cf}. Appendix C). Also, the positions of the peak
approximatively coincide in the four cases, and correspond  to $Z_{\rm
  peak}\simeq 4.6$, so $k_{\rm peak}\simeq 4.6
\tilde{\beta}/\vout=4.6/R(\eta_{\rm fin})\simeq \pi/R(\eta_{\rm
  fin})$. This is comparable to the peak of the anisotropic
stress power spectrum $k\simeq 3/R(\tau)$, due to the fact that GW
production accounts for the full evolution of the bubbles, and
$R(\tau)\leq R(\eta_{\rm fin})$ for all times $\tau<\eta_{\rm
  fin}$. The peak of the GW spectrum is
independent of time and corresponds to the mean bubble radius at the
final stages of the phase transition, close to $R(\eta_{\rm fin})$. The
amplitude at the peak is roughly the same for the two top-hat
approximations and for the coherent case, with the value of the integral at the peak of the order of $0.08\pm 0.02$.
Within the precision of our analytical evaluation this
difference is negligible. On the other hand, the incoherent case
overestimates the amplitude by a factor 5. The small scale
part of the spectrum is also different in the incoherent case with
respect to the others: it decays slower, as $Z^{-0.8}$ as opposed to
$Z^{-\beta}$ with $\beta =2\pm 0.2$. The reason for this behaviour is
apparent by looking at the 
integral in Eq.~(\ref{psGWinco}): once $Z$ has overcome the value at
which the integrand peaks, the contribution from the integral becomes
small and the function decays nearly like $Z^{-1}$. Conversely, in the cases of
Eqs.~(\ref{psGWco},\ref{psGWmix2},\ref{psGWmix3}), at sufficiently
large scales the spectrum scales nearly like $Z^{-2}$. We
explain the small scale decay on the basis of dimensional arguments in
Appendix C. The two top-hat approximations are in very good agreement
up to $Z\simeq 10$. For larger $Z$, the spectrum obtained for the
top-hat in the velocity correlation becomes negative. This is due to
the contribution from the first integral in Eq.~(\ref{psGW}), which is
negative after $sZ> 8$. Its absolute value is between the totally
coherent approximation and the top-hat in wave-number 
approximation. We consider the top-hat in wave-number ansatz which is given in
Eqs.~(\ref{psGWmix2},\ref{psGWmix3}) to be the best approximation for
the unequal time correlators.
\begin{figure}[htb!]
\begin{center}
\includegraphics[height=7cm]{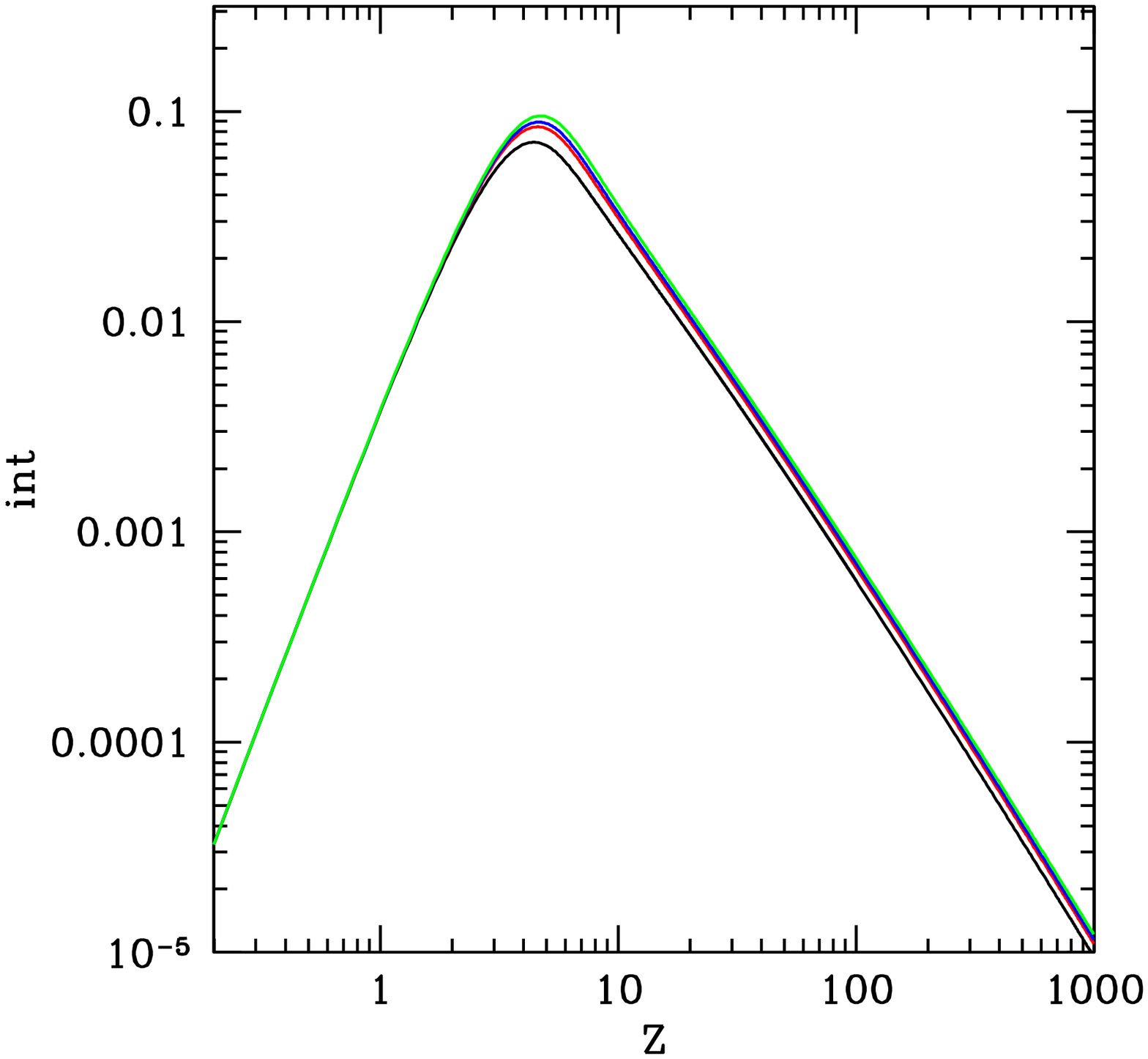}
\includegraphics[height=7cm]{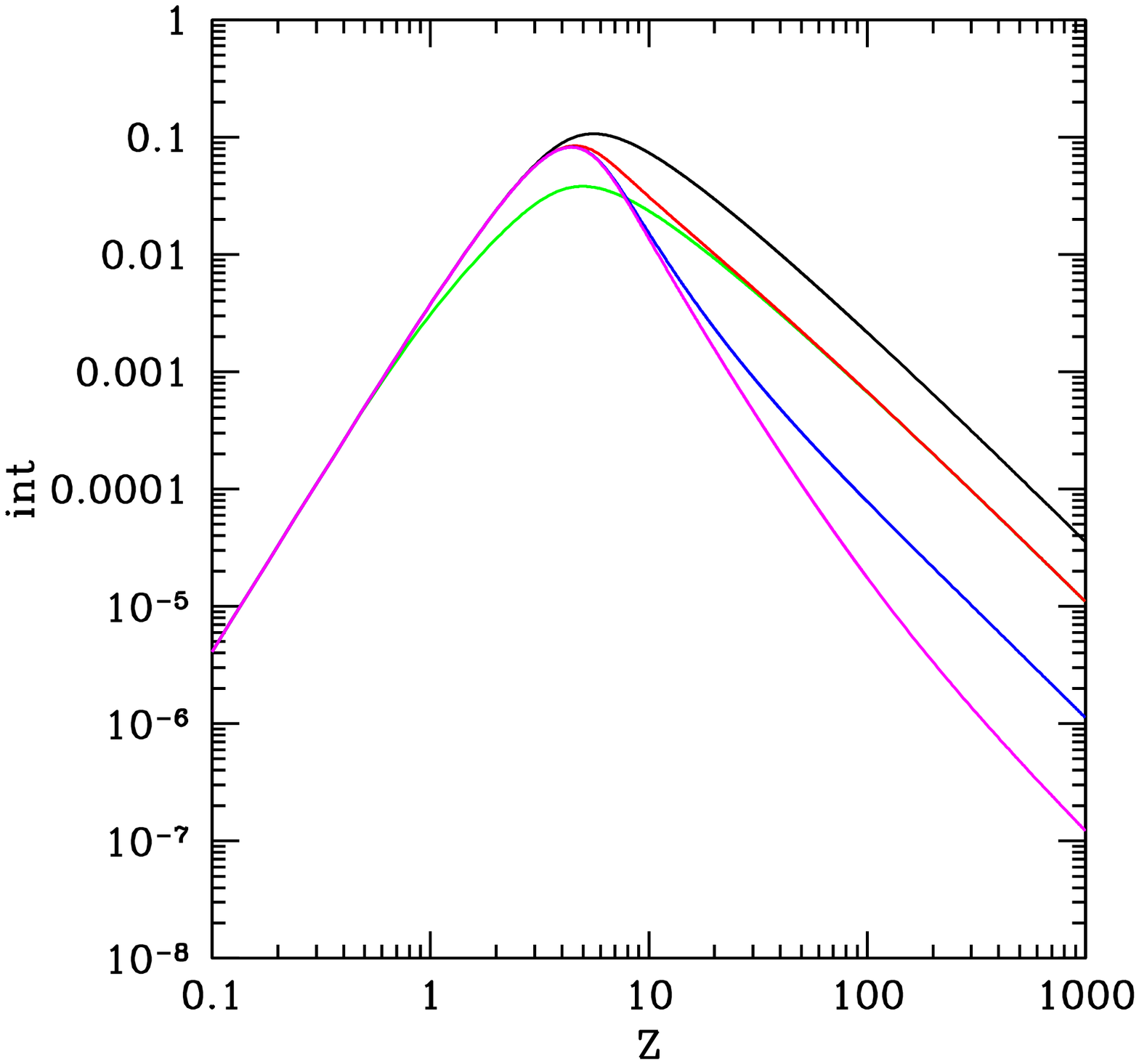}
\includegraphics[height=7cm]{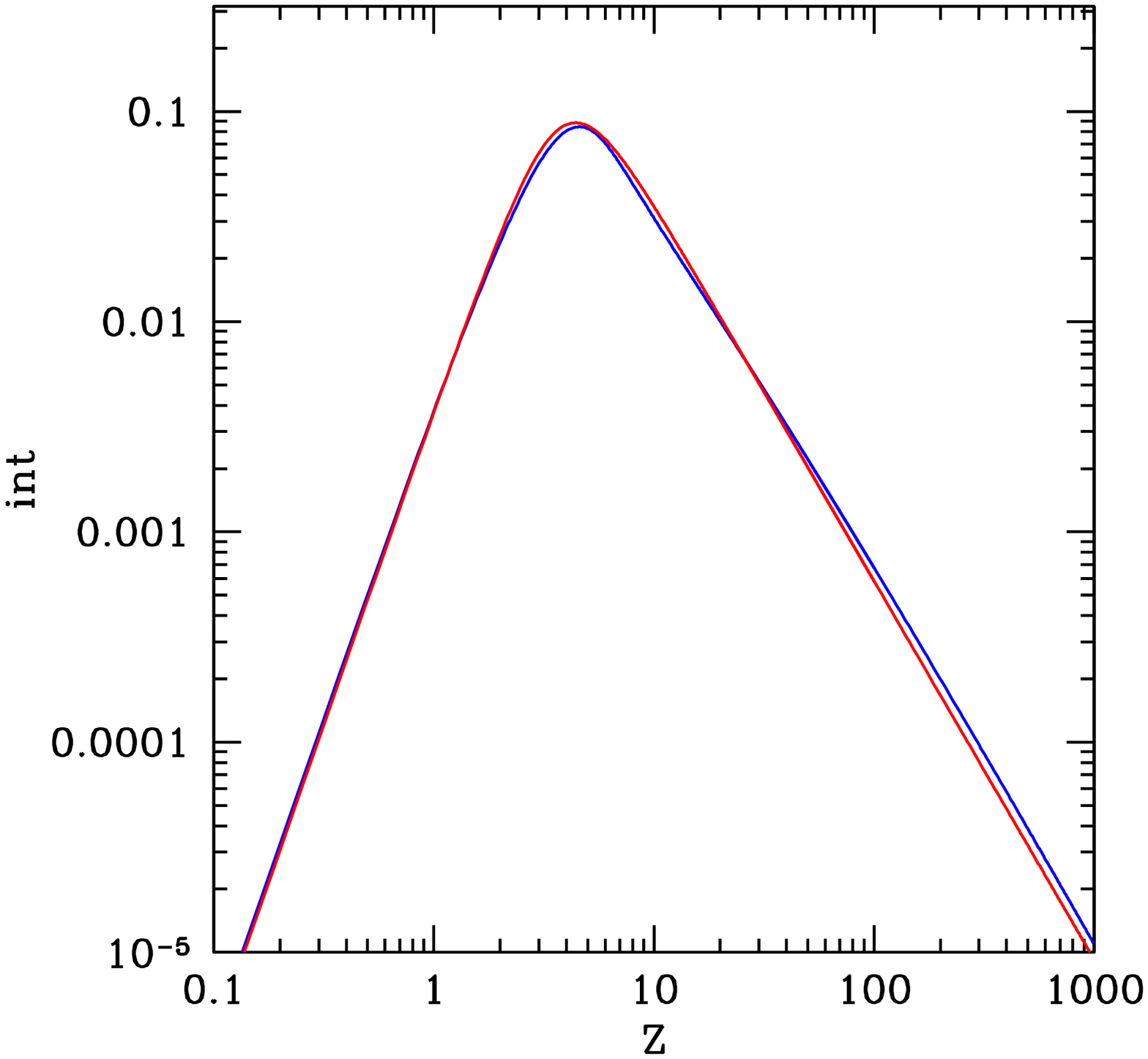}
\caption{\label{fig7}  The top left panel shows the weak
  $\al$-dependence of our result Eqs.~(\ref{psGWmix2},\ref{psGWmix3}):
  the different (black, red, blue and green) curves are respectively for  for
  $\al=0.1, 1/2, 1, 10$. The top right panel shows  the dependence on
  $x_c$, for fixed $\al=1/2$. We remind that $x_c$ defines the time
  interval $x_c/k$ beyond which the correlator of the anistropic
  stress tensor vanishes. The lines from top to bottom (black, red
  green, blue and magenta) plot the integral in
  Eqs.~(\ref{psGWmix2},\ref{psGWmix3}) evaluated at $x_c=\pi/2,
  0.9\pi, 0.1\pi, 0.99\pi, 0.999\pi$  respectively. The lower panel shows
  the GW spectrum of Eqs.(\ref{psGWmix2},\ref{psGWmix3}) and its
  approximation  Eq.(\ref{psGWdet}) for $\al=1/2$ and $x_c=0.9\pi$.}
\end{center}
\end{figure}

In the top left panel of Fig.~\ref{fig7} we show that the integral of
Eqs.~(\ref{psGWmix2},\ref{psGWmix3}) is only weakly dependent on
$\al$. The parameter $\al$ plays a greater role in determining the
overall amplitude of the GW signal (which will be discussed later
on). On the other hand, the shape of the spectrum depends significantly on the choice of
$x_c$ once  $x_c$ approaches $\pi$. We remind that $x_c$ defines the
time interval $x_c/k$ beyond which the correlator of the anistropic
stress tensor vanishes.  In the top right panel of Fig.~\ref{fig7} we
fix $\al=1/2$ and vary $x_c$. Values of $Z$ around the peak correspond
to the region $Z>x_c\vout$, where the result is dominated, for a wide
range of values of $x_c$, by the first integral in
Eq.~(\ref{psGWmix3}). The pre-factor of this integral decays like
$Z^{-2}$ at high 
enough values of $Z$. This is roughly the decay we see in Fig.~\ref{fig7} for
low $x_c$. When $x_c$ becomes very close to $\pi$, the factor
$\sin(x_c)$ multiplying the first integral in Eq.~(\ref{psGWmix3})
becomes so small that the contribution from the second integral in
(\ref{psGWmix3}) takes over. For high values of $Z$, the second
integral decays much faster than $Z^{-2}$. However, because of the
integration limits, it is more and more suppressed as $Z$ increases:
therefore, at some given $Z$ value, the first integral takes over
again and the spectrum decays roughly like $Z^{-2}$. More precisely we
find a $Z^{-1.8}$ decay at large $Z$. This behavior is reached for higher
and higher values of $Z$ as $x_c$ approaches $\pi$, as is shown in
Fig.~\ref{fig7}. For our final results, we choose the value
$x_c=0.9\pi$. This may underestimate the signal somewhat at high
frequencies compared to values $x_c\sim \pi/2$, but this is a
reasonable conservative choice.  Values $x_c$ still closer to $\pi$,
 would be unjustifiably fine-tuned.

The following approximate gravitational wave power spectrum  is fairly
general. We chose $\alpha=1/2$ but as shown in Fig.~\ref{fig7}, the
shape of the spectrum is almost insensitive to the value of $\alpha$;
the $\alpha$ dependence is essentially only in the prefactor
$({\Om_{\rm kin}^*}/{\Om_{\rm rad}^*})^2$ and implicitly in
${(1-s^3)^2}/{s^4}$ :

\bigskip

\fbox{\vbox{
\bea
\left.\frac{d \Omega (k,\eta_0)h^2}{d \ln k} \right|_{\rm deto}&\simeq& \frac{3}{2\pi^3}
\left(\frac{g_0}{g_*}\right)^{\frac{1}{3}}\Om_{\rm rad}h^2
\left(\frac{\Om_{\rm kin}^*}{\Om_{\rm rad}^*}\right)^2 
\left(\frac{\mathcal{H}_*}{\tilde{\beta}}\right)^2 
\frac{(1-s^3)^2}{s^4}\times {\rm int}|_{\rm deto}(Z) \nonumber \\
\mbox{ where } ~~~~{\rm int}|_{\rm deto}(Z) &=& \frac{0.21
  \left(\frac{Z}{Z_{\rm m}}\right)^3}{1+\frac{1}{2}
  \left(\frac{Z}{Z_{\rm m}}\right)^2 + \left(\frac{Z}{Z_{\rm m}}\right)^{4.8}} ~,
\qquad Z_{\rm m} =3.8~.
\label{psGWdet}
\eea}}

\bigskip

\noindent 
This approximation for int$(Z)$ is shown in the lower panel of
Fig.~\ref{fig7}, and we remind that $\Om_{\rm kin}^*/\Om_{\rm rad}^*$
is defined as a function of $s$ and $\vf$ in Eq.~(\ref{Omkin}). 
\begin{figure}[htb!]
\begin{center}
\includegraphics[height=7cm]{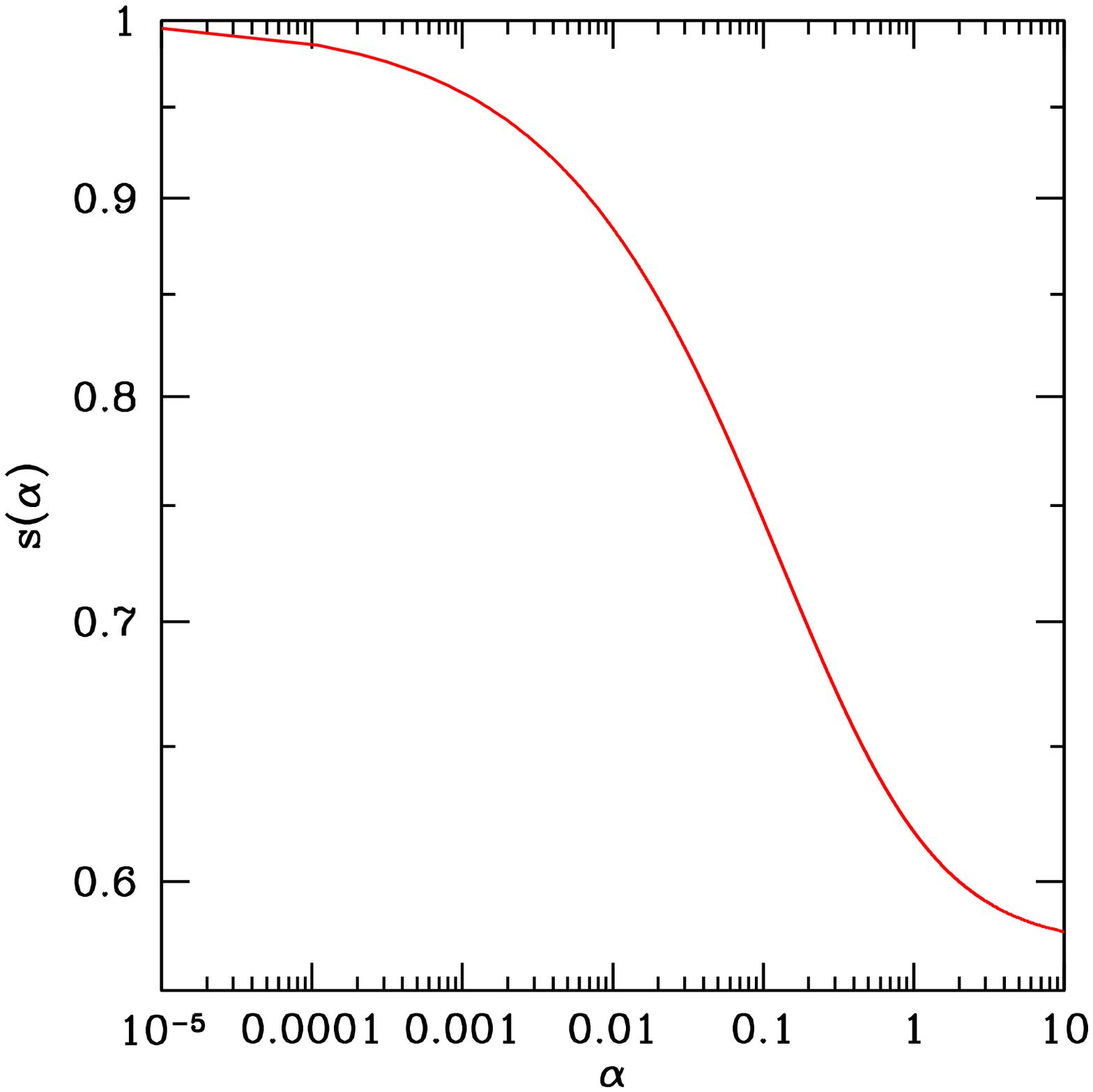}
\includegraphics[height=7cm]{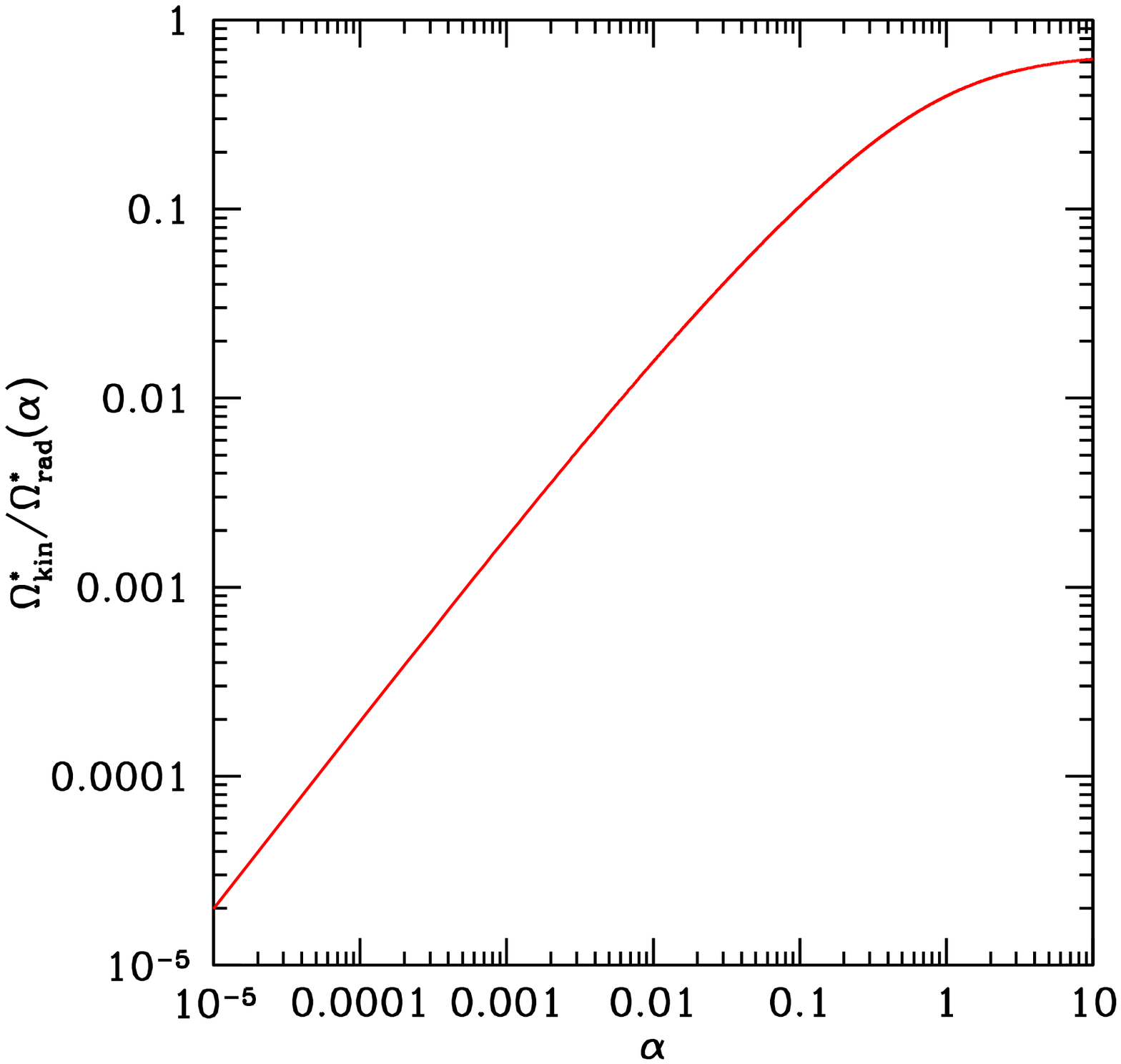}
\caption{\label{fig9} In the left panel we show
  $s(\al)=c_s/v_b(\alpha)$ and in the right panel 
  $\Om_{\rm kin}^*/\Om_{\rm rad}^*$, as functions of $\al$, for the Jouguet
  detonation case.} 
\end{center}
\end{figure}
\begin{figure}[htb!]
\begin{center}
\includegraphics[height=7cm]{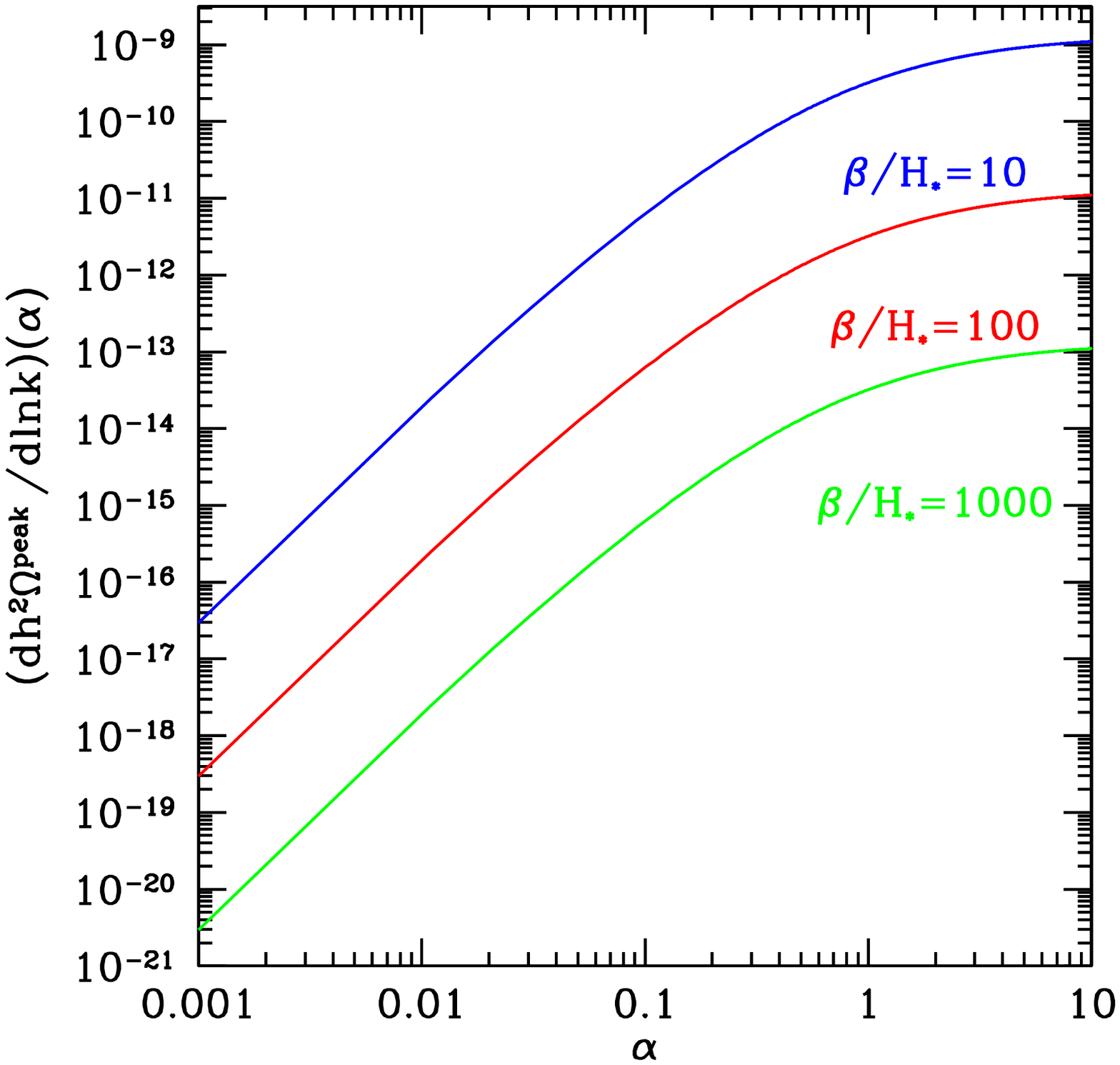}
\includegraphics[height=7cm]{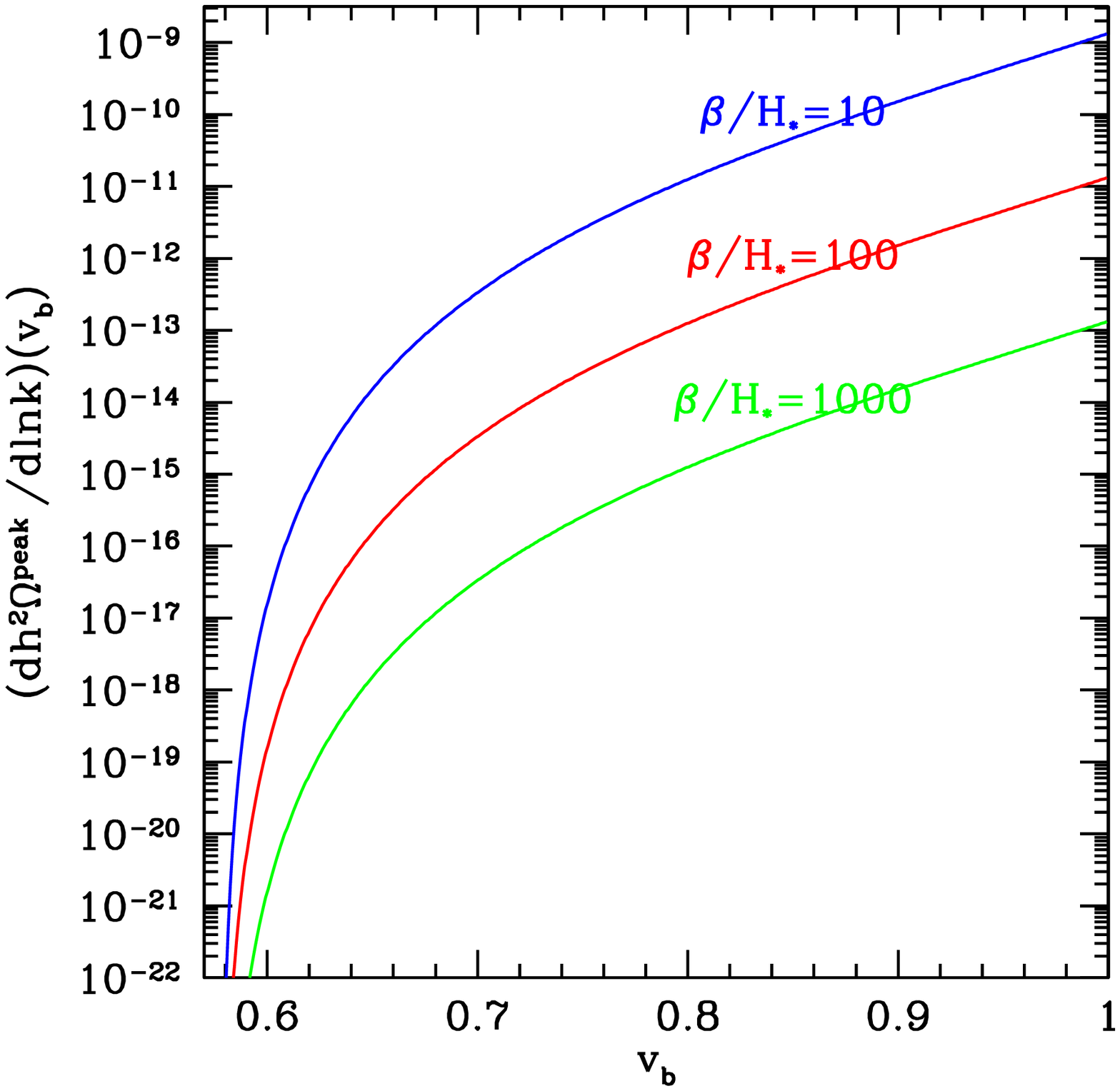}
\caption{\label{fig10} Amplitude of the GW signal at the peak
  frequency from Jouguet detonations for
  $\tilde{\beta}/\mathcal{H}_*=10,100,1000$, as a function of $\al$
  (left) and as a function of $v_b$. 
 The signal reaches a plateau at large $\al$ since it is bounded by the
  maximal possible value of the bubble wall velocity, $v_b<1$.}
\end{center}
\end{figure}

Let us now investigate the overall amplitude of the signal at the peak
frequency. 
\be
\left.\frac{d \Omega (k,\eta_0)h^2}{d \ln k}\right|_{\rm peak}\simeq
\frac{3}{2\pi^3}\left(\frac{g_0}{g_*}\right)^{\frac{1}{3}}\Om_{\rm rad}h^2 
\left(\frac{\Om_{\rm kin}^*}{\Om_{\rm rad}^*}\right)^2 
\left(\frac{\mathcal{H}_*}{\tilde{\beta}}\right)^2 
\frac{(1-s^3)^2}{s^4}\, 0.084~,
\label{amplitude}
\ee
where  the factor $0.084$ accounts for the contribution of the integral
at the peak frequency $k\simeq 4.6\tilde{\beta}/\vout$ ({\it cf.} low 
panel of Fig.~\ref{fig7}). In order for the bubbles to percolate
and convert the entire universe to the broken phase, the phase
transition must occur much faster than one Hubble time. We study the
values of $\tilde{\beta}/\mathcal{H}_*\sim
\mathcal{O}(10),\mathcal{O}(100),\mathcal{O}(1000)$. Since the
velocity of the bubbles $\vb=\vout$ is fully determined by the
parameter $\al$, so is the fluid velocity $\vf$ and, in turn, the
factor $s=c_s/\vout$ and the mean kinetic energy parameter $\Om_{\rm
  kin}^*/\Om_{\rm rad}^*$, see Eqs.~(\ref{vbdet},\ref{vfdet},\ref{sdet}).
We plot $s$ and $\Om_{\rm kin}^*/\Om_{\rm rad}^*$ as a function of
$\al$ in Fig.~\ref{fig9}. Specifying $\al$ and $\tilde{\beta}$ fully
determines the amplitude. Substituting  in (\ref{amplitude}), and
setting $a=\sqrt{3\al^2+2\al}$, we find:
\be
\left.\frac{d \Omega (k,\eta_0)h^2}{d \ln k}\right|_{\rm peak}\simeq
1.7\cdot 10^{-6}\left(\frac{\mathcal{H}_*}{\tilde{\beta}}\right)^2
\frac{(\al-a)^4 (-3a-(3+2a)\al-3(2+a)\al^2+\al^3)^2}{(1+a)^2(1+a+
5a\al-(5-3a)\al^2-3a\al^3+6\al^4)}
\label{amplia}
\ee
which we plot in Fig.~\ref{fig10}.

The peak amplitude depends quadratically on the duration of the phase
transition, $\tilde{\beta}^{-1}$ and grows steeply (also roughly
quadratically) with $\al$ as long as $\al <1$. At $\al>1$ the
dependence on $\al$ becomes rather weak, since the bubble velocity
approaches its maximum, $v_b\ra 1$. The maximum gravity wave density
parameter which can be achieved by bubble collisions in a strongly
first order phase transition $\al > 1$, $\vb\simeq 1$ and $\tilde{\beta} \sim
10\mathcal{H}_*$ is of the order of $5\times 10^{-10}$. Note that our
approximations break down for smaller values of $\tilde\beta$, since
we have neglected the factor $\ln(M/m)$ so that $\tilde\beta^{-1}$ becomes
the duration of the phase transition which must be significantly smaller than
a Hubble time.

\subsection{GW from deflagrations}
\label{GWdef}

In the case of deflagrations, $R(t)=\vout t$ coincides with the shock
front of the shock wave taking place in the symmetric
phase. Therefore, we set $\vout=v_{\rm shock}$. The inner radius is
the bubble wall, so that $\vint=\vb=v_2$. The fluid velocity is given
by 
\be
\vf=\frac{\vb-v_1}{1-\vb v_1}
\ee
where $v_1$ is the incoming velocity of the symmetric phase fluid into
the front, in the rest frame of the discontinuity. 

In order to estimate the gravitational wave production, we follow
the analysis of deflagrations presented in Appendix A of
Ref.~\cite{Kamionkowski:1993fg}. There, the authors numerically
integrate the energy-momentum conservation equations at the
deflagration front with different initial conditions. The velocity
profile they find is in principle different both from the constant one
occurring in planar deflagrations, and from the linear one that we
took in our simplified analysis. However, we choose two different
cases which they have analyzed, for which the deflagration is quite
strong and the velocity profile is actually almost constant. In the
first case, they set $\vb=0.1$ and $\vf=0.09$, and find $v_{\rm
  shock}\simeq 0.59$ (and consequently $s=\vint/\vout=\vb/v_{\rm
  shock}=0.17$). In the second case, close to Jouguet deflagrations,
they set $\vb=0.5$ and $\vf=0.45$, and find $v_{\rm shock}\simeq 0.73$
(and consequently $s=0.68$). The velocity profiles for these values
are shown in Fig.~9 of \cite{Kamionkowski:1993fg}. Since in the case
of deflagrations there is no simple relation among the velocity of the
shock front and the parameter $\al$ denoting the strength of the phase
transition, we simply leave $v_{\rm shock}$ as a free parameter in our
analysis, for which we take the two values given above.

\begin{figure}[htb!]
\begin{center}
\includegraphics[height=7cm]{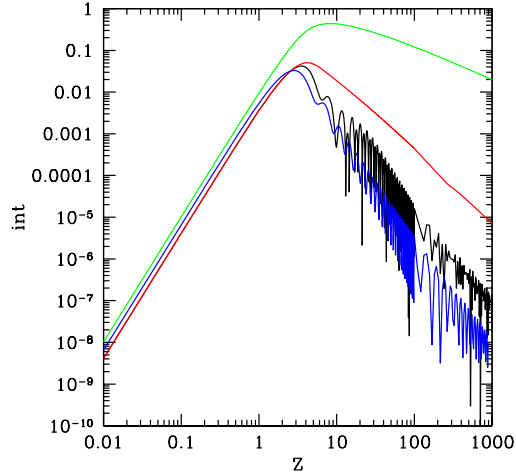}
\caption{\label{fig11} Same as Fig.~\ref{fig6} but for the case of
  deflagrations. We fix  $v_{\rm shock}=0.59$. Compared to
  detonations, the oscillations are enhanced in approximation
  Eq.~(\ref{psGW}) and (\ref{psGWco}) corresponding to top-hat in the velocity
  correlation.}
\end{center}
\end{figure}

The analysis of the shape of the spectrum is on the same footing as for
detonations. The results of the unequal time correlator for the
anisotropic stress are shown in Fig.~\ref{fig11} for fixed values of
$v_{\rm shock}=0.59$ and $x_c=0.9\pi$. The results for the incoherent
and top-hat  in the anisotropic stress cases are similar to the case of
detonation. We recover the expected $k^3$ behaviour at large scales,
and the amplitudes are comparable at low wave number. The peak is
located at $Z_{\rm peak}\simeq 4.2$, corresponding to $k_{\rm
  peak}\simeq 4.2\, \tilde{\beta}/\vout=4.2/R(\eta_{\rm fin})\simeq
\pi/R(\eta_{\rm fin})$. On the other hand, the coherent and top-hat in the
velocity correlation cases are slightly different in the high frequency
range, because the frequency of the oscillations is increased,
due both to the smaller value of $\vout=v_{\rm shock}=0.59$, and to
the smaller value of $s=0.17$. These parameters in fact appear in the
arguments of the Green functions in Eq.~(\ref{psGW}) and (\ref{psGWco}). In the region
where the power spectrum becomes negative, in the top hat in velocity case, the solution is no longer
reliable. We therefore discard it and take as best approximation the
unequal time correlator of the anisotropic stress power spectrum, as
we did for detonations.

\begin{figure}[htb!]
\begin{center}
\includegraphics[height=8.cm]{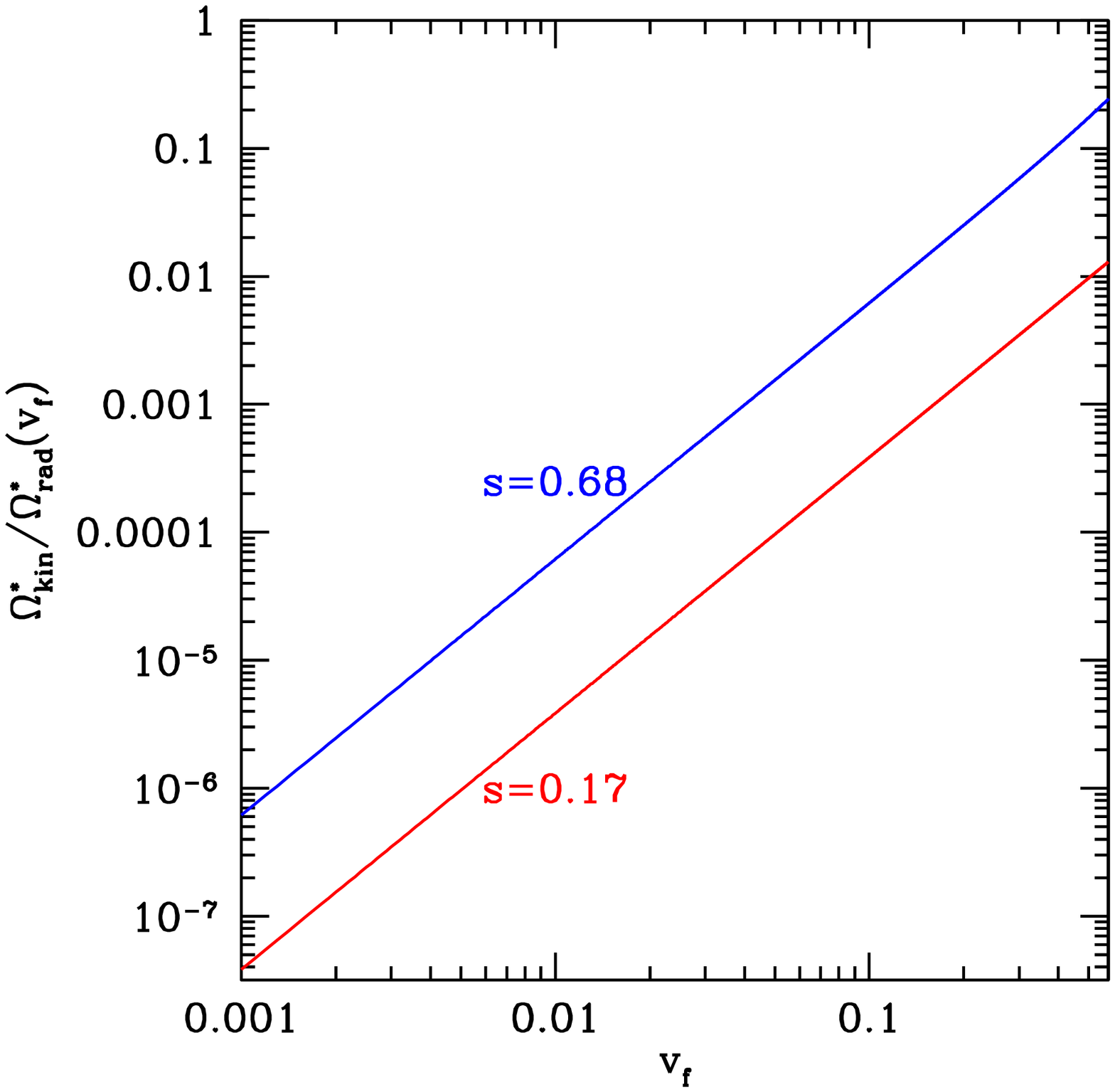}
\includegraphics[height=8.cm]{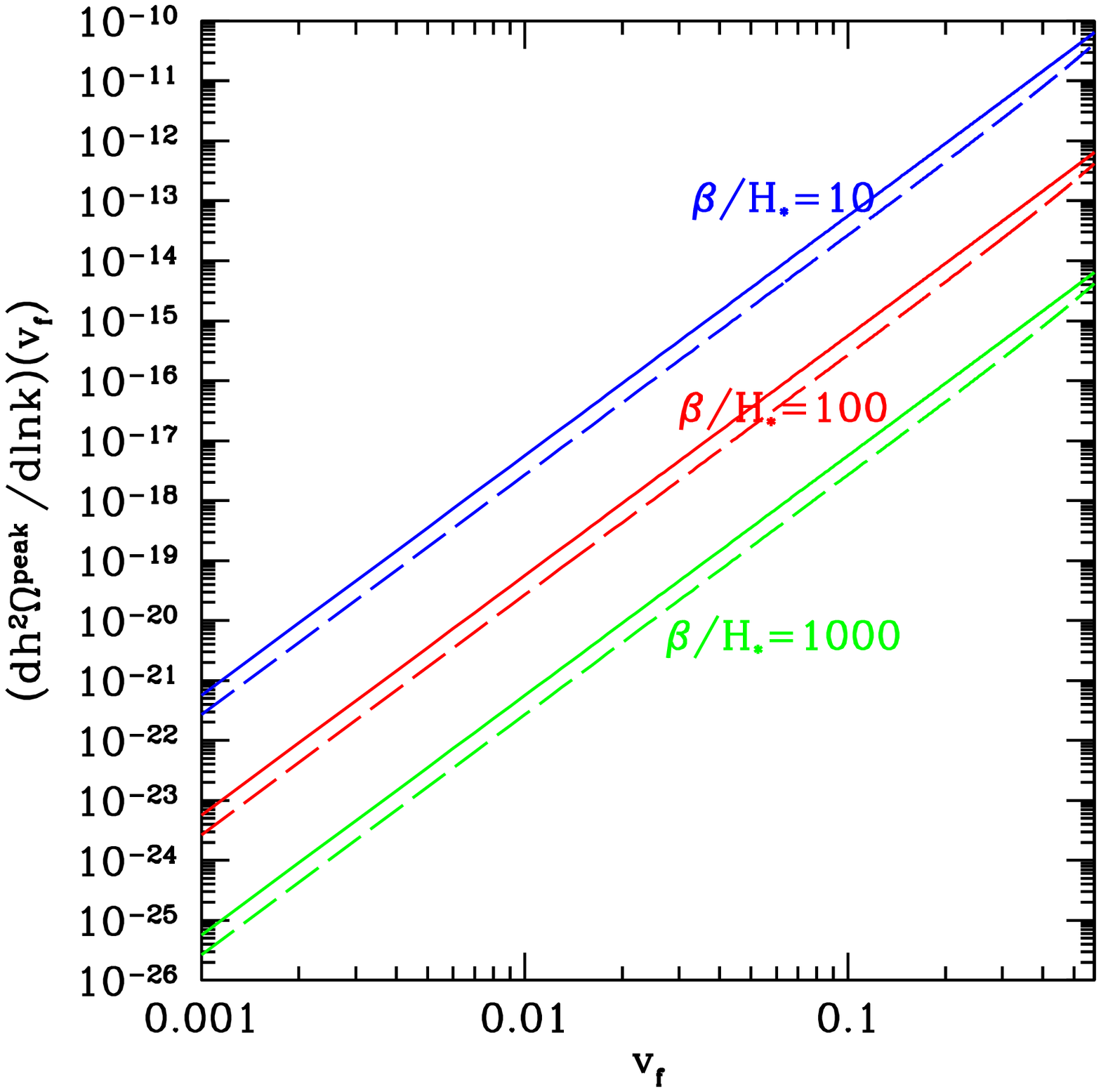}
\caption{\label{fig14} The case of deflagrations: The left plot shows
  $\Om_{\rm kin}^*/\Om_{\rm rad}^*$
as a function of $\vf$. The red curve corresponds to the first case
  analyzed in \cite{Kamionkowski:1993fg}, for which $s=0.17$, with
  $\vb=0.1$ and $v_{\rm shock}\simeq 0.59$
and the blue one to the second case with $s=0.68$, $\vb=0.5$ and
  $v_{\rm shock}\simeq 0.73$. The right plot shows the amplitude of
  the GW signal at the peak frequency given in
  Eq.~(\ref{amplitudedef}) as a function of $\vf$ for different values
  of $\tilde\beta/\mathcal{H}_*$. Solid lines are for  $s=0.68$
  corresponding to $\vb=0.5$ and $v_{\rm shock}\simeq 0.73$, and
  dashed ones for $s=0.17$ corresponding to $\vb=0.1$ and $v_{\rm
  shock} \simeq 0.59$. As long as $sv_{\rm
  f}\ll 1$, we have $\Om_{\rm kin}^*/\Om_{\rm rad}^* \propto v_{\rm
  f}^2$ and correspondingly $d\Om^{\rm peak}h^2/d\ln k\propto v_{\rm
  f}^4$. }
\end{center}
\end{figure}

The result of Eqs.~(\ref{psGWmix2},\ref{psGWmix3}) is very weakly
dependent on the value of the velocity $v_{\rm shock}$. This is
analogous to the weak dependence on $\al$ found for detonations.
We do not display this dependence as it is very similar to the top
left panel of Fig.~\ref{fig7} where the role of $\al$ is played by
$v_{\rm shock}$ which is varied between 0.59 and 0.73. Correspondingly,
when varying $x_c$, we obtain a dependence similar to the one shown in
Fig.~\ref{fig7}. For fixed $x_c=0.9\pi$ and $v_{\rm shock}=0.59$ we
can approximate the result as follows:

\bigskip

\fbox{\vbox{
\bea
\left.\frac{d \Omega (k,\eta_0)h^2}{d \ln k}\right|_{\rm defla} &\simeq& \frac{3}{2\pi^3}
\left(\frac{g_0}{g_*}\right)^{\frac{1}{3}}\Om_{\rm rad}h^2 
\left(\frac{\Om_{\rm kin}^*}{\Om_{\rm rad}^*}\right)^2 
\left(\frac{\mathcal{H}_*}{\tilde{\beta}}\right)^2 
\frac{(1-s^3)^2}{s^4}\times {\rm int}|_{\rm defla}(Z) \nonumber\\
\mbox{where} ~~~
{\rm int}|_{\rm defla}(Z) &=& \frac{0.23 \left(\frac{Z}{Z_{\rm m}}\right)^3}{1+2
  \left(\frac{Z}{Z_{\rm m}}\right)^2 + \frac{3}{2}
  \left(\frac{Z}{Z_{\rm m}}\right)^{4.8}}~, \qquad Z_{\rm m}=3.6\,.
\label{psGWdef}
\eea}}

\bigskip

\noindent
This is essentially identical to Eq.~(\ref{psGWdet}) except for the
values of $s$ and $\vf$ in ${\Om_{\rm kin}^*}/{\Om_{\rm rad}^*}$: In
the case of deflagrations we cannot reduce the dependence exclusively
to the two parameters $\al$ and $\tilde\beta$, since no direct relation
between the velocities of the shock front, the velocity of the bubble
wall and the strength of the phase transition $\al$ is known in
general. In this case, the 
velocity of the shock front is expected to depend also on the properties of the
ambient plasma, like friction, and not only on the strength of the phase
transition determined by $\al$. The amplitude at the peak frequency is
the same as the detonation formula (\ref{amplitude}) except that  0.084
is replaced by 0.050. This factor accounts for the contribution of the
integral at the peak frequency $k\simeq 4.2\tilde{\beta}/\vout$ ({\it
  cf.} Fig.~\ref{fig11}): 
\be
\left.\frac{d \Omega (k,\eta_0)h^2}{d \ln k}\right|_{\rm peak}\simeq
\frac{3}{2\pi^3}\left(\frac{g_0}{g_*}\right)^{\frac{1}{3}}
\Om_{\rm rad}h^2 \left(\frac{\Om_{\rm kin}^*}{\Om_{\rm rad}^*}\right)^2 
\left(\frac{\mathcal{H}_*}{\tilde{\beta}}\right)^2\frac{(1-s^3)^2}{s^4}\, 0.050
\label{amplitudedef}
\ee

 The first plot of  Fig.~\ref{fig14} shows $\Om_{\rm kin}^*/\Om_{\rm
 rad}^*$  as a function of the fluid velocity $\vf$ for two values of
 $s=\vb/v_{\rm shock}$, corresponding to the cases analyzed in
 \cite{Kamionkowski:1993fg}. Obviously, the value of $\vf$ is in
 principle fixed once $\al$, $\vb$ and $v_{\rm shock}$ are given;
 however, the relation among these parameters is not known
 explicitly. Therefore, we have decided to keep $\vf$ as a free
 parameter. The fluid velocity $\vf$ induces the biggest variation in
 the amplitude of the signal, and is therefore the most relevant
 parameter determining $\Om_{\rm kin}^*/\Om_{\rm rad}^*$. This appears
 clearly in the second plot of Fig.~\ref{fig14}, where we show the
 amplitude of the GW signal Eq.~(\ref{amplitudedef}) as a function of
 $\vf$, for the same two values of $s$ and varying
 $\tilde\beta/\mathcal{H}_*$. The dependence on $s$ is negligible
 compared to the dependence on $\vf$.

\section{Some comments on our approach}
\label{section:comments}

In numerical simulations, the stochastic background of GWs arises from
averaging over several deterministic realizations of bubble
collisions. In our approach instead,  to account for the intrinsic
randomness of the nucleation process, we define the bubble velocity as
a random variable. The source of the stochastic background of GWs is
the tensor part of the anisotropic stress of the stochastic,
homogeneous and isotropic velocity field. 

The calculation of the velocity correlation function in
Sec.~\ref{velocity_power}  implicitly assumes that the bubbles, and
consequently the velocity configuration, are all independent from each
other: the velocity correlation function is different from zero only
if the points $\bx$ and $\by$ are in the same bubble. Therefore, one
may wonder in which sense this procedure is a model of bubble
collisions. 
Indeed, in our model the only non-vanishing correlation coming from a
(possibly) collided region is given by the sum of correlations  from
two independent bubbles: for every $\bx$ and $\by$ we can find a
bubble center position such that a bubble encompasses the two points,
giving a non-zero result, provided that the probability to find a
center in that position is not zero. This is accounted for by
multiplying the correlation with the probability that a given point is
in the broken phase at a given time, {\it cf.}
Eq.~(\ref{correlationintegral}).

Therefore, even though we do not model in a deterministic way the
resulting velocity field from the collision of bubbles, we do account
for their overlap; it is the overlap from several bubbles which gives
us a non-zero correlation function for the tensor part of the
anisotropic stress, as opposed to the spherically symmetric situation
described in Appendix A. The anisotropic stress correlation function,
in fact, involves the four-point correlation function of the velocity
field, {\it cf.} Eq.~(\ref{Tpowspec}). This quantity, in contrast with
the two-point correlation function, is non-zero also for points $\bx$
and $\by$ belonging to different bubbles\footnote{$ \langle
  v_a({\bf x})v_b({\bf x})v_c({\bf y})v_d({\bf y})\rangle \supset  
\langle v_a({\bf x})v_b({\bf x})\rangle \langle v_c({\bf y})v_d({\bf y})
\rangle$.}.  
 In other words, when we calculate the two-point correlation function
 of the velocity field Eq.~(\ref{correlationintegral}), we break
 spherical symmetry since every point in space becomes equivalent to
 another and there is no longer a center of symmetry. The center of
 symmetry $\bx_0$ has become a random variable, and we average over
 all its possible positions. Using the ergodic theorem, this is
 equivalent to an average over several realizations for the center
 positions, {\it i.e.} several realizations of the nucleation process,
 {\it i.e.} several realizations of the velocity field
 distribution. In this way, we account for the overlap of several
 bubbles, which breaks spherical symmetry and leads to a non-zero
 power spectrum of the tensor anisotropic stress, generating
 gravitational radiation.

\section{Summary}

Since the GW signal from bubble collisions was already evaluated in
Refs.~\cite{Kosowsky:1992rz,Kosowsky:1991ua,Kosowsky:1992vn,Kamionkowski:1993fg},
in this section we gather our final formulas and compare them with the
ones given so far in the literature. We conform to the notation used
previously, for the comparison to be straightforward. The relevant
formulas are compiled for instance in Ref.~\cite{Nicolis:2003tg}:
there, Eq.~(4) gives the frequency at which the GW spectrum peaks and
Eq.~(5) shows the dependence of the amplitude of the GW signal at the
peak frequency on the relevant parameters of the bubbles
evolution. These formulas are valid in the case of Jouguet
detonations, and do not show how the GW spectrum depends on other than
the peak frequency. Before proceeding with the comparison, we recap
the main assumptions in our calculation: 
1) we assume radiation domination; 2) the fluid velocity profile
inside the bubble is linear; 
3) $v_i$ is our stochastic variable via the bubble center as we
average over all possible positions of the bubble center. Therefore,
we do not model collisions in a deterministic way. We rather account
for all possible configurations of the velocity field which include
the case where bubbles overlap, even though we use the velocity
profiles of {\it uncollided} bubbles; 
4) we use the Wick theorem to express the 4-point correlation function
in terms of the 2-point correlation function, even if the velocity is
presumably not a gaussian random variable; 
5) all bubbles have the same size $R=\vb (\eta-\eta_{\mbox{\tiny in}})$;
6) we use the top hat approximation for the correlator of the
anisotropic stress tensor at different times; 
7)  the enthalpy $w$ and the Lorentz factor $\gamma$ do not depend on $\bf{x}$.

Using the notation 
\be
h^2\Omega_{\mbox{\tiny coll}} \  (f)\equiv 
\frac{d \Omega (k,\eta_0)h^2}{d \ln k} \ \ \mbox{where} \ \ \ f=\frac{k}{2\pi}
 \ \ \ \mbox{and}\ \ \
 h^2\Omega_{\mbox{\tiny peak}} \equiv \left.\frac{d \Omega
   (k,\eta_0)h^2}{d \ln k}\right|_{\rm peak} ~,
 \ee
from the results given in Section \ref{subsection:DETO},
Eqs.~(\ref{psGWdet}, \ref{amplitude}) and Section \ref{GWdef},
Eqs.~(\ref{psGWdef}, \ref{amplitudedef}), we find that 
$h^2\Omega_{\mbox{\tiny coll}} (f)$ increases at small $f$ as $f^3$ (compared to $f^{2.8}$ in Kosowsky et al.
\cite{Kosowsky:1991ua}), and at large $f$ it scales as $f^{-1.8}$ (as in Kosowsky et al).
The full spectrum is given by

\bigskip

\fbox{\vbox{
\bea
 \hspace*{-0.8cm}\left.h^2\Omega_{\mbox{\tiny coll}} (f)\right|_{\rm deto} &= & 
\left. h^2 \Omega_{\mbox{\tiny
      peak}}\right|_{\rm deto}\,\frac{2.5 \left(\frac{f}{f_{\rm m}}\right)^3}{1 +
  \frac{1}{2}\left(\frac{f}{f_{\rm m}}\right)^2 +
  \left(\frac{f}{f_{\rm m}}\right)^{4.8}}~,~~~
f_{\rm m} = 0.87f_{\mbox{\tiny peak}} \\
 \hspace*{-0.8cm}\left.h^2\Omega_{\mbox{\tiny coll}} (f)\right|_{\rm defla} &= & 
 \left.h^2 \Omega_{\mbox{\tiny
      peak}}\right|_{\rm defla}\,\frac{4.5\left(\frac{f}{f_{\rm m}}\right)^3}{1 +
           2 \left(\frac{f}{f_{\rm m}}\right)^2 +
 \frac{3}{2}  
\left(\frac{f}{f_{\rm m}}\right)^{4.8}}~,~~~ 
f_{\rm m} = 0.86f_{\mbox{\tiny peak}}
\eea}}

\bigskip

\noindent 
Let us first discuss the peak frequency. It is given by
\be
f_{\mbox{\tiny peak}} = \frac{k_{\mbox{\tiny peak}}}{2 \pi}  \ \ \ 
\ \ k_{\mbox{\tiny peak}}\simeq 4.5 \frac{\tilde{\beta}}{\vout}  
\ \  \mbox{where} \ \  \tilde{\beta}=a_* \beta \ \ \mbox{and}
\label{eq:peak_frequency}
\ee

\be
\vout=\vb \ \  \mbox{ for detonations, }  \  \ \vout=v_{\rm shock} \ \
\mbox{ for deflagrations.}
\ee
We remind that in our notations $\beta^{-1}$ expresses the duration of
the phase transition: 
bubbles are generated at the beginning of the phase transition and
collide after a time given approximately by $\beta^{-1}$. We have in
fact set to 1 the logarithm relating $\tilde\beta$ and
$\etafin-\etain$, {\it c.f.} Eq.~(\ref{withlog}) of section
\ref{timedep} (this logarithm can be easily inserted in all the
following equations: it will introduce a multiplying factor in the
ratio $\beta/H$ and will also change the value of the integral
(\ref{psGW})). Moreover, $\vout$ 
corresponds to either  the characteristic bubble velocity (in
detonations) or the shock wave velocity (in deflagrations). The
characteristic frequency at the time of emission is  $f_{\mbox{\tiny
    peak}} /a_*\approx  \beta/\vout$.
As expected, \emph{it is associated with the maximal size of the spherical 
fluid velocity configuration which generates the GW signal}. This can
be either the bubble itself, in the case of detonations, or the
spherical shock front preceding the bubble, in the case of
deflagrations. Since $v_{\rm shock}<\vb$, the peak frequency for
detonations is smaller than that for deflagrations (the size of the
velocity configuration is bigger). Using $\mathcal{H}_*=a_*H_*$ and
$a_*=a_*/a_0=T_0/T_*(g_0/g_*)^{1/3}$, we can rewrite the peak frequency as
\bea
f_{\mbox{\tiny peak}} &\simeq & \frac{4.5}{2
  \pi}\frac{1}{\vout}\frac{\tilde{\beta}}{{\cal
    H}_*}\frac{T_0}{T_*}\left(\frac{g_0}{g_*}\right)^{1/3}H_*
\qquad \mbox {which yields}
\eea
\be
\boxed{
f_{\mbox{\tiny peak}} \simeq
 1.12 \times 10^{-2}\mbox{ mHz} \  {\left(\frac{g_*}{100}\right)}^{1/6} \ \frac{T_*}{100 \mbox{ GeV}}
\ \frac{\beta}{H_*} \ \frac{1}{\vout}
}
\ee
where of course $\tilde{\beta}/{\cal H}_*=\beta/H_*$. 

The above peak frequency is bigger than the one in
Refs.~\cite{Kamionkowski:1993fg,Nicolis:2003tg} by a 
factor $\sim 2/\vout$. The factor $1/\vout$, that shifts the peak to
higher frequencies for low velocities, is absent in
\cite{Kamionkowski:1993fg,Nicolis:2003tg}, and is related to the fact that the
characteristic frequency we find is determined by the size of the
bubbles instead of the duration of the phase transition \cite{Witten:1984rs}. 
The property
of causality of the source directly determines this characteristic
frequency. Since the source is causal, the velocity correlation
function goes to zero at the length-scale corresponding to the size of
the bubbles. The same length-scale determines the peak in the velocity
power spectrum, and from there it is transferred to the GW power
spectrum (\emph{c.f.} also the discussion in
Ref.~\cite{Caprini:2006rd}, where it is explained that gravity waves
emitted during short cosmological events such as phase transitions
typically inherit the wave number and not the frequency of the
source)~\footnote{Note that, being in a cosmological context, we do
  not perform a 
time Fourier transform of the energy momentum tensor in contrast to Eq. (23) of
Ref.~\cite{Kamionkowski:1993fg}. In~\cite{Kamionkowski:1993fg} the 
authors {\it assume} that ``the frequency dependence of the spectrum is set
by the timescale $\beta^{-1}$" (as written before eq. (28)
of~\cite{Kamionkowski:1993fg}).  
Our finding, on the other hand, is that the frequency dependence of the
spectrum is set by the bubble size $R$.  
Since in the detonation regime $\vout$ approaches 1, in the simulations of
Ref.~\cite{Kamionkowski:1993fg} one does not really see this difference.}.

The
factor $2$, instead, comes from the details of our modeling of the source. 
It follows from the factor 4.5 in Eq.~(\ref{eq:peak_frequency}) (see
Fig.~\ref{fig6}  and Fig.~\ref{fig10} and discussion thereafter). 
We found that the
velocity power spectrum has a characteristic wave number $\sim 2.5/R$,
corresponding quite well to that coming from the bubble diameter
$2\pi / 2R$. The time-dependent anisotropic stress power spectrum
instead changes slope around $3/R$ (see Fig.~\ref{fig5}). 
The characteristic wave number of the GW spectrum is
associated to $R_{\mbox{\tiny fin}}$, the typical radius of bubbles at
the end of the phase transition when bubbles collide, and is found to
be  $k_{\mbox{\tiny peak}} \sim 4.5/ R_{\mbox{\tiny fin}}$. This 
differs from the value obtained in Kamionkowski {\it et al.} $k_{\mbox{\tiny
    peak}} \sim 2 \beta$ (see Fig.~7 of
Ref.~\cite{Kamionkowski:1993fg} and the associated uncertainty). 
For $\beta/H_*=100$, $T_*=100$ GeV and $g_*\sim 100$, corresponding to a
typical (first order) electroweak phase transition, we find 
$f_{\mbox{\tiny peak}}\sim 1\mbox{ mHz}/\vout$ to be compared with
LISA's peak sensitivity which is estimated to be 2 mHz. The
increase in the peak frequency that we obtain relative  to
Ref.~\cite{Kamionkowski:1993fg}  is actually welcome for probing the
electroweak phase transition with LISA\footnote{This result should be taken with caution, see the ``Note added" at the end of this Section.}.

Let us now discuss the peak amplitude. A simple order of magnitude
estimate shows how the result depends on the  
duration and the energy density of the source. From the perturbed
Einstein's equations $\de G_{\mu\nu}=8\pi G\, T_{\mu\nu}$, one gets
the following order of magnitude estimate for the amplitude of the
tensor perturbation $h$ (we drop indices for simplicity): 
\be
\beta^2 h \sim 8\pi G \,T
\label{est1}
\ee
 where we inserted $1/\beta$ as the characteristic time on which the
 perturbation is evolving, and $T$ denotes the energy momentum tensor
 of the source. From Eq.~(\ref{Tab}) and definition (\ref{Omkin}), we
 can write 
\be
T\sim \rho_{\rm rad}\, \frac{\Om_{\rm kin}^*}{\Om_{\rm rad}^*}\,.
\ee 
We want to estimate the energy density in gravitational waves, defined in Eq.~(\ref{definition}). The above equation (\ref{est1}) suggests that $\dot h \sim 8\pi G \,T/\beta$, and so we obtain   
\be
\rho_{GW}\sim \frac{\dot h^2}{8\pi G}\sim \frac{8\pi G}{\beta^2}\,T^2 
 \sim 8\pi G\left(\frac{H}{\beta}\right)^2 \frac{\rho_{\rm rad}}{H^2} \, \rho_{\rm rad} \left(\frac{\Om_{\rm kin}^*}{\Om_{\rm rad}^*}\right)^2\,.
\ee
Substituting Friedmann equation in the radiation dominated era
 $H^2=\frac{8\pi G}{3} \rho_{\rm rad}$, and considering that the GW
 energy density evolves like radiation, we obtain for the GW energy
 density today the simple expression 
\be
\Om_{GW} \sim \Om_{\rm rad} \left(\frac{H}{\beta}\right)^2 
\left(\frac{\Om_{\rm kin}^*}{\Om_{\rm rad}^*}\right)^2\,.
\ee
This shows that the GW energy density scales like the square of the
 ratio between the time duration of the source and the Hubble time,
 and the square of the energy density in the source.  

\begin{figure}[htb!]
\begin{center}
\includegraphics[height=5cm]{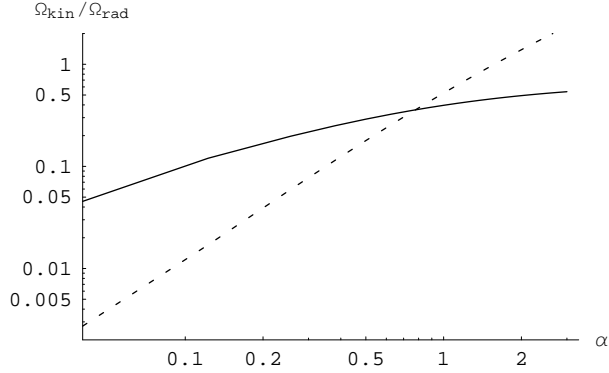}
\caption{\label{comparaison1}
Comparison of $\alpha\kappa(\alpha)$ defined in
Ref.~\cite{Kamionkowski:1993fg} (dashed line) with ${\Om_{\rm
    kin}^*}/{\Om_{\rm rad}^*}$ (solid line), where $v_b(\alpha)$ is
given by Eq.~(\ref{vbdet}). In the case of Jouguet detonations, both
parameters are fully determined in terms of $\alpha$. We remind that
$\alpha \kappa(\alpha)$ and ${\Om_{\rm kin}^*}/{\Om_{\rm rad}^*}$ are
not defined exactly in the same way (see text), but they reflect the
same physical quantity.}  
\end{center}
\end{figure}

\begin{figure}[htb!]
\begin{center}
\includegraphics[height=5cm]{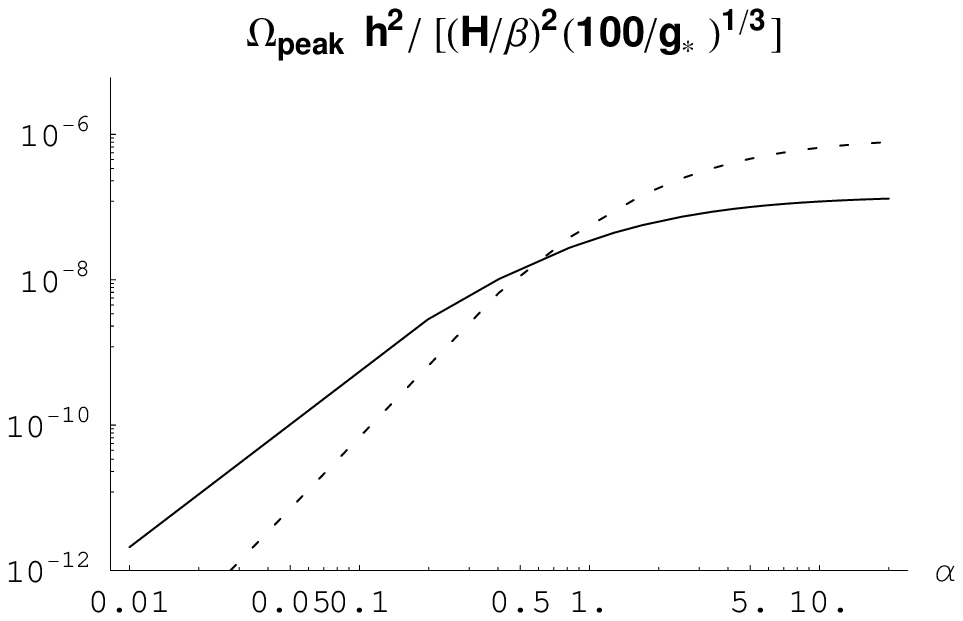}
\includegraphics[height=5cm]{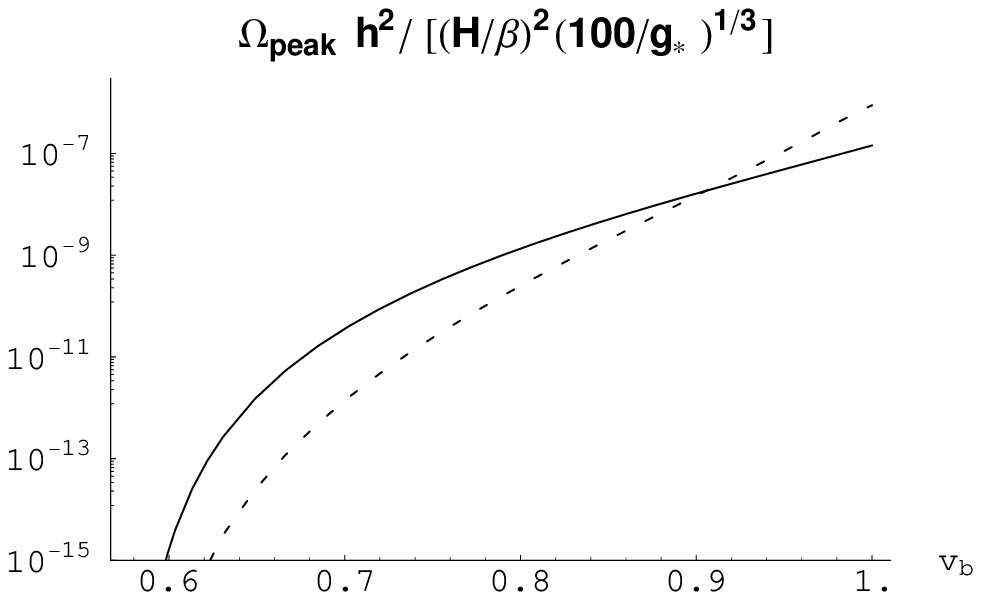}
\caption{\label{comparaison2}
The  case of Jouguet detonation. Comparison of the peak amplitude used in
Ref.~\cite{Nicolis:2003tg} (dashed line, Eq.~(\ref{nicolisexpression}) ) with our value (solid line, Eq.~(\ref{ourexpression}) ) as a function of $\alpha$ in the left hand panel and as a function of $v_b$ in the right hand panel.  }
\end{center}
\end{figure}

More precisely, from Eqs.~(\ref{amplitude},\ref{amplitudedef}) we get
the following result: 
\bea
h^2\Omega_{\mbox{\tiny peak}}& \simeq&
\frac{3}{2\pi^3}\left(\frac{g_0}{g_*}\right)^{\frac{1}{3}}\Om_{\rm
  rad}h^2 \left(\frac{\Om_{\rm kin}^*}{\Om_{\rm rad}^*}\right)^2
\left(\frac{\mathcal{H}_*}{\tilde{\beta}}\right)^2
\frac{(1-s^3)^2}{s^4}\, 0.084
\eea
where 0.084 is replaced by 0.050 for deflagrations, and
\bea
\frac{\Om_{\rm kin}^*}{\Om_{\rm
    rad}^*}&=&\frac{4}{3}\frac{(s\vf)^2}{1-(s\vf)^2} 
\eea
with $s=\rint/R$:
\bea
&s=\frac{c_s}{\vb} \ , \ \vf=\frac{\vb-c_s}{1-\vb c_s} \ \
v_{\rm b} =\frac{\frac{1}{\sqrt{3}} +\sqrt{\al^2+2\al/3}}{1+\al} \quad
&\mbox{for Jouguet detonations,}  \label{vba} \\
&s =\frac{\vb}{v_{\rm shock}}   &\mbox{for deflagrations.}
\eea
There is no  simple analytic relation between $\vb$, $\vf$ and $v_{\rm
  shock}$ in the case of deflagrations. For detonations, $\vb(\alpha)$
is given above, taken from
Refs.~\cite{Steinhardt,Kamionkowski:1993fg}. Using $g_0=3.36$ and
$\Om_{\rm rad}h^2=4.2 \times 10^{-5}$ we find the peak amplitude:

\be
h^2\Omega_{\mbox{\tiny peak}}
\simeq  5.4 \times 10^{-8} \  \frac{(1-s^3)^2}{s^4} \
 \left(\frac{\Om_{\rm kin}^*}{\Om_{\rm rad}^*}\right)^2\  
\left(\frac{{H}_*}{\beta}\right)^2  \  
\left(\frac{100}{g_*}\right)^{\frac{1}{3}} ~,
\ee
\be
\label{ourexpression}
\boxed{
h^2\Omega_{\mbox{\tiny peak}}
\simeq  9.8 \times 10^{-8} \ \vf^4 \
 \frac{(1-s^3)^2}{(1-s^2 \vf^2)^4} \
 \left(\frac{{H}_*}{\beta}\right)^2  
     \  \left(\frac{100}{g_*}\right)^{\frac{1}{3}}~,}
\ee
where 9.8 is replaced by 5.8 for deflagrations. 
The signal vanishes if the fluid velocity is zero. It  also vanishes
when the shell's thickness is zero ($s=1$), this happens if $\vb=c_s$
(corresponding to $\vf=0$) for detonations and for $\vb=v_{\rm shock}$
for deflagrations. 
Our result is to be compared with the expression reported in Eq.~(5)
of Ref.~\cite{Nicolis:2003tg}, which is only valid for Jouguet
detonations: 
\bea
\label{nicolisexpression}
h^2\Omega_{\mbox{\tiny peak}}
&\simeq& \frac{1.1 \times 10^{-6}}{(1+\alpha)^2} \  \frac{v_b^3}{0.24+v_b^3} \
 \al^2 \ \kappa^2
   \left(\frac{{H}_*}{\beta}\right)^2  \  
\left(\frac{100}{g_*}\right)^{\frac{1}{3}}
\eea
Here $\kappa(\al)=(0.715\alpha+4(3\alpha/2)^{1/2}/27)/(1+0.175\alpha)$
is the parameter defined in Eq.~(22) of
Ref.~\cite{Kamionkowski:1993fg}, denoting the fraction of vacuum
energy which goes into kinetic energy of the fluid (rather than
thermal energy). Therefore, the combination $ \alpha \ \kappa$ is
equivalent in our notation to the parameter ${\Om_{\rm
    kin}^*}/{\Om_{\rm rad}^*}= {\rho_{\rm kin}}/{\rho_{\rm rad}}$. The
above expression for $\kappa(\al)$ is a fitting formula that the
authors of \cite{Kamionkowski:1993fg} determined by integrating
numerically the energy momentum tensor of the velocity profile
corresponding to a Jouguet detonation, for different values of
$\alpha$. On the other hand, we have defined ${\Om_{\rm
    kin}^*}/{\Om_{\rm rad}^*}$ in terms of the fluid velocity at the
inner boundary of the velocity shell, starting from the approximated
velocity profile given Eq.~(\ref{velocity}). As it should be, the
dependence on $\al$ of the two parameters is comparable, see
Fig.~\ref{comparaison1} where we used $s={c_s}/{\vb}$, $
\vf=({\vb-c_s})/({1-\vb c_s})$ and
$\vb(\alpha)=(c_s+\sqrt{\alpha^2+2\alpha/3})/(1+\alpha)$. In
Fig.~\ref{comparaison2}, we show the comparison between our  peak
amplitude (\ref{ourexpression}) and the result used in
\cite{Nicolis:2003tg} (Eq.~\ref{nicolisexpression}). 
\\
\begin{figure}[htb!]
\begin{center}
\includegraphics[height=6cm]{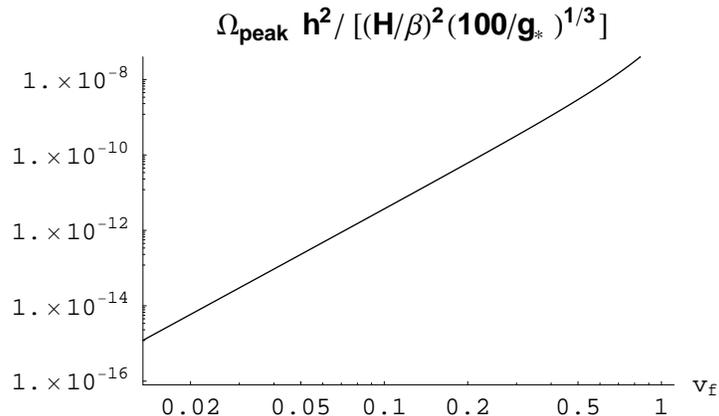}
\caption{\label{Total}
The GW signal at the peak frequency, given in
Eq.~(\ref{ourexpression}), as a function of the maximal value of the
fluid velocity $\vf$, for fixed $s=0.68$. One can have $\vf<c_s$ both
for deflagrations and detonations, while $\vf>c_s$ is possible only
for detonations. One should keep in mind that in principle once $s$ is
fixed, $\vf$ is not a free parameter. In the case of Jouguet detonations
its value is well-known,  $\vf=c_s(1-s)/(s-c_s^2)$; for deflagrations
there is no analytic formula, and in order to derive $\vf$ one needs
to know $\vb$ as well as $v_1$, the incoming fluid velocity in the
frame of the bubble discontinuity ({\it c.f.} section
\ref{velocity_profile}). Nonetheless, the aim of this figure is just
to show the order of magnitude of the GW signal when the fluid
velocity is taken as a free parameter. This gives a reasonable
estimate, since the dependence on $s$ is small, {\it c.f.}
Fig.~\ref{fig14}. } 
\end{center}
\end{figure}

In summary, our amplitude for the signal is comparable (but can differ
by one order of magnitude for some values of the parameters) to the
one obtained with an inherently different method, that of numerical
simulations in the envelope approximation
\cite{Kosowsky:1992rz,Kosowsky:1991ua,Kosowsky:1992vn,Kamionkowski:1993fg}.
This confirms that the details of the collision's modeling are not so
crucial and what really matters at the end is the size of the
velocities involved in the process. Note also  that we computed the GW
energy density spectrum without using the Weinberg formula {\it i.e.}
without making the wave zone approximation. The latter assumes that
the observer is at a distance much larger than the dimension of the
source, while our source is spread over the entire universe. 

Kamionkowski {\it et al.} obtain a non-vanishing signal in the limit of
vanishing thickness. This can be understood as follows.
In Ref.~\cite{Kosowsky:1991ua,Kosowsky:1992rz,Kosowsky:1992vn}, they
first study bubble collisions taking place in vacuum, so that the
source of GW was the energy momentum tensor of the scalar field: being
given by the spatial gradient of the scalar field, this is non-zero
only at the bubble wall.
In their later work Ref.~\cite{Kamionkowski:1993fg}, they consider
bubble collisions in a thermal bath and in this case they use as the
GW source the energy-momentum tensor of the relativistic fluid rather
than that of the scalar field. Nevertheless, they keep using the
envelope approximation. This is why they obtain a large signal even if
the shell of fluid velocity has vanishing thickness. In contrast, by
construction, there is no signal in this limit in our model, meaning 
that the kinetic energy of the fluid (which is our only source as we did 
not include the gradient energy of the scalar field) is different from zero
only over a finite volume. On the other hand, we can extrapolate our
results to the deflagration regime. Typically, for the detonation
case, in the limit $\vb\rightarrow c_s$
Ref.~\cite{Kamionkowski:1993fg} finds a non-zero GW signal, since in
the envelope approximation this is just the lower bound of the bubble
wall velocity. Within our model, in this limit, the thickness of the
non-zero velocity shell goes to zero and the GW signal as well;
however, this only shows the break down of the detonation regime, and
the necessity of treating the problem in the deflagration approach.

In Figure~\ref{Total}, we plot the peak amplitude as a function  of
the maximal fluid velocity $\vf$. In contrast with the case of Jouguet
detonations, for deflagrations, we cannot express the signal as a
function of $\alpha$ only.  Indeed, for small velocities,  the
relation between $\alpha$ and the bubble velocity will depend on the
interactions between the bubble wall (the Higgs field) and the
particles in the thermal bath. In any given model of a first-order phase
transition, one can in principle compute the bubble wall velocity and
the consequent fluid velocity profile, see {\it e.g.}
Ref.~\cite{Moore:2000wx} for the case of a weakly first-order EW phase
transition, and Ref.~\cite{Bodeker:2004ws} for the strongly first-order phase
transition presented in Ref.~\cite{Grojean:2004xa}. Provided that the
released latent heat is large, one can obtain a large bubble wall velocity.
In addition, one would have to look  carefully at the physics of
deflagrations to derive the relation between the fluid velocity and
that of the bubble wall, see for instance Ref.~\cite{Ignatius}.

To conclude, our main new results can be summarized as follows:
\begin{itemize}
\item Our description applies to both detonation and deflagration (the
  fluid velocity is non-zero over a finite volume rather than on an
  infinitely thin wall and $v \neq 1$). 
\item Our peak frequency is parametrically larger ($\propto 1/v$). 
\item We provide an analytic expression for the shape of the spectrum.
\end{itemize}

The possibility that the signals discussed here, if they are produced 
at the electroweak phase transition, could be detected with LISA will be 
discussed in an upcoming publication~\cite{CDS2}. We confirm that the 
GW signal coming just from bubble
collisions is observable only for very large fluid velocities.
Indeed, LISA's best sensitivity is not much below $\Omega h^2\sim 10^{-12}$.
Such a value can be reached  if $\beta/H_*\sim 10$  and $\vf \sim
0.2$. More realistic values  $\beta/H_*\sim 10^2$ would require
$\vf\sim 0.5$ in order to lead to an observable signal. As mentioned
in the introduction and
as will be presented in \cite{CDS2}, there are other sources (magnetic field and turbulence) of GW
during phase transitions and the signal from bubble collisions is just one contribution.

\section*{Note added}
Our results  for the position of the peak and the slope of the spectrum at large 
frequencies strongly depend on the time structure of the anisotropic stress. In our modelization, it grows with time and is suddenly switched off at the end of the transition. This discontinuity is somehow unphysical. Indeed, our Eq.~\ref{prob} does not take into account the fact that towards the end of the transition, there is no more intersecting region but only one single bubble. 
As discussed in details in \cite{CDKS}, if the source is switched off smoothly and we consider the coherent case as indicated by numerical simulations \cite{Huber:2008hg} and if the relevant correlation length is taken to be the size of the uncollided region rather than the bubble size, then
the peak position does not depend on the velocity and is given by $k_{\tiny{peak}}\sim \beta$. 
Besides, numerical simulations \cite{Huber:2008hg} which are carried out under the thin-wall approximation also indicate that the slope of the spectrum at high frequencies typically scales as $1/k$.

\section*{Acknowledgments}
We thank Riccardo Sturani for reading the manuscript.
CC and RD acknowledge support by the Swiss National Science Foundation.

\appendix
\section{One single bubble}
\label{Appen:onesinglebubble}

If there was only one bubble, there would not be any gravitational radiation since a single bubble
corresponds to a spherically symmetric distribution of energy and
momentum which cannot emit gravitational waves. In this Appendix we show
that the anisotropic stress $\Sigma_{ij}$ from a single bubble is purely scalar:
this means that there exists a function $f$ such that
\be
\Sigma_{ij}=T_{ij}-\frac{1}{3}T\de_{ij} = (\dd_j\dd_j-\frac{1}{3}\de_{ij}\De)f  \label{ea:Pif}~,
\ee
where $T_{ij}$ denotes the energy momentum tensor of a bubble.
Such a scalar component is projected out by the tensor projection
operator which is given in Fourier space in Eq.~(\ref{projected}), and does
therefore not contribute.
In this case, from the above equation the function $f$ must satisfy the condition
\be
\De^2f = \frac{3}{2}\dd_j\dd_i\Sigma^{ij}~.
\ee
We now demonstrate that this condition is always satisfied for a single, spherically symmetric bubble.
For one spherically symmetric bubble we have, up to the constant enthalpy
$\rho+p$,
\bea
\Sigma_{ij} & =& v_i(r)v_j(r) -\frac{1}{3}\de_{ij}v^2(r) \qquad \mbox{ so
  that} \label{ea:Piv}\\
\dd_j\dd_i\Sigma^{ij} &=& \dd_i\dd_j\left[ v_i(r)v_j(r)
  -\frac{1}{3}\de_{ij}v^2(r)\right] \\
  &=& \frac{4}{3}\left[ vv''+v^{\prime 2} +5\frac{vv'}{r}
  +\frac{3}{2r^2}v^2\right]  \\
 &=&\left(\dd_r +\frac{2}{r}\right)\left(\frac{4}{3}v'v
+\frac{2}{r}v^2\right)  ~.
\eea
For the third equal sign we have used spherical symmetry, we work
with the Ansatz $\bv = v(r){\mathbf e}_r$, and the prime denotes the
derivative w.r.t.~$r$.
On the other hand, for a spherically symmetric function $f$ we have
$$ \De^2f =\left(\dd_r +\frac{2}{r}\right)\left(f'' +\frac{2}{r}
f'\right)'~.$$
Therefore, if we find a function $f$ which satisfies
$$ \frac{4}{3}v'v +\frac{2}{r}v^2 = \frac{2}{3}\left(f'' +\frac{2}{r}
   f'\right)' ~,$$
or, equivalently
\be\label{ea:fv}
 f''-\frac{1}{r}f' = r\left(\frac{f'}{r}\right)' = v^2~,
\ee
the anisotropic stress (\ref{ea:Piv}) is equal to the expression given
   in Eq.~(\ref{ea:Pif}), therefore, it is a pure scalar. This is indeed the case, since
   Eq.~(\ref{ea:fv}) is an ordinary linear differential equation
which always has a solution for a given radial velocity $v$.

Hence a single spherically symmetric bubble only generates scalar
perturbations and does not contribute to the tensorial part of the energy momentum tensor
$\Pi_{ij}$, which is the source of gravity waves. This is of
course not surprising given that spherically symmetric configurations only
have scalar degrees of freedom. Nevertheless, we have added this
brief calculation here since one might
draw wrong conclusions from the fact that the anisotropic stress
of a single, spherically symmetric  bubble does not vanish.

Another way to arrive at the same result is to show that the tensor
projection operator given in Eq.~(\ref{projected}),
\bea M_{ijkl} &\equiv& P_{ik}P_{jl} -\frac{1}{2}P_{ij}P_{kl} \qquad
\mbox{ with}\\
P_{ik} &\equiv&   \De^{-1}(\De\de_{ik}-\dd_i\dd_k)
\eea
vanishes when applied to a spherically symmetric stress tensor.
For a single bubble one has then
$$  \Pi_{ij}= M_{ijkl}T_{kl}= M_{ikjl}\left(v_k(r)v_l(r)\right) = 0~. $$

\section{Calculation of the velocity field power spectrum}
\label{Appen:velocity}

In this appendix we explain how to evaluate the velocity correlation function Eq.~(\ref{correlationintegral}) and Fourier transform it to obtain the velocity power spectrum Eq.~(\ref{finalspectrum}). The quantity we need to calculate is the tensor
\be
I_{ij}(r,R,\rint)=\int_{V_i}d^3x_0 (\bx-\bx_0)_i (\by-\bx_0)_j
\label{intapp}
\ee
The intersection volume $V_i$ varies with the distance between $\bx$ and $\by$, as shown in Fig.~\ref{fig2}, with $r=|\bx-\by|$. We choose an orthonormal basis with $\hat{e}_2\parallel \widehat{\bx-\by}$. One identifies four different regions: setting $a=|\bx-\by|/2=r/2$, they are given by the limiting values
\bea
&&0\leq a \leq \frac{R-\rint}{2} \,,\hspace{2cm} \frac{R-\rint}{2}\leq a \leq \rint \,, \nonumber\\
&&\rint\leq a \leq \frac{R+\rint}{2}\,,\hspace{2cm} \frac{R+\rint}{2}\leq a \leq R \,.
\eea
The intersection volume is symmetric under rotations around $\hat{e}_2$, so in order to perform the integral in (\ref{intapp}) we choose cylindrical coordinates with $z\parallel \hat{e}_2$ and $\rho\parallel \hat{e}_3$. We have
\be
d^3x_0=\rho \,d\rho\, dz \,d\vph\,,~~~\bx_0=(\rho\cos(\vph), z, \rho\sin(\vph))\,,~~~
\bx=(0,-a,0)\,,~~~\by=(0,a,0)\,.
\ee
Substituting the above formulas in (\ref{intapp}) and performing the integration in $d\vph $, one sees that, because of cylindrical symmetry, the tensor $I_{ij}$ is diagonal. Evaluating (\ref{intapp}) reduces simply to calculate the two integrals
\be
I_{11}=I_{33}=\pi \int_{V_i} dz \,d\rho \,\rho^3~,~~~~~~I_{22}=2\pi \int_{V_i} dz \,d\rho \,\rho\,(z^2-a^2)~.
\ee
The limits of integration, here generically denoted with $V_i$, depend in fact on the variable $a$. As an example, in the region $0\leq a \leq (R-\rint)/2$, the first integral becomes:
\bea
& &I_{11}=\pi \left[ \int_{-z_2}^{-z_1}dz \int_{0}^{\rho_1} d\rho \,\rho^3 +
\int_{-z_1}^{0}dz \int_{\rho_2}^{\rho_1} d\rho \,\rho^3 +
\int_{0}^{z_1}dz \int_{\rho_3}^{\rho_4} d\rho \,\rho^3 +
\int_{z_1}^{z_2}dz \int_{0}^{\rho_4} d\rho \,\rho^3\right]\,,\nonumber \\
& &z_1=\rint+a\,,~~z_2=R-a \,,\nonumber \\
& &\rho_1=\sqrt{R^2-(a-z)^2}\,,~~\rho_2=\sqrt{\rint^2-(a+z)^2}\,,~~
\rho_3=\sqrt{\rint^2-(z-a)^2}\,, \nonumber\\
& & \rho_4=\sqrt{R^2-(a+z)^2}\,. \nonumber
\eea
Analogous expressions can be found for the remaining three regions, and similarly for $I_{22}$. The final result of the integrations are the two continuous functions $I_{11}(a,\rint, R)$ and $I_{22}(a,\rint,R)$, as a function of the variable $0\leq a \leq R$, which are too long expressions to be written explicitly here.

Knowing $I_{11}=I_{33}$ and $I_{22}$, we impose the condition of statistical homogeneity and isotropy for the velocity field, meaning that we impose the tensorial structure
\be
I_{ij} \equiv f\,\de_{ij}+g \,\hat{r}_i\hat{r}_j=I_{11}\,\de_{ij}+(I_{22}-I_{11})\,\hat{r}_i\hat{r}_j
\ee
where the second equality is a straightforward consequence of our choice $\hat{r}=\widehat{\bx-\by}\parallel \hat{e}_2$. The two-point correlation function for the velocity field takes the final form ({\it cf.} Eq.~\ref{vcorrfunct})
\be
\langle v_{i}(\bx,t)  v_{j}(\by,t) \rangle=\phi(t)\,\frac{\vf^2}{R^2}\, \left[ \frac{I_{11}}{V_c}\de_{ij}
+\frac{I_{22}-I_{11}}{V_c}\,\hat{r}_i\hat{r}_j \right]
\ee
where the pre-factor is independent of $a=r/2$ and the volume $V_c$ is
given in Eq.~(\ref{Vc}).

In order to know the power spectrum, we have to Fourier transform the above equation with respect to the variable $r$. Because of homogeneity and isotropy, we get a delta function in momentum. We rewrite the Fourier transform of the second term in the sum, which is direction dependent, in terms of derivatives with respect to the wave vector, and we obtain:
\be
\langle v_{i}(\bk,t)  v_{j}^*(\bq,t) \rangle =
\de(\bk-\bq)\phi(t)\frac{\vf^2}{R^2}\left[
\mathcal{F}\left(\frac{I_{11}}{V_c}\right)\delta_{ij}-\partial_{k_i}\partial_{k_j} \mathcal{F}\left(\frac{I_{22}-I_{11}}{r^2V_c}\right)\right]
\ee
The Fourier transform of a function only of $r$ gives a function only of wave number $k$. Knowing this,
we can re-express the partial derivatives, to obtain expression (\ref{powv}) and followings. The Fourier integrals, as for example
\be
\frac{4\pi}{k}\int_0^\infty dr\,r\,\sin(kr) \frac{I_{11}(r,\rint,R)}{V_c(r)}
\ee
need to be further divided in the sum over the four integrals corresponding to the regions described above, depending on the value of $r$, since in each of these regions the integrand takes a different form. Again, we do not write the complicated full expression of the result, for which we found the fitting formulas (\ref{Ak}) and (\ref{Bk}) shown in Fig.~\ref{fig3}.

\section{Large and small scale part of the GW power spectrum}
\label{Appen:largeandsmall}

We have seen that the gravitational wave power spectrum, independently of
the different time approximations, always grows like $k^3$ at scales
larger than the peak scale $k_{\rm peak}^{-1}\simeq R(\eta_{\rm
  fin})/4.5$. The reason for this general behaviour is  that the source
is uncorrelated at these scales: therefore, the power spectrum of the
anisotropic stress source is simply  the incoherent sum of
uncorrelated regions, and is white noise. The white noise behaviour
for the anisotropic stress in turn determines the $k^3$ increase for
the GW power spectrum.

On the other hand, we also saw that for the
small scale part of the spectrum we always recover roughly a $k^{-2}$
decrease, with the only exception of the completely incoherent case
({\it cf.} Eqs.~(\ref{Pitotin},\ref{psGWinco})). The small scale
decrease can be understood from dimensional arguments. In this
Appendix we present  general arguments for the origin of the above
mentioned power laws for the large and small scale part of the GW
power spectrum.

Let us first concentrate on the large scale limit. We start with a
generic velocity power spectrum showing a peak at a characteristic
scale $k=L^{-1}$. The scale $L$, corresponding in our case to the bubble diameter
$L=2R$, may depend on time. We assume that the large and small scale behaviors are given
by two power laws,
\bea
P(k)\propto v^2 L^3 \left\{\begin{array}{ll}
(Lk)^n & {\rm for}~Lk < 1 ~, \\
(Lk)^m & {\rm for}~Lk > 1 ~, \end{array}\right.
\eea
satisfying the conditions $n>-3$, $m<-3$ so that the energy density is
dominated by the contribution at the peak. The pre-factor $v^2$
denotes the average energy per unit enthalpy of the source. For bubbles,
we have $n=0$ for the function $A(k)$ and $n=2$ for  $B(k)$, {\it cf.}
Eqs.~(\ref{Ak},\ref{Bk}) and the discussion thereafter (we will treat
the small scale decrease of these functions later on). The anisotropic
stress power spectrum is given by the convolution of the velocity
power spectrum ({\it cf.} Eq.~\ref{Pik}). Setting $Q=qL$ and $K=kL$,
and neglecting angular
dependencies, we have
\be
\Pi(k)\propto\int_0^\infty dq\,q^2\, P(|\bk-\bq|)P(q)\propto
v^4L^3\left[ \int_0^1 dQ Q^{n+2} P(|{\bf K}-{\bf Q}|)+
\int_1^\infty dQ Q^{m+2} P(|{\bf K}-{\bf Q}|)\right]\,.
\label{Piappend}
\ee
In the large scale limit $k \ll L^{-1}$, in the second integral we can
safely neglect ${\bf K}$ with respect to ${\bf Q}$ and simply set
$P(|{\bf K}-{\bf Q}|)\simeq Q^m$. In the first integral we have to be
a little more careful. If $n<-3/2$, the main contribution to the
integral comes from the divergence at ${\bf Q}\rightarrow{\bf K}$ and
the integral picks a behaviour $K^{2n+3}$. In our physical bubble
situation, we have $n=0$, $n=2$; we therefore choose to analyze only the
case $n\geq -3/2$, for which we find:
\be
\Pi(k\rightarrow 0)\propto v^4L^3\left[ \int_0^{1} dQ \,Q^{2n+2}
  +\int_1^{\infty} dQ\, Q^{2m+2} \right]
= v^4L^3\left( \frac{1}{2n+3}-\frac{1}{2m+3}\right)\,.
\ee
Therefore, the anisotropic stress at large scales is white
noise. Going back to Eqs.~(\ref{GWenergy}) and (\ref{h'2}), we see
that for a constant anisotropic stress power spectrum, the GW power
spectrum behaves like
\be
\left. \frac{d\Om(k)}{d\ln(k)}\right|_{k\rightarrow 0}\propto k^5
|h'(k)|^2\propto k^3 \Pi(k)
\ee
Here we have re-expressed the conformal time derivative appearing in
Eq.~(\ref{GWenergy}) in terms of a derivative with respect to
$x=k\eta$, and we have taken into account that converting the double
integral over the Green function in Eq.~(\ref{h'2}) into a double
integral with respect to conformal time, induces an additional factor $k^2$.
We therefore recover the observed large scale behavior  $\propto k^3$.

We now turn to the small scale limit, $k\gg L^{-1}$. In this case, our
fits to the functions $A(k)$ and $B(k)$ decrease like $k^{-4}$. As
argued in Section~\ref{sec3.3}, this
power law is transferred to the small scale behaviour of
the anisotropic stress power spectrum, {\it cf.}
Eqs.~(\ref{Pint},\ref{alk}), which takes the form
\be
\langle \Pi_{ij}(\bk,\tau) \Pi_{ij}^*(\bq,\tau)\rangle \propto
\de(\bk-\bq) R(\tau)^3 {\cal I}(K(\tau))
\ee
where ${\cal I}(K)$ decreases  like $K^{-4}$ for $K\gg 1$, and we
neglect all the other time dependent factors, which are irrelevant for
the argument presented here. Inserting the above expression in
Eq.~(\ref{h'2}), one finds
\be
|h'(k)|^2\propto \frac{1}{k^4} \int dy \int dz\, R^3(\tau) \cos(y-z)
{\cal I}(K(\tau)) \propto \frac{1}{k^7} \int dy \int dz \,y^3
\cos(y-z) {\cal I}(y) ~,
\ee
where for the second equality we have used the fact that
$R(\tau)\propto \tau$. Substituting the above formula into
Eq.~(\ref{GWenergy}) rewritten in terms of the derivative with respect
to $x$, one finally obtains
\be \label{120}
\left. \frac{d\Om(k)}{d\ln(k)}\right|_{k\gg L^{-1}}\propto k^5
|h'(k)|^2\propto \frac{1}{k^2} \int dy \int dz \,y^3
\cos(y-z) {\cal I}(y) ~,
\ee
where the double integral simply causes a modulation in the
spectrum. We recover, in the large wave number limit, the $k^{-2}$
decrease, which is a simple consequence  of
dimensional analysis since ${\cal I}$ is a function of $y=k\eta$ only
in the relevant regime. This power law usually acquires small
corrections due to the fact that the integral in Eq.~(\ref{120}) is
not completely independent of $k$. 
Depending on the approximation for the unequal time anisotropic stress power spectrum, we
actually found  power laws  $k^{-\beta}$ with $\beta$ in the range
$1.8<\beta< 2.2$. Within our approximation, these may well also be
logarithmic corrections to the slope $\beta=2$.

In the case of the totally incoherent unequal
time approximation, the argument is changed because of the delta
function in time: the form of the anisotropic stress power spectrum is
now ({\it cf} Eq.~\ref{Pitotin})
\be
\langle \Pi_{ij}(\bk,\tau) \Pi_{ij}^*(\bq,\zeta)\rangle \propto
\de(\bk-\bq) \de(\tau-\zeta) R(\tau)^2 {\cal I}(K(\tau))~.
\ee
Repeating the above argument, one easily finds that the large wave
number power law is now changed to $k^{-1}$ ($k^{-0.8}$ with the corrections from the integral).

\end{document}